\documentclass[a4paper,11pt]{article}
\pdfoutput=1 

\usepackage{xcolor}
\usepackage{ulem}

\usepackage{jinstpub} 
\usepackage{subfigure}
\title{Particle Identification at MeV Energies in JUNO}

\author[b,c]{L.~Ludhova}
\author[a,1]{H.~Rebber\note{Corresponding authors.}}
\author[a]{B.~Wonsak}
\author[b,c,1]{Y.~Xu}

\affiliation[a]{University of Hamburg, Institute of Experimental Physics, \\ Luruper Chaussee 149, 22761 Hamburg, Germany}
\affiliation[b]{Forschungszentrum J\"ulich IKP, \\ Wilhelm-Johnen-Strasse, D-52428 J\"ulich, Germany}
\affiliation[c]{III. Physikalisches Institut B, RWTH Aachen University, \\ Aachen, Germany}

\emailAdd{henning.rebber@desy.de}
\emailAdd{y.xu@fz-juelich.de}

\abstract{
JUNO is a multi-purpose neutrino experiment currently under construction in Jiangmen, China. 
It is primary aiming to determine the neutrino mass ordering. 
Moreover, its 20\,kt target mass makes it an ideal detector to study neutrinos from various sources, including nuclear reactors, the Earth and its atmosphere, the Sun, and even supernovae. 
Due to the small cross section of neutrino interactions, the event rate of neutrino experiments is limited. 
In order to maximize the signal-to-noise ratio, it is extremely important to control the background levels. 
In this paper we discuss the potential of particle identification in JUNO, its underlying principles and possible areas of application in the experiment.
While the presented concepts can be transferred to any large liquid scintillator detector, our methods are evaluated specifically for JUNO and the results are mainly driven by its high optical photon yield of 1,200 photo electrons per MeV of deposited energy.
In order to investigate the potential of event discrimination, several event pairings are analysed, i.e.~$\alpha/\beta$, $p/\beta$, $e^+/e^-$, and $e^-/\gamma$.
We compare the discrimination performance of advanced analytical techniques based on neural networks and on the topological event reconstruction keeping the standard Gatti filter as a reference. 
We use the Monte Carlo samples generated in the physically motivated energy intervals. 
We study the dependence of our cuts on energy, radial position, PMT time resolution, and dark noise. 
The results show an excellent performance for $\alpha/\beta$ and $p/\beta$ with the Gatti method and the neural network.
Furthermore, $e^+/e^-$ and $e^-/\gamma$ can partly be distinguished by means of neural network and topological reconstruction on a statistical basis. 
Especially in the latter case, the topological method proved very successful.
}


\usepackage{natbib}
\usepackage{graphicx}

\begin{document}
\flushbottom
\maketitle

\section{Introduction}

Liquid scintillator (LS) technology has a key role in the detection of low energy neutrinos. 
The almost linear relation between energy deposition and light emission enables calorimetric measurements even in the sub-MeV regime.
Present and future experiments instrument large target masses in the order of kilotons in unsegmented tanks in order to address the unsolved issues in neutrino physics, which include the neutrino mass ordering~\cite{JUNOyellowbook}, CP violation in neutrino oscillations \cite{theia19,jinping17}, and neutrinoless double beta decay~\cite{Kraus10,Gando16}.  
Furthermore, the determination of low energy neutrino fluxes offers a unique way to study energy production in the Earth and Sun, as well as the dynamics of supernovae.   

Various channels enable neutrino detection in a LS detector, e.g.

\begin{itemize}
\item inverse beta decay (IBD): $\bar{\nu}_e + p \rightarrow e^+ + n$,  
\item elastic scattering (ES) with electrons: $\nu + e \rightarrow \nu + e$,
\item elastic scattering with protons: $\nu + p \rightarrow \nu + p$. 
\end{itemize}
Identifying signal events is crucial to all neutrino studies, since background is usually dominating the event rates. 
The IBD channel features a characteristic coincidence signature.
A prompt signal arises from energy deposition of the $e^+$ and is followed after $\sim$200\,$\mu$s by a gamma emission as a consequence of the $n$ being captured onr hydrogen.
Although most backgrounds can be suppressed by time and space coincidence requirements, the $n$-accompanied $\beta^-$ decay of cosmogenic isotopes mimics the signal pattern. 
The ES channels on the other hand cause single energy depositions and thus cannot be distinguished from point-like background events due to, for example, radioactive contaminants of the construction materials. Usually, an optimization of the signal-to-background ratio can only be achieved by selecting a proper energy range as well as the so called fiducial volume, a wall-less region of the LS, defined through the reconstructed vertex.

The identification of particle type (PID) is an appealing concept since it offers an independent way for background reduction and can hence lead to an enhancement of the detector sensitivity. 
Characteristic decay sequences and event topologies, as well as differences in the scintillation processes, can affect the topology of the emitted light, and thus, provide handles for PID. 
Pulse shape analysis, investigating the temporal and/or spatial distribution of detected photons, has proven to be a powerful discrimination tool in several neutrino LS experiments.
Borexino established a reliable discrimination of $\alpha$ and $\beta$~\cite{borexinoGeo19,borexinoSeasonalMod17,PhysRevD.89.112007} and also a statistical discrimination of $e^+$ and $e^-$ events~\cite{PhysRevD.89.112007}.
Double Chooz used pulse shape analyses to tag ortho-positronium events~\cite{doubleChooz2014} and developed discriminators based on Fourier power spectra with PID sensitivity towards stopping $\mu$, $\alpha$, and $e^+$ and $e^-$ events~\cite{Abraho18}.
Furthermore, Double Chooz used PID also for the discrimination between $e^+$ and protons so as to reject fast neutron background from IBD samples~\cite{Wagner:2014zea}.
A discrimination between neutron and gamma events was successfully studied for the proposed LENA experiment~\cite{M_llenberg_2015}. 

As a general rule, the potential for PID in a LS experiment is determined by the degree of information contained in a physical event. 
Three detector-specific parameters determine the information yield to first order: these are quantity, precision, and purity of the event data.  
The first is controlled by the photo electron (p.e.) yield, which in turn depends on the absorption length of the LS, the optical coverage of the detector, and the detection efficiency of its light sensors.
The second mainly refers to timing uncertainties of the used light sensors and electronics.
The third comprises both the amount of dark noise on the sensors and and the ratio of direct photons, which reached the sensors without having been deflected by processes like Rayleigh scattering or re-emission. 

This work focuses on four general discrimination categories: $\alpha/\beta$, $p/\beta$, $e^+/e^-$, and $e^-/\gamma$. 
Our studies are based on MC simulations for the upcoming JUNO detector, making use of its extraordinarily high yield of detected photons of $\sim$1,200 per MeV of deposited energy. 
However, the concept is universal and can be adapted to similar LS based detectors with detailed time information such as SNO+, Borexino, KamLAND-Zen, and also next-generation experiments like Theia~\cite{theia19} and Jinping~\cite{jinping17}. 

The paper is structured as follows. 
In Section~\ref{sec:juno}, we shortly describe the JUNO experiment and discuss how PID can be helpful for its various physics purposes.
In Section~\ref{sec:methods}, we explain two different PID methods, which is a single parameter cut based on a topological event reconstruction on the one hand, and a pure machine learning approach on the other hand. We introduce our MC data sets and the figures of merits, which we use to evaluate the PID performance.
In Section~\ref{sec:results}, we present and discuss our results. 
Finally, we give the conclusions and outlook in Sec.~\ref{sec:conclusion}.

\section{The JUNO Experiment}\label{sec:juno}

\begin{figure}[htb]
\centering
\includegraphics[width=\textwidth]{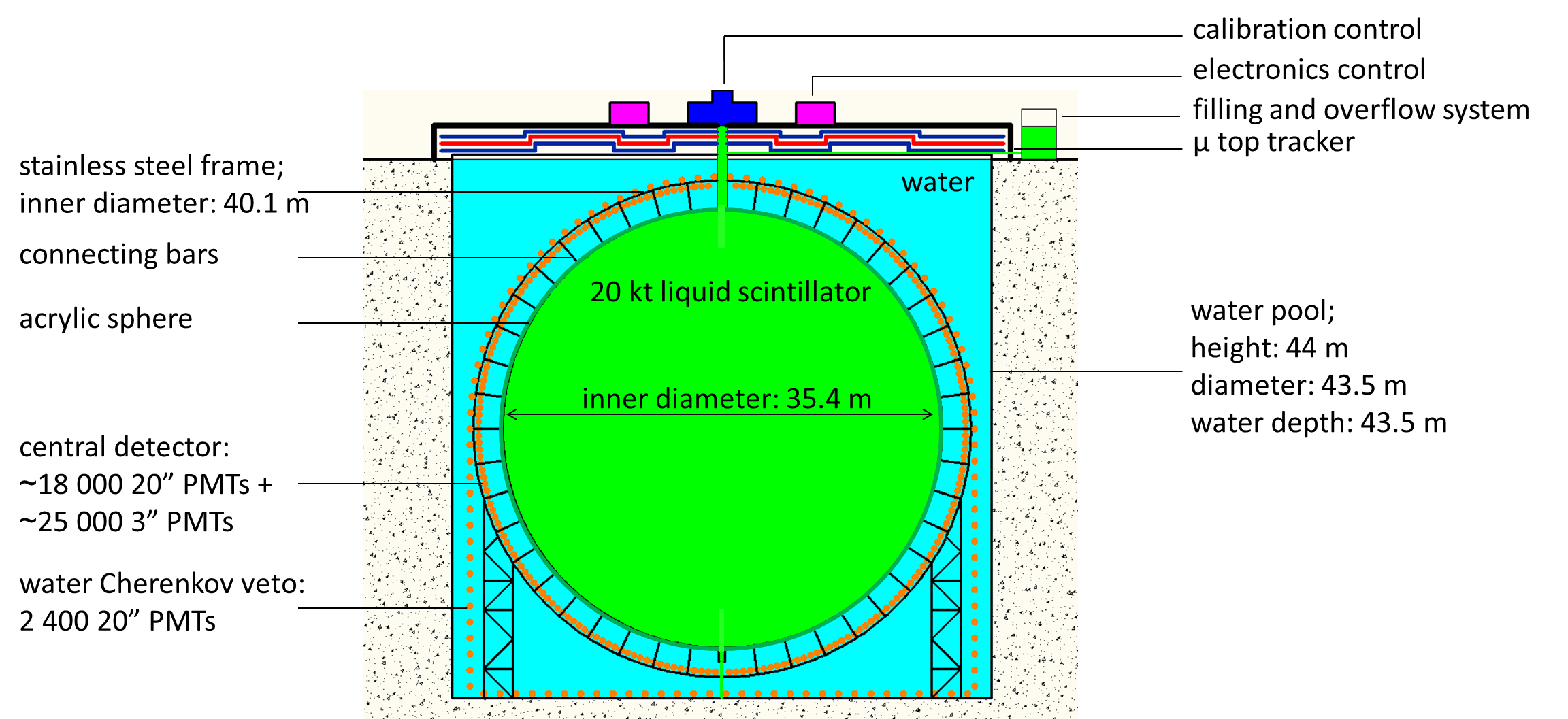}
\caption{Schematic view of the JUNO detector.}
\label{fig:JUNO_detector}
\end{figure}

The Jiangmen Underground Neutrino Observatory (JUNO)~\cite{JUNOyellowbook} is a next generation neutrino experiment currently being built $\sim$680\,m underground in Jiangmen in south China. Its large target mass and excellent energy resolution offer exciting opportunities for research in neutrino physics.
Figure~\ref{fig:JUNO_detector} shows a schematic view on the setup. 
The heart of the experiment is the central detector (CD), an acrylic sphere with a radius of 17.7\,m holding 20\,kt of LAB-based LS with admixtures of PPO and Bis-MSB. 
The characteristic length for light attenuation exceeds 20\,m in order to make up for the huge CD dimensions~\cite{JUNOyellowbook,Zhou15}.
For light detection, $\sim$18,000 large (20 inch) PMTs and $\sim$25,000 small (3 inch) PMTs facing the sphere are mounted on a surrounding stainless steel frame, adding up to a total optical coverage of 78\,\%. 
The CD is placed in a cylindrical water pool containing ultra-pure water. 
The water pool acts both as a shield against external fast neutrons and gammas and, through the equipment with another $\sim$2,000 20" PMTs, as a Cherenkov veto for cosmic muons.
Additionally, muons can be tracked precisely with an array of plastic scintillator modules placed on top of the water pool.

\begin{figure}[htb]
\centering
\includegraphics[width=0.65\textwidth]{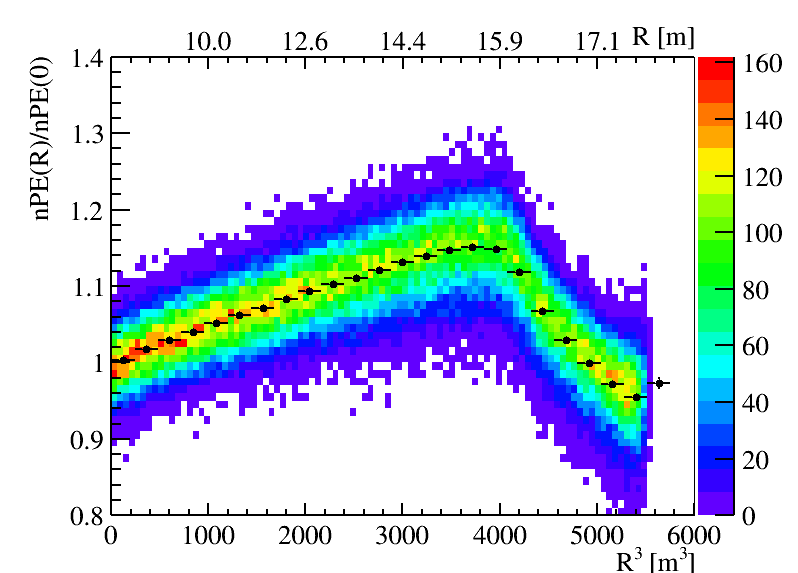}
\caption{Variation of the relative number of registered photons within the detector volume as a function of radius $R$ (top horizontal axis) and $R^3$ (bottom horizontal axis) with respect to the detectors center. The abrupt change at the radius of about 16\,m is due to the total reflection at the acrylic sphere separating the liquid scintillator and water volumes.}
\label{fig:nPEoverRadius}
\end{figure}

Two types of large PMTs are used in the CD, differing principally in the mechanism for p.e. multiplication. 
About 5,000 units feature a common dynode structure as it is used, e.g.~in Super-Kamiokande~\cite{Fukuda03}.
Their time resolution, measured as transit time spread (TTS), is 3\,ns FWHM. 
The p.e.~amplification in the remaining $\sim$13,000 units is carried out by microchannel plates (MCPs).
The different structure results in an increased TTS of 18\,ns FWHM. 
However, both types achieve high photon detection efficiencies (PDEs) of almost 30\% on average.
The mean rate of "dark", i.e.~spurious, counts (DCR) for large PMTs is expected to be 30\,kHz.\footnote{
PDE, DCR, and TTS are determined in PMT mass tests. The given values are based on intermediate testing results.}
The small PMT system builds on dynode devices with a TTS of 4.5\,ns FWHM and PDEs around 25\%. The smaller size results in low DCRs below 2\,kHz.  
In total, JUNO's CD yields the high number of at least 1,200\,p.e.~per MeV of deposited energy, depending mildly on the event location.
This behaviour can be studied in Fig.~\ref{fig:nPEoverRadius} (a), showing the relative p.e.~yield as a function of detector radius $R$. 
Light attenuation has the strongest effect in the detector center, where 1,200\,p.e.~are yielded. From here the curve rises towards the detector edge, whereas total reflection diminishes the yield in the outermost region above the peak observed around $R=16$\,m.  
The overall high p.e.~statistics, 1,300\,p.e./MeV on average, result in an unprecedented energy resolution for large LS detectors of $3\,\%/\sqrt{\textnormal{energy}/\textnormal{MeV}}$.

Background in JUNO is mostly assigned to one of the four main categories: {\it internal background} due to the decays of radioactive contaminants of LS (natural U and Th chains, $^{210}$Pb/$^{210}$Bi, $^{210}$Po, $^{14}$C,$^{85}$Kr), {\it external background} from gammas penetrating inside the LS volume from outside (e.g.~from stainless steel frame, PMTs, acryllic vessel), {\it cosmogenic background} induced by cosmic muon interactions, and in some particular cases, {\it neutrino events} from other than the envisaged sources. 
In the following, we give an overview on JUNO's various physics goals and the respective main backgrounds.
Especially, the potential contributions to background reduction with PID are pointed out.

\begin{itemize}
    \item 
    The main purpose of JUNO is the determination of \textit{neutrino mass ordering} (MO). 
    Electron antineutrinos from two nuclear power plants will be detected via IBD with visible energies from the $e^+$ component ranging from 1\,MeV to 10\,MeV. 
 Given the good energy resolution, the MO follows from the measurement of a subdominant oscillation imprinted on the $e^+$ energy spectrum. 
    With rates comparable to the MO signal, the most serious background is expected from the cosmogenic spallation products $^8$He and $^9$Li, both having the potential to undergo ($\beta^- +  n)$ decays. 
    The resulting signals coincide temporally and spatially and hence, mimic the IBD signature.
    Muon vetoes can suppress such events by the cost of roughly 15\% exposure loss~\cite{JUNOyellowbook}.
    An $e^+/e^-$ discrimination, even if not feasible on event-by-event basis, would mean valuable input for the direct measurement of $^8$He and $^9$Li production rates. 
    The latter has been carried out in KamLAND~\cite{Abe10}, Borexino~\cite{Bellini13}, Daya Bay~\cite{An17}, and Double Chooz~\cite{DC_Li9He8yield}.
    For $^8$He, only KamLAND could measure a rough yield value, while Double Chooz and Borexino provided upper limits~\cite{DC_Li9He8yield}.
    JUNO would be able to use the combined potential of very large exposure and a statistical PID. 
    The results can in turn find use in the optimised design of more efficient muon vetoes.
    
    \item 
    Although suffering from an overburden which is relatively low compared to other underground experiments such as LS-based Borexino~\cite{Agostini:2018uly} or water-based SuperKamiokande~\cite{Abe:2016nxk}, JUNO can contribute to \textit{solar neutrino} measurements. 
    Combining the large volume and high light yield, it has a large potential to observe the $^8$B solar neutrinos with decreased energy threshold and high statistics. 
    The measurement of other solar neutrino species below 2\,MeV ($^7$Be, $pep$, $pp$) will strongly depend on the internal contamination of the LS. 
    Since the signal is given by ES with target electrons, all kinds of single events in the energy range of interest represent background. 
    A highly efficient $\alpha/\beta$ discrimination is indispensable to suppress $\alpha$ decays from internal $^{210}$Po contamination. 
    The decay energy is 5.4\,MeV~\cite{atomicmasses17}. 
    $\alpha$ energies up to 10\,MeV will cause visible energies below 1.5\,MeV in JUNO due to quenching.
    Besides internal radioactivity, the external background demands a fiducial volume cut several meters deep into the CD sphere.
    A key measurement will be of $^8$B neutrinos down to $\sim$2\,MeV, since it is capable of probing the unexplored upturn region in the MSW paradigm~\cite{Wolfenstein:1977ue,Mikheev:1986gs}. 
    Here, the dominant background comes from the cosmogenic $^{10}$C and from the $\gamma$'s of external background and from neutron captures~\cite{B8Borex2019}. 
    The $e^-/\gamma$ discrimination has thus a potential to expand the exposure significantly, although high reliability is required due to the exponentially growing rate of $\gamma$ events towards the CD edge. 
    $^{10}$C undergoes $\beta^+$-decay, followed after 1\,ns by a $718\,$keV $\gamma$-transition, and is reducible by $e^+/e^-$ discrimination.
        
    \item 
    \textit{Geo-neutrinos}, $\bar{\nu}_e$ created in natural $\beta^-$ decays inside the Earth's crust and mantle, are a unique tool to asses the Earth's radiogenic heat, a key parameter to global understanding of our planet. With roughly 400 events per year, JUNO is expected to collect the world's largest sample of geo-neutrinos within one year of measurement. Since the detection channel is IBD with $e^+$ signals below 3\,MeV, reactor antineutrinos are inevitable background.
    Cosmogenic $^8$He and $^9$Li contribute as discussed above.
    Furthermore, it is known from the experience in KamLAND \cite{KamLAND_geo} that $^{13}$C$(\alpha,n)^{16}$O reactions constitute another background for IBD due to various ways to create a prompt signal, one of which is neutron elastic scattering on a proton  \cite{Zhao_2014}. $\beta/p$ discrimination can reject such events.
    $\alpha/\beta$ discrimination can further be used to tag $^{210}$Po decays in order to estimate the amount of $(\alpha,n)$ reactions. 

    \item 
    The rare occurrence of a core-collapse \textit{supernova} in our galaxy would flush JUNO with all kinds of neutrinos and antineutrinos, triggering a whole bunch of detection channels.
    Among these, IBDs will make up the highest signal rate, exceeding by far the rates from reactor antineutrinos and associated backgrounds.
    One particular channel open to all neutrino species is given by the ES off protons.
    Being singles with visible energies mainly below 1\,MeV, the signals are hard to distinguish from radioactivity background. 
    Main contributions come from the $\beta^-$ emitters $^{85}$Kr and $^{210}$Bi, and below 0.2\,MeV especially from $^{14}$C. 
    All could be rejected with $e/p$ discrimination.
    PID would further help to distinguish the signal from supernova channels like electron ES.

    \item 
    The diffuse supernova neutrino background (DSNB), a low isotropic flux of neutrinos expected from the cumulated supernova rate in our universe, has never been detected, yet.
    With its large target mass, JUNO could find between one and two DSNB events per year as IBDs~\cite{Sawatzki20}.
    High event rates of reactor antineutrinos rule out detections below 10\,MeV, while charged current interactions of atmospheric neutrinos start to dominate over DSNB above 30\,MeV. 
    In between, LENA studies~\cite{M_llenberg_2015} show that the remaining backgrounds from fast neutrons and neutral current interactions with atmospheric neutrinos can be reduced below the expected signal level with the help of PID.

\end{itemize}

\section{Methods for Particle Identification }\label{sec:methods}

\begin{table}
\centering
\begin{tabular}{|l|c|c|c|c|}
     \hline
     Particle type & Fast  & Intermediate & Slow\\
    & $\tau_1$ / $w_1$ & $\tau_2$ / $w_2$ & $\tau_3$ / $w_3$  \\
    & [ns] / [\%] & [ns] / [\%] & [ns] / [\%] \\
     \hline
     $\gamma$, $e^+$, $e^-$  & 4.93 / 79.90 & 20.6 / 17.10 & 190 / 3.00  \\
      p & 4.93 / 65.00 & 34.0 / 23.10 & 220 / 11.90 \\
     $\alpha$  & 4.93 / 65.00 & 35.0 / 22.75 & 220 / 12.25 \\
     \hline 
\end{tabular} 
\caption{Time constants $\tau_i$ and relative weights $w_i$ assumed for the three exponential contributions to the light emission curves (Eq.~\ref{eq:lsc_timedecay}) for different particle types assumed in the JUNO MC simulation.}
\label{tab:timeConstant}
\end{table}

\begin{figure}[htb]
\centering
\subfigure[] {\includegraphics[width=0.45\textwidth]{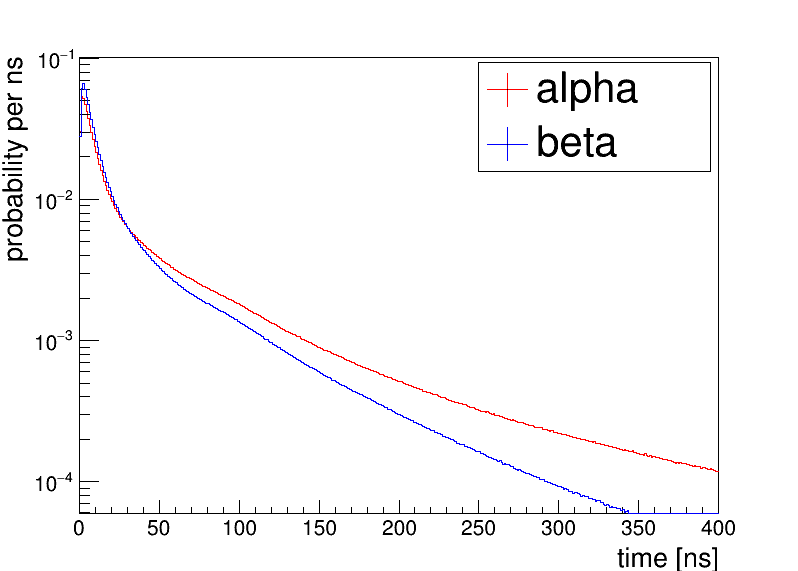}}\quad
\subfigure[] {\includegraphics[width=0.45\textwidth]{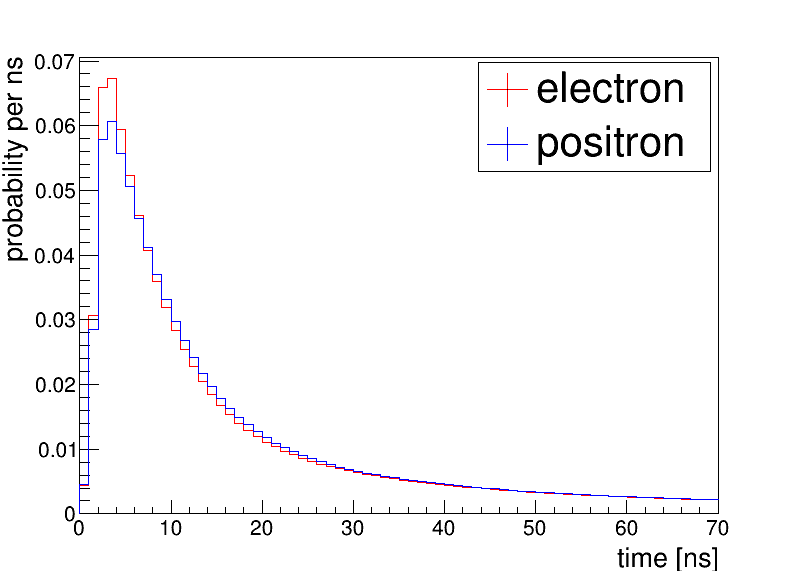}}
\caption{The MC-based time profiles of light emission expected for different particles in JUNO, based on the parameters from Table~\ref{tab:timeConstant}. 1000 normalised pulses were superimposed for each curve. TTS and dark noise were not considered. Comparison of (a): $\alpha$ and $\beta$ and (b): $e^+$ and $e^-$ time profiles.} 
\label{fig:timeProfiles}
\end{figure}

The signal formation in LS detectors is mainly induced by ionising particles causing an excitation of LS molecules along their path. 
The subsequent de-excitation goes along with isotropic light emission in the shortwave part of the optical spectrum.  
Several particle-related effects alter the detected pulse shape and can be exploited for PID. 
In the case of $\alpha/\beta$ and $p/\beta$ discriminations, the most striking difference can be traced back to the time curves representing the emission of scintillation photons. 
The general behaviour of this process can be described by a superposition of typically $n=3$ exponential decay curves:

 \begin{equation}
 \label{eq:lsc_timedecay}
 \phi_\textnormal{em} (t)
 =\sum_{i=1}^{n}\frac{w_i}{\tau_i}\textnormal{e}^{-\frac{t-t_0}{\tau_i}} 
 \qquad \textnormal{with} \qquad 
 \sum_{i=1}^{n}w_i =1,
 \end{equation}
each parametrised with a weight $w_i$ and time constant $\tau_i$.
The de-excitation of singlet states leads to a dominant fast emission component. 
Excited triplet states lose their energy mainly via non-radiative processes rather than light emission~\cite{parkerHatchard61,parkerHatchard62,Laustriat68}. 
However, interactions with excited triplet states may create further excited singlet states which then decay, leading to a suppressed and hence slower emission component. 
Furthermore, the interaction of excited singlet states with each other favours ionisation quenching~\cite{birks64}, i.e.~the light yield per unit of deposited energy is being reduced.
As a consequence, the ratio between the emission components depends strongly on the concentration of excited states. 
Particles like $\alpha$ and $p$ entail higher ionisation rates along their path and cause more quenching compared to $e^+$, $e^-$, and $\gamma$. 
Accordingly, the $w_i$ and $\tau_i$ take characteristic values for certain groups of ionising primary particles as can be seen in Table~\ref{tab:timeConstant}, which lists corresponding values expected for the JUNO LS mixture. 
The effect on the pulse shape is compared exemplarily for $\alpha$ and $\beta$ particles in Fig.~\ref{fig:timeProfiles} (a). 
Both hit time profiles are constructed as a superposition of 1000 MC simulated events with visible energies between 0.2\,MeV and 1.5\,MeV and disregarding TTS of PMTs.
The hit times were corrected by the photon time of flight (ToF) between the vertex point and PMT.
One observes considerably higher expectation for late emission times ($>$200\,ns) for $\alpha$ events. 
We note, that the time profiles for protons are expected to be very similar to those of $\alpha$'s (Table~\ref{tab:timeConstant}).   


\begin{figure}[htb]
\centering
\includegraphics[width=0.65\textwidth]{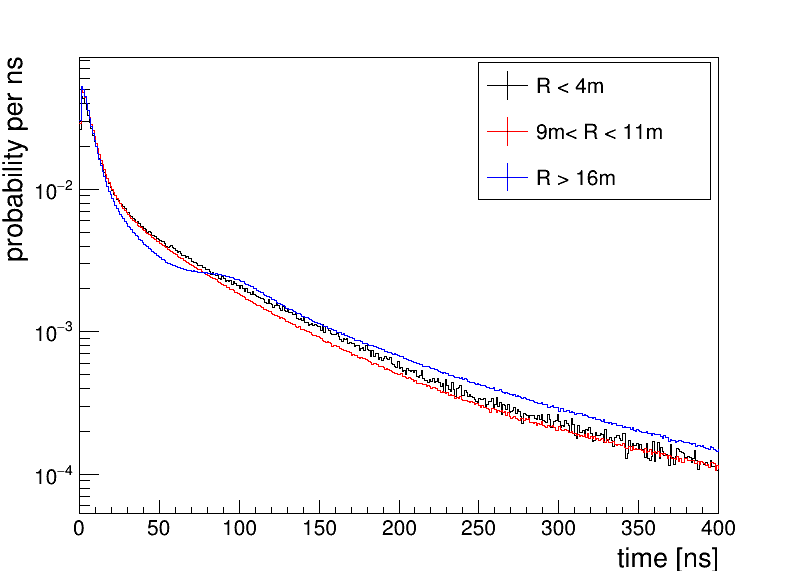}
\caption{The MC-based time profiles of light emission curves for $\alpha$'s expected at different radii in the central detector. The largest difference is observed at large radii above 16\,m, in the region of total reflection. }
\label{fig:timeoverRadius}
\end{figure}

We note that the particles' time profiles exhibit some level of radial dependence. This is demonstrated in Fig.~\ref{fig:timeoverRadius} on the example of $\alpha$s by showing the time profiles for events selected in the central part of the detector, at radii 9\,m to 11\,m, and at large radii above 16\,m. As expected, the largest change is observed for the latter case due to the total reflection. At the base of this radial dependence of the time profiles is the fact that the relative ratio between the direct and the scattered/reflected light does depend on radius in a large detector like JUNO. This property justifies also the radial dependence of the cut values applied in the particle identification methods as it will be described in the following subsections.

Regarding $e^+$ and $e^-$, weights and time constants are almost identical. PID is instead based on processes including positronium formation and positron annihilation.
While the ionisation losses per track length are almost equal for both particles, the $e^+$ will most probably form a short-lived meta-state with a local electron called positronium (Ps) before finally annihilating into two 511\,keV $\gamma$. 
Depending on the spin configuration, the decay time in LS is either 125\,ps (para-Ps) or 3\,ns (ortho-Ps). 
The fraction of ortho-Ps formation was reported to lie around 50\% \cite{franco11,Consolati13,schwarz19}.
Since the decay time for ortho-Ps is comparable to the dominating fast time constant  for scintillation (4.93\,ns) and to the time resolution of the PMTs, its pulse shape, influenced by the emission of delayed annihilation photons, could be recognized. 
Borexino uses this for a statistical removal of positron background~\cite{PhysRevD.89.112007}.
Double Chooz is able to recognize ortho-Ps formation on event-by-event basis~\cite{doubleChooz2014}.

Moreover, $e^+$ events feature a characteristic topology: in contrast to the $e^\pm$ track, which ends after a few cm for kinetic energies below 10\,MeV, the $\gamma$ particles typically undergo several Compton scattering processes, each of which with a mean free path of tens of cm. 
Since optical photons travel $\sim$20\,cm per ns, the spread of $e^+$ topology should also leave tiny detectable traces on the pulse shape compared to a point-like $e^-$ event.
Figure~\ref{fig:timeProfiles}(b) displays how both effects slightly shift the peak position of the $e^+$ time profile to higher times. 
1000 pulse shapes with visible energies ranging from 1\,MeV to 10\,MeV were normalised and superimposed  for each particle type.
Ortho-Ps was considered (see Sec.~\ref{sec:methods:data}).

In a similar way, topology can serve as the key to a direct $e^-/\gamma$ discrimination. 
Here, the delocalised energy deposition results only from the multiple Compton scattering processes needed to release the gamma energy into the scintillator.
None of the resulting points of energy deposition has a fixed energy.
This means that the topology is on the one hand not as well defined as for positrons, but will on the other hand not be dominated by one single deposition spot when moving to higher particle energies.
Instead, a higher event energy will only increase the spread of the topology further.

Although all PID methods introduced in the following are based on the hit times measured by the PMTs, they fall into two categories. 
Firstly, the methods based on the pulse shape (Sec.~\ref{subsec:pules_shape_methods}) evaluate the difference of time profiles between different particles directly from the time-of-flight subtracted hit times. Secondly, the method based on the topological reconstruction (Sec.~\ref{subsec:topological_method}) is using the hit times to create a topological event map prior to further analyses. In the following, the principle of both types of methods is described. The distributions of the characteristic parameters of each method are shown on example of electrons and positrons.

\subsection{Methods Based on Pulse Shape}
\label{subsec:pules_shape_methods}

As discussed above, one can discriminate two event classes based on their characteristic time profiles.
Accordingly, it is required to know the vertex point of the event in order to do a ToF-correction.
For the actual discrimination, Gatti filters~\cite{Gatti62} are commonly used, e.g.~in Borexino~\cite{PhysRevD.89.112007}. 
In the Gatti analysis, it is required to know the expected time profile $P_i(t)$ for each particle $i$.
The profiles serve as density distributions of the probability $r_i(t_n)$ for a particle $i$ to register a PMT hit between two times $t_n$ and $t_{n+1}$ as of

\begin{equation}
r_i(t_n) 
= \int_{t_n}^{t_{n+1}} P_i(t)dt.
\end{equation}
Given the binned pulse shape $r'(t_n)$ of an actual event to be categorized as a particle of the type 1 or 2, the Gatti parameter $G$ is defined as:
\begin{equation} 
\label{eq:gatti}
G 
= \Sigma_n r'(t_n)w(t_n)
\quad \textnormal{with} \quad w(t_n) 
= \frac{r_1(t_n) - r_2(t_n)}{r_1(t_n) + r_2(t_n)}
\end{equation}
and can be used for discrimination. 
The probability $r_i(t_n)$ is determined through averaging of the time profiles over the whole detector volume. Since the time profiles slightly change with radius (Fig.~\ref{fig:timeoverRadius}), we have verified that considering this dependence in the calculation of the weight $w(t_n)$ does not bring to any substantial improvement in the performance of the Gatti method.

\begin{figure}[h]
\centering
\subfigure[] {\includegraphics[width=0.45\textwidth]{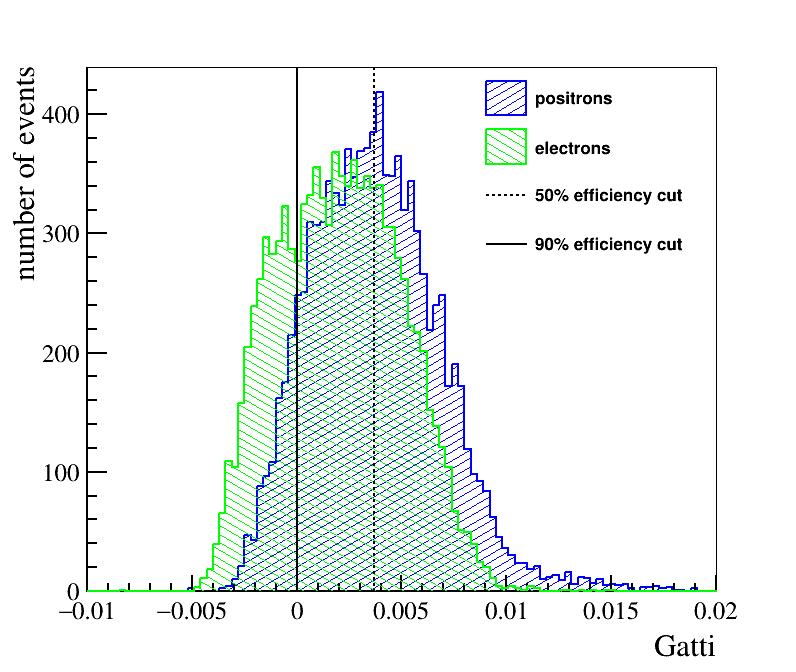}}\quad
\subfigure[] {\includegraphics[width=0.45\textwidth]{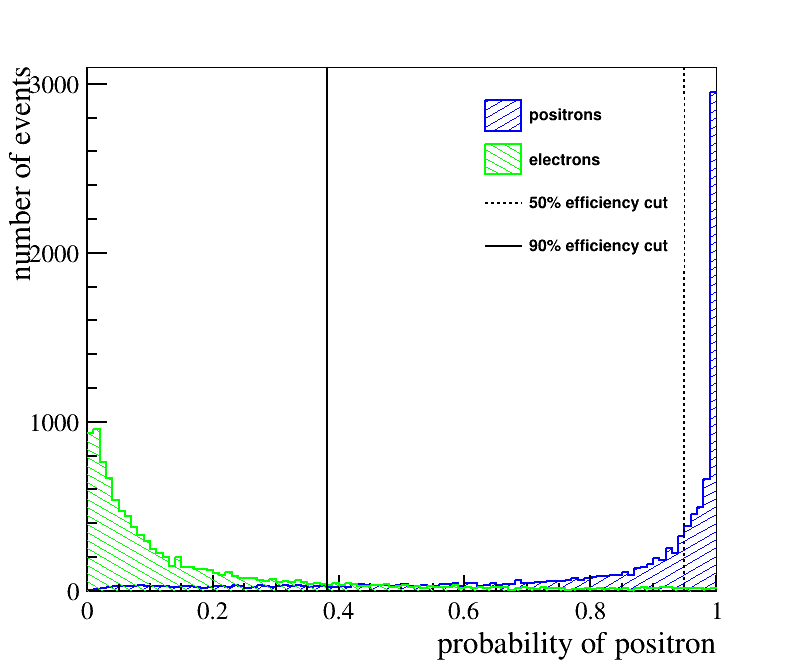}}
\caption{
Distribution of discrimination parameters for the Gatti (a) and NN (b) method. The examples represent electron and positron events at detector radii between 9.5\,m and 10.5\,m and with visible energies between 2\,MeV and 4\,MeV.
TTS, vertex smearing, and dark noise are not considered. Vertical lines indicate exemplary cut values at 90\,\% (solid) and 50\,\% (dashed) efficiency for electron signal.}
\label{fig:positronExampleDR}
\end{figure}

Due to the simplicity and stability of the Gatti analysis, we will use it as our baseline in this paper. 
An example for the distribution of $G$ can be found in Fig.~\ref{fig:positronExampleDR} (a) for simulated electrons and positrons with visible energies between 2\,MeV and 4\,MeV. The events were distributed in the radii between 9.5\,m and 10.5\,m. 
TTS smearing, vertex uncertainty, and dark noise were not considered. Further examples for other particle types were included in the Appendix.

\begin{figure}[htb]
\centering
\includegraphics[width=0.4\textwidth]{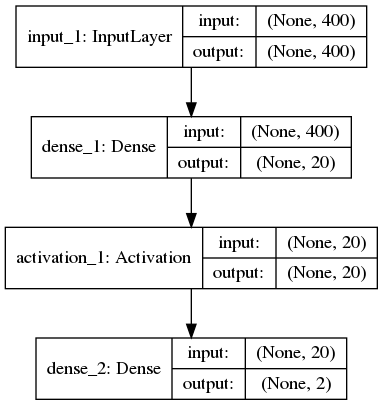}
\caption{Structure of the neural network applied for the particle identification.}
\label{fig:NN}
\end{figure}

Additionally, also in order to address more subtle problems like $e^+/e^-$ and $e^-/\gamma$ discrimination, we compare our Gatti results to a neural network (NN) analysis.
NNs have manifold applications, one of which is data classification.
A NN can be seen as a structured array of numerous computing nodes, typically arranged in a hierarchical sequence of layers.
During training process the NN optimises free parameters in the single nodes by itself. 
Artificial learning can develop its full potential only when the NN reaches a level of complexity which is appropriate to the amount of information being processed, thus the success typically depends on the chosen network architecture.
In principle, the Gatti filter is a linear signal transformation and can be seen as a NN reduced to only input and output layer, with the time profile replacing the training process.
The structure of the implemented NN is shown in Fig.~\ref{fig:NN}.
Analogous to the Gatti method, the input of the NN is a $1\times400$ array, representing the time profile of the particle with 400 bins of 1\,ns in size. 
Hence, the event information is already reduced to a one dimensional representation at the input stage.
Only one hidden dense layer with 20 neurons followed by a {\it{softmax}}-activation layer was added as it turned out that additional hidden layers did not improve the results.
The output layer gives a $1\times2$ array and assigns an affiliation probability to each particle type as shown in Fig.~\ref{fig:positronExampleDR} (b) for the example of electrons and positrons. 
Further examples for other particle types can be seen in the appendices.

\subsection{Method Based on Topological Reconstruction}
\label{subsec:topological_method}

The topology of an event in a large, unsegmented LS detector can be partly recovered from the PMT hit information using a method described in~\cite{Wonsak18}. 
In addition to hit times and respective charges, the topological reconstruction (TR) requires a knowledge of the reference parameters $\textbf{r}_\textnormal{ref}$ and $t_\textnormal{ref}$, which denote one point in space and time, respectively, which the primary particle must have traversed. The parameters
$\textbf{r}_\textnormal{ref}$ and $t_\textnormal{ref}$ can be obtained e.g.~from an independent vertex reconstruction. 
The detection time $t_\textnormal{hit}$ of a scintillation photon produced at a position $\textbf{r}$ along the particle track and observed as the $k^{\textnormal{th}}$ hit on the $j^{\textnormal{th}}$ PMT at position $\textbf{r}_j$ can be expressed by 

\begin{equation}\label{eq:hitTime}
    t_\textnormal{hit}
    = t_\textnormal{ref} \pm \frac{|\textbf{r} - \textbf{r}_\textnormal{ref}|}{c_0} + \frac{|\textbf{r}_j -  \textbf{r}|}{v_\textnormal{g}} + t_\textnormal{s} .
\end{equation}
The second term represents the flight time of the particle under the assumption that it moves with vacuum speed of light $c_0$, being subtracted or added depending on the particle reaching $\textbf{r}$ before or after traversing $\textbf{r}_\textnormal{ref}$, respectively. 
The third term considers the time of flight of the scintillation photon, whose group velocity $v_\textnormal{g}$ depends on its wavelength and the refractive index of the surrounding medium. 
Light attenuation as caused e.g.~by Rayleigh scattering or photon absorption is not considered in the mathematical description, thus scattered photons will be treated as direct messengers just like direct photons.
The non-deterministic contributions from the statistical scintillation process and the timing uncertainty of the PMTs are merged in the summand $t_\textnormal{s}$, which can thus also be negative. 

Solving Eq.~\ref{eq:hitTime} for $\textbf{r}$ yields an isochronic surface centered around the PMT at $\textbf{r}_j$. 
However, the fact that the exact $t_s$ is unknown but instead emanates from a probability density function (PDF) of time causes the isochrone to smear out perpendicularly to the surface. 
The profile of this smearing is mostly (ignoring dispersion affects during propagation) given by the scintillation time profile convoluted with the time response of the PMTs. We always use the scintillation time profile expected for electrons, although the scintillation time profile depends on the interacting particle.
\footnote{That is why we expect different reconstruction results for particle with other scintillation time profiles such as alphas and protons. This is where the discrimination power in these cases stems from.}
 In addition, a filter is applied to this 3D distribution in order to take into account the local probability $\varepsilon_j(\textbf{r})$ of light to be detected at $\textbf{r}_j$, considering light attenuation and the detector geometry. 
The result, when normalized to 1, is a 3D PDF for the origin of the detected photon, in the following referred to as $\phi_{j,k}(\textbf{r})$. 
Adding up the contributions from all hits and PMTs, i.e.~$\sum_{j,k}\phi_{j,k}(\textbf{r}) $, yields a rough impression of the spatial origin of all detected light. 
The actual local density $\Gamma_\textnormal{em}(\textbf{r})$ of the number of emitted photons can be gained from re-weighting $\sum_{j,k}\phi_{j,k}(\textbf{r}) $ with the inverse of the local detection efficiency $\varepsilon(\textbf{r})$. 
The latter is gained from summing $\varepsilon_j(\textbf{r})$ over all PMTs, i.e.

\begin{equation}\label{eq:iterationResult}
    \Gamma_\textnormal{em}(\textbf{r})
    = \frac{\sum_{j,k}\phi_{j,k}(\textbf{r})}{\sum_j \varepsilon (\textbf{r})}.
\end{equation}

The mere superposition of $\phi_{j,k}(\textbf{r})$ contributions treats photon emissions as independent incidents. 
In fact all emissions share a common event topology and are thus correlated. 
This can be utilised by treating the previous result as prior information in further iterations. 
While re-evaluating $\phi_{j,k}(\textbf{r})|_n$ in the $n^{\textnormal {th}}$ iteration, $\Gamma_\textnormal{em}(\textbf{r})|_{n-1}$ is introduced as weighting mask before normalisation, ideally minus the contribution from $\phi_{j,k}(\textbf{r})|_{n-1}$ in order to prevent self enhancement.

For high energy $\mathcal{O}$(GeV) events on the one hand, the TR can reveal regions along the particle track, where an excess of energy deposition has occurred, e.g.~due to a hadronic shower. 
On the other hand, for the discussed $\mathcal{O}$(MeV) low energy regime, the TR can, given the $\mathcal{O}$(ns) time resolution of the PMTs, by no means resolve topological structures on scales below 10\,cm.

\begin{figure}[h]
\centering
\subfigure[] {\includegraphics[width=0.45\textwidth]{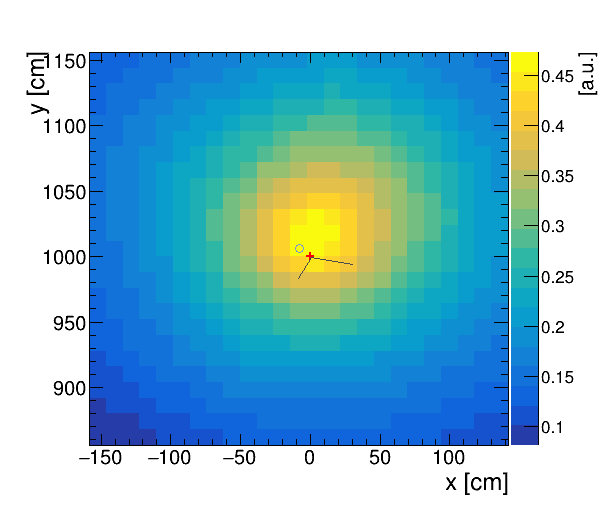}}\quad
\subfigure[] {\includegraphics[width=0.45\textwidth]{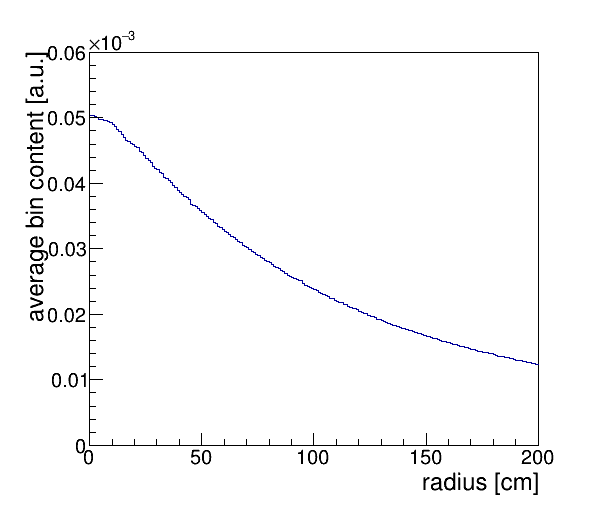}}
\caption{Topological reconstruction of a simulated positron event with a visible energy of 3.6\,MeV: (a) projection of the emission density $\Gamma_\textnormal{em}(\textbf{r})$ on the x-y-plane in arbitrary units and (b) its corresponding radial dependence around the reference point $\textbf{r}_\textnormal{ref}$. Details in text.}
\label{fig:positronExampleTR}
\end{figure}

\begin{figure}[htbp]
\centering
\subfigure[Reconstructed alpha.] {\includegraphics[width=0.45\textwidth]{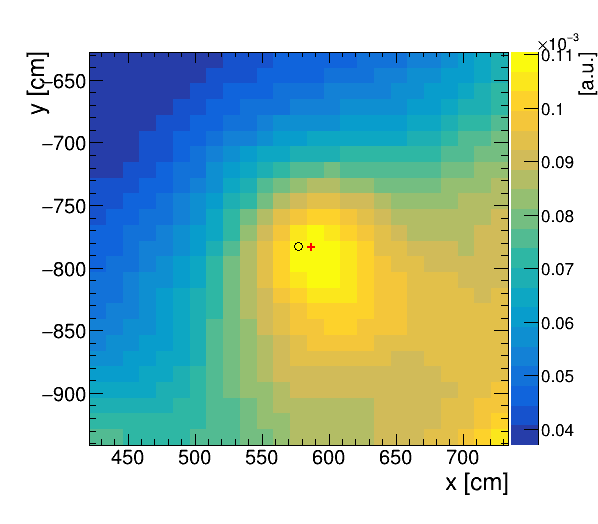}}\quad
\subfigure[Alpha radial profile.] {\includegraphics[width=0.45\textwidth]{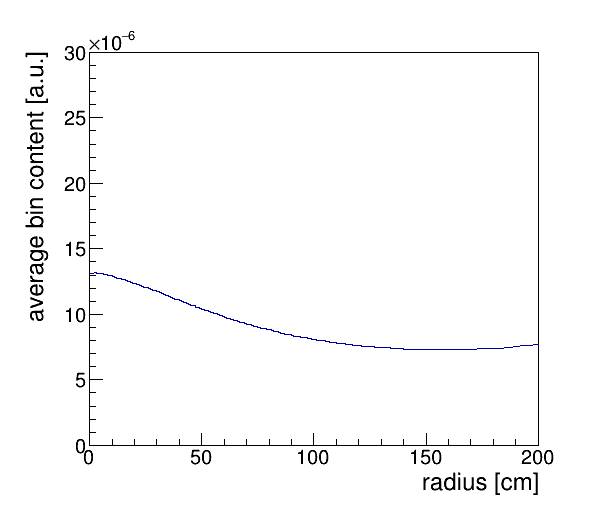}}\\
\subfigure[Reconstructed electron.] {\includegraphics[width=0.45\textwidth]{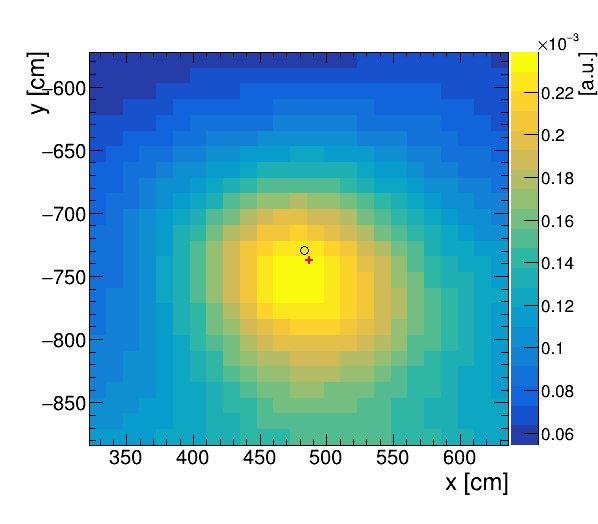}}\quad
\subfigure[Electron radial profile.] {\includegraphics[width=0.45\textwidth]{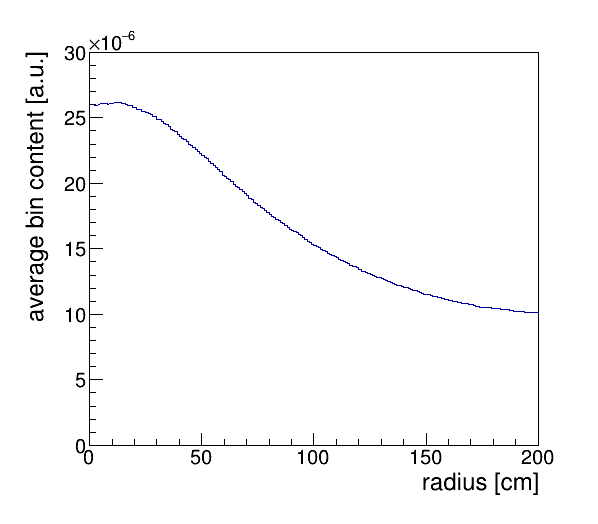}}\\
\subfigure[Reconstructed gamma.] {\includegraphics[width=0.45\textwidth]{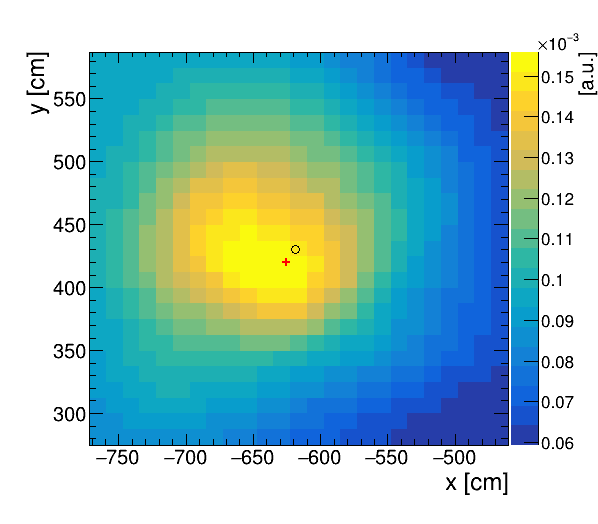}}\quad
\subfigure[Gamma radial profile.] {\includegraphics[width=0.45\textwidth]{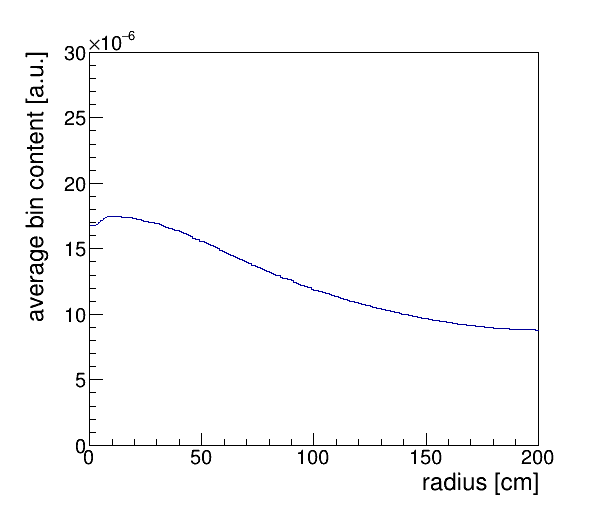}}
\caption{Topological reconstruction of different simulated particles, all with a visible energy of 1.5\,MeV and near a detector radius of 10\,m: left plots represent projections of the emission density $\Gamma_\textnormal{em}(\textbf{r})$ on the x-y-plane in arbitrary units and right plots show the corresponding radial dependence around the reference point $\textbf{r}_\textnormal{ref}$.}
\label{fig:SeveralExamplesTR}
\end{figure}

Fig.~\ref{fig:positronExampleTR} (a) shows a typical example for a low energy TR event in JUNO with the colour code representing a projection of the emission density $\Gamma_\textnormal{em}(\textbf{r})$ on the x-y-plane.
The units are arbitrarily scaled.
The TR was carried out in 9 iterations for a simulated positron event with a visible energy of 3.6\,MeV. 
A red cross and a black ring mark the true and reconstructed vertex point, respectively. 
Two black straight lines indicate the simulated tracks of the annihilation gammas. 
The reconstructed topology resembles a cloud around the reference point, coming from which the density gradually decreases. 
However, the energy depositions from the gammas do not appear as distinct features in the topology.
Instead, the off-centered emissions of scintillation photons cause the cloud to become more diffuse and spread a little wider compared to a more point-like electron event. 
In case of an alpha or proton event, a similar effect takes place since the increased number of late-photon emissions is associated less closely with the reference point.
An example for a 1.5\,MeV alpha can be studied in Fig.~\ref{fig:SeveralExamplesTR} (c).
However, it has to be noted that pulse features in regions of low intensity take effect upon the TR result only marginally. 
The reason is that the gradual increase in contrast which is attained during the iteration process goes along with fading of less pronounced topology regions. 
The TR method in its current state is thus optimised to expose near-peak variations as anticipated in $e^+/e^-$ and $e^-/\gamma$ discrimination. 
Fig.~\ref{fig:SeveralExamplesTR} provides a selection of TR results for different particles, all located near a detector radius of 10\,m and having visible energies around 1.5\,MeV. 
The obtained topologies for the sample electron (c) and gamma (e) resemble in structure very much the previous example of a positron.
The alpha event (a) on the other hand, appears more divergent towards the bottom right-hand corner of the plot.

The compactness of the reconstructed topology can be studied when building the radial profile $f(r)$ as shown in the right plots of Fig.~\ref{fig:positronExampleTR} and \ref{fig:SeveralExamplesTR}, i.e.~plotting the bin content found on average in a radius $r$ around $\textbf{r}_\textnormal{ref}$. 
The gradient defined as 
\begin{equation}
\label{eq:gradient}
    g(r)
    =\frac{f(r) - f(r+\Delta r)}{\Delta r}
\end{equation}
over a window with constant size $\Delta r$ takes higher values for more compact topologies. 
Accordingly, the highest value $g_\textnormal{max}$ found along $r$ was chosen to be used as a discrimination parameter.

\begin{figure}[htb]
\centering
\includegraphics[width=0.5\textwidth]{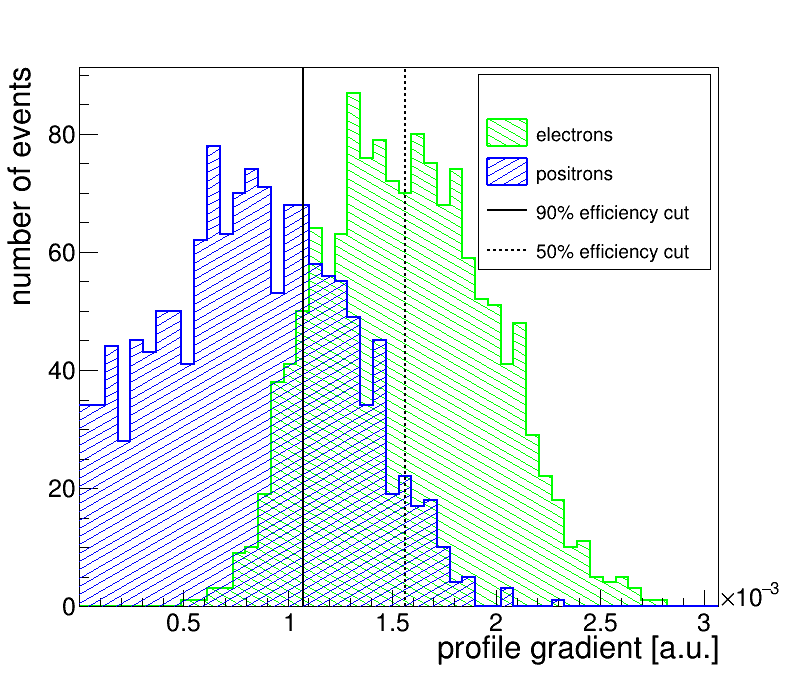}
\caption{Distribution of discrimination parameter $g_\textnormal{max}$ based on the topological event reconstruction for electron and positron events at detector radii between 9.5\,m and 10.5\,m and with visible energies between 2\,MeV and 4\,MeV. TTS, vertex smearing, and dark noise are not considered. Vertical lines indicate exemplary cut values at 90\,\% (solid) and 50\,\% (dashed) efficiency for electron signal.}
\label{fig:paramDistributione+e-}
\end{figure}

Figure~\ref{fig:paramDistributione+e-} shows the direct comparison of $g_\textnormal{max}$ values for simulated electrons (green) and positrons (blue). 
The depicted events were picked at detector radii between 9.5\,m and 10.5\,m. This region was chosen in order to avoid advantages from potential symmetry effects in the detector centre and be unaffected by edge effects.
The visible energies range from 2.0\,MeV to 4.0\,MeV. Not only does the positron distribution peak at a lower value, corresponding to the topology being less point-like, but also does it exhibit a shoulder along its rising edge, caused by the delayed annihilation in ortho-Ps events.

\subsection{Datasets} \label{sec:methods:data}

Event simulations were carried out with the official Geant4-based JUNO simulation.
Apart from the detailed detector geometry, a full optical model is implemented, which includes in particular the wavelength-dependent descriptions of the parameters determining the optical transport in all materials.
All significant optical processes, i.e.~scintillation, Cherenkov radiation, refraction, Rayleigh scattering, absorption, and re-emission are considered.
At 430\,nm, which marks the peak of the scintillation spectrum, the characteristic lengths for photon absorption and Rayleigh scattering are assumed to be 27.0\,m and 79.7\,m,respectively, which inversely adds up to an attenuation length of 20.2\,m. 
The JUNO simulation was used up to the stage of photon detection by the PMT channels.
Dark noise could be switched on or off at will.
Timing response of the PMTs was added in the form of TTS for certain datasets.


All analyses were performed on three distinct datasets:

\begin{itemize}
\item Dataset 1: the pure MC truth data. 
A full simulation of the detector was done, implying the kinematics during energy deposition, the emission of scintillation light, and the passage of optical photons through the detector media. 
However, exact knowledge was assumed for detected hittimes and for the reference point and reference time used in our methods. 
The reference point was chosen as the barycentre of energy deposition, the reference time as the time of first energy deposition. 
Note that we sometimes refer to this point and time as vertex point and vertex time, although it is only identical to the primary vertex in case of point-like events and not so for gamma events.
This idealised dataset is used in order to explore the absolute limits of our method.
The results of PID will here depend mainly on p.e.~yield and attenuation length. 
This implies that the performance in other LS experiments can partly be deduced from an adequate shift along the energy axis (with the exception of $e^+/e^-$ discrimination, where the deciding topological features from the two 511\,keV gammas become less distinct at higher positron energies).

\item Dataset 2: smearing of vertex and hit times.
The consideration of a finite timing resolution of PMTs is strongly related to the resolution in vertex reconstruction, which furthermore depends on the number of measured photons, i.e.~on visible energy.
Based on the events from Dataset 1, the hit times were smeared with a Gaussian of the width $\sigma_\textnormal{TTS}$ corresponding to the TTS values in JUNO.
The TTS values, quoted as FWHM in Sec.~\ref{sec:juno}, translate into a $\sigma_\textnormal{TTS}$ of 7.6\,ns for 13k large PMTs\footnote{
  Actually, the distribution of transit times of a single MCP PMT is not exactly Gaussian but rather has a large substructure. However, the positions of peaks in this substructure are individual for each PMT. Since the peaks average out over the totality of tubes, a Gaussian representation is a fair approximation.} and 1.3\,ns for 5k large PMTs.
Gaussians were also used to smear the vertex point and time.
The standard deviations were estimated from the current efforts for vertex reconstruction in JUNO to follow a $\sigma/\sqrt{E/\textnormal{MeV}}$-rule, with $E$ denoting the visible energy: the values of 10\,cm and 0.7\,ns were chosen for the smearing in each vertex dimension and time, respectively, both underlying conservative assumptions.\footnote{The spatial smearing was deduced from a fit of the results from JUNO vertex reconstruction published in \cite{junoVertex18}. The standard deviation for time smearing was determined from an internal reconstruction algorithm. Our experience with this algorithm points at even lower time uncertainty, which is in accordance with results from LENA~\cite{DominicusPhdthesis}.} 

\item Dataset 3: adding of dark noise.
Based on Dataset 2, dark noise is added with a rate of 30\,kHz for all large PMTs.
With 18k large PMT channels this leads to an expectation of little more than 200 dark hits in a 400\,ns time window, as opposed to the 1200 signal hits per MeV of deposited energy.
The direct comparison between Datasets 2 and 3 can reveal the impact of dark noise on our discrimination methods.

\end{itemize}

\begin{table}
\caption{Energy range and position distribution of the simulated data samples. The energy ranges were chosen with regard to the physics applications indicated in Sec.~\ref{sec:juno}.}
\centering
\begin{tabular}{|l|c|c|c|c|}
     \hline
     Particle & $\alpha/\beta$ & $p/\beta$ & $e^+/e^-$ & $e^-/\gamma$  \\
     \hline
     Energy [MeV] & [0.2, 1.5]  & [0, 2.0] & [1.0, 10.0] & [0, 3.0] \\
     \hline
     Position  & \multicolumn{4}{|c|}{uniformly in the whole central detector} \\
     \hline 
\end{tabular} 

\label{tab:datasets}
\end{table}

For each particle type in the discrimination categories $\alpha/\beta$, $p/\beta$, $e^+/e^-$, and $e^-/\gamma$, 120k events were simulated, 100k of which were taken as training sample and the remaining 20k events for validation.  
The events were spread uniformly over the whole CD with energies according to the intervals quoted in Table~\ref{tab:datasets}, selected according to the expected physics applications, as discussed in Sec.~\ref{sec:juno}.
For $e^+$ events, ortho-Ps was considered at a fraction of 54.5\% and with a lifetime of 3.08\,ns.
Since the small PMTs in the CD account for less than 4\% of the optical coverage from the large PMTs, they only play a minor role in PID. 
It was decided to generally ignore hits on small PMTs in favour of computation time.
This applies to all datasets.

\subsection{Figures of Merit}

In order to compare and analyse our methods, the results will be presented based on a fixed scheme. 
A selection of the figures of merit introduced here will be shown for each event pairing in Sec.~\ref{sec:results}. 
A full collection of the plots can be found in the appendix.

We define {\it (i) discrimination efficiency} $\epsilon_{\textnormal{sig}}$ as the ratio of the number of signal events passing a cut and total amount of signal events and {\it (ii) impurity} $\epsilon_{\textnormal{bkg}}$ as the ratio of the number of remaining background events after the cut and total background events. 
This implies that neither $\epsilon_{\textnormal{sig}}$ nor $\epsilon_{\textnormal{bkg}}$ depend on the actual ratio of signal to background events.

Impurity will be plotted over efficiency in a fixed energy and radius range.
The choice of mean energy is in each case motivated by the energy region in which we observe the best discrimination performance for the respective event pairing.
Three plots, representing our discrimination methods, will be shown, each containing three curves for the analysed datasets.

Efficiency and impurity will both be plotted as a function of energy at a fixed level of $\epsilon_{\textnormal{bkg}}$ and $\epsilon_{\textnormal{sig}}$, respectively. 
Note that this configuration allows to draw equivalent conclusions for switching signal and background by simply switching the labels efficiency and impurity and reversing both their axes. 
All datasets will be presented in order to analyse the differences between ideal and realistic data. 
Besides, the radius range in the CD was also fixed, since the cut parameters were found to change with detector radius $R$.
The specific region with $R=(10 \pm 0.5)$\,m was chosen in order to neither profit from potential symmetry effects in the detector centre nor distort the results by edge effects which occur in the outermost detector regions.

Efficiency and impurity will also be plotted over the CD volume, parametrised by $R^3$, in a defined energy range.
An additional horizontal axis indicates the corresponding $R$-values for orientation.
The discussion within Sec.~\ref{sec:results} is limited to the most realistic Dataset 3. 

All binned values for efficiency (impurity) are presented with error bars, determined by $1/\sqrt{N_i}$, with $N_i$ representing the total number of signal (background) events being considered in bin $i$. 
This implies that errors are purely statistical and do not consider systematic impact.

\section{Performance in Particle Identification} \label{sec:results}

\subsection{$\alpha/\beta$ and $p/\beta$ Discrimination}

\begin{figure}[hbt]
\centering
	\subfigure[Gatti] {\includegraphics[width=7cm]{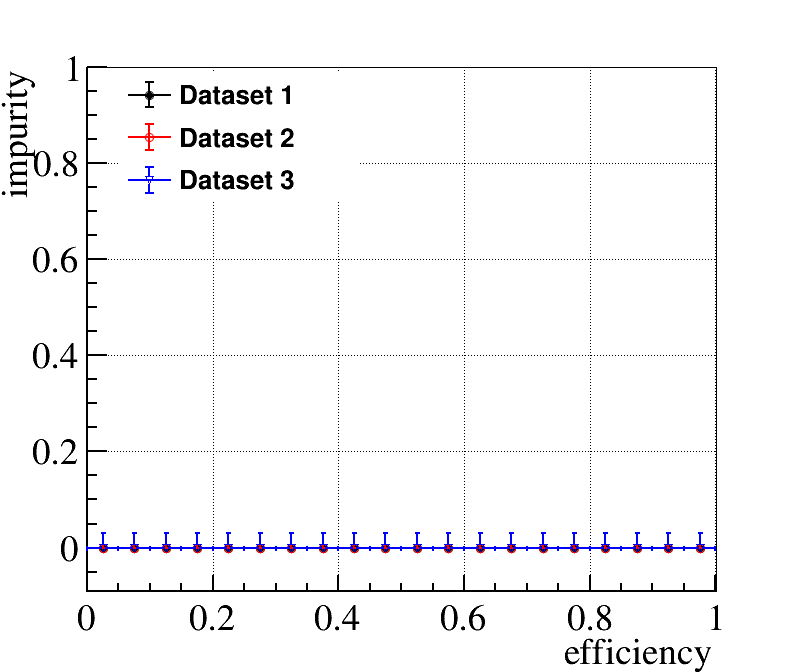}}\quad
	\subfigure[NN] {\includegraphics[width=7cm]{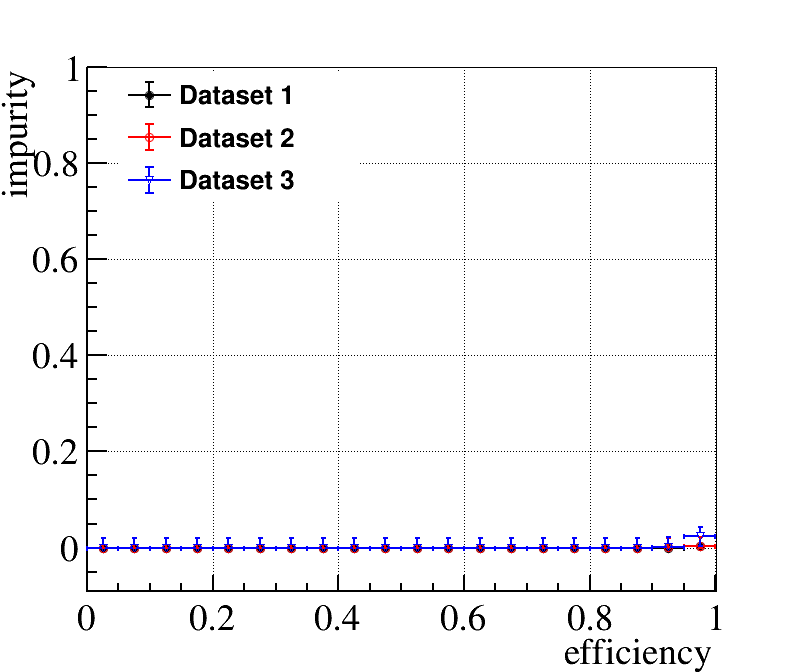}}\quad
	\subfigure[TR] {\includegraphics[width=7cm]{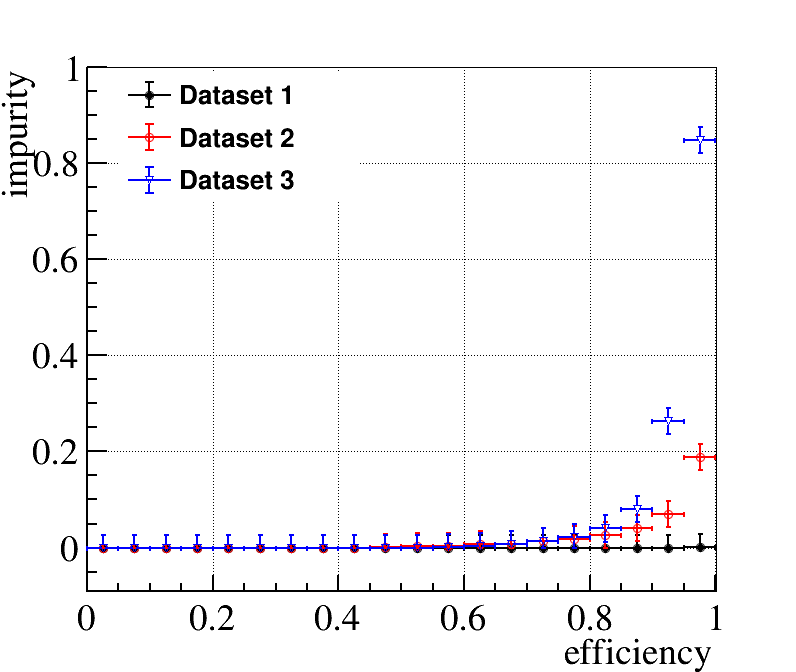}}
\caption{Impurity as a function of efficiency for $\alpha/\beta$ and $p/\beta$ discrimination. The results were obtained for visible energies between 1.5\,MeV and 2.0\,MeV, and with detector radii between 9.5\,m and 10.5\,m.}
\label{fig:ep_efficiency}
\end{figure}

\begin{figure}[hbt]
	\centering
	\subfigure {\includegraphics[width=7cm]{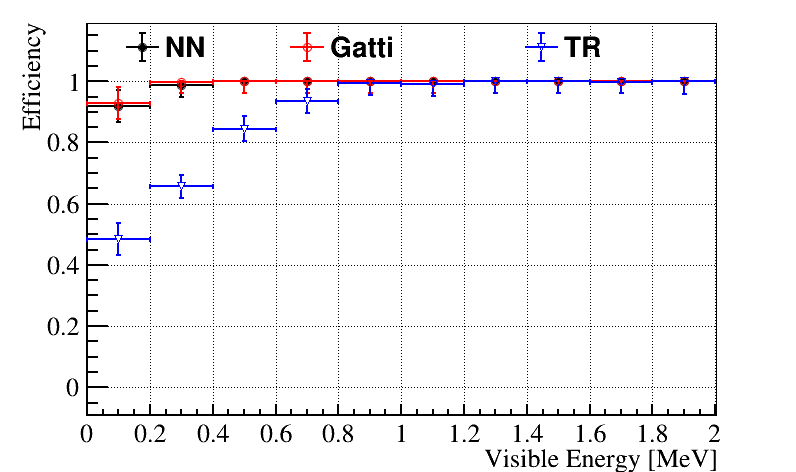}} 
	\subfigure {\includegraphics[width=7cm]{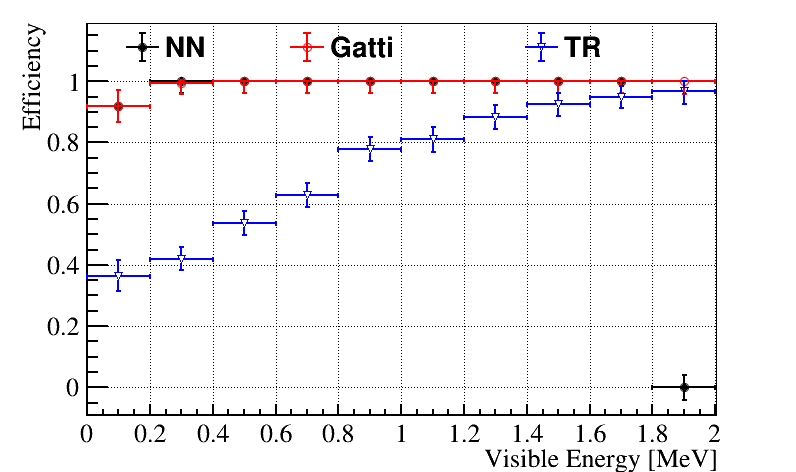}}\\
	\subfigure[Dataset 1]{\includegraphics[width=7cm]{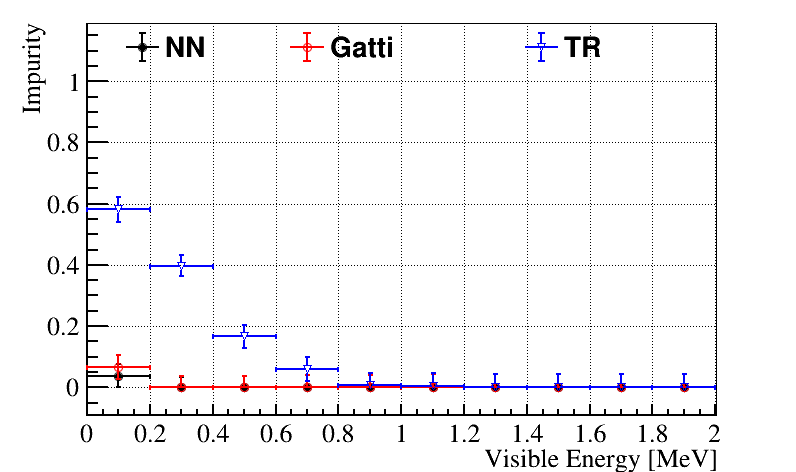}}\quad
	\subfigure[Dataset 2] {\includegraphics[width=7cm]{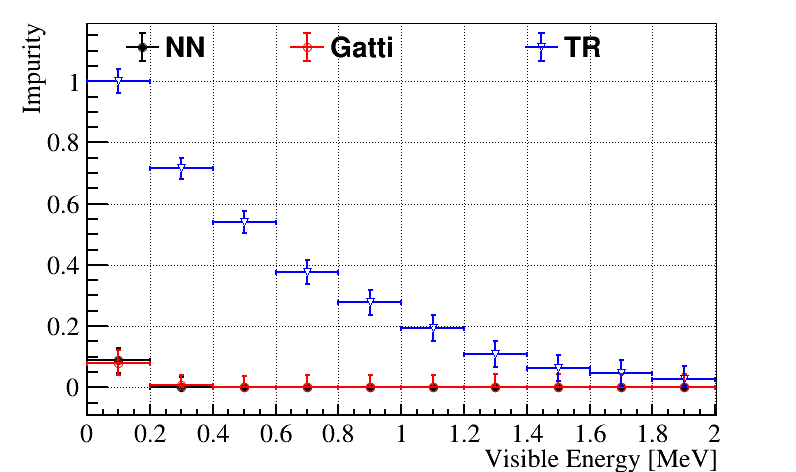}}\\
	\subfigure {\includegraphics[width=7cm]{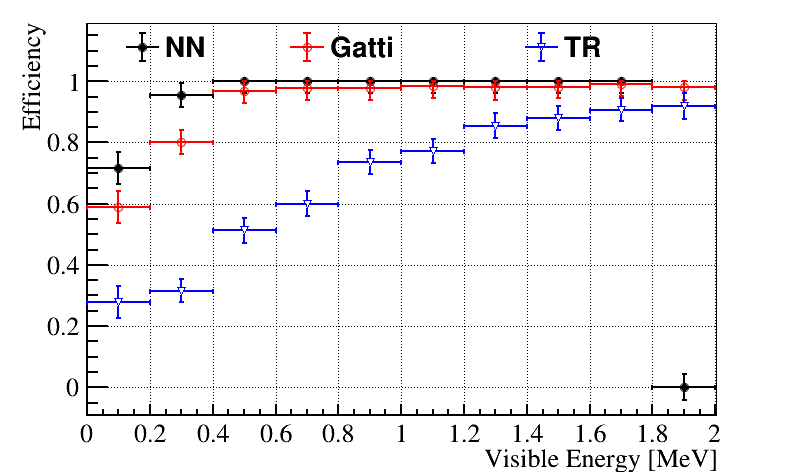}}\quad
	\subfigure {\includegraphics[width=7cm]{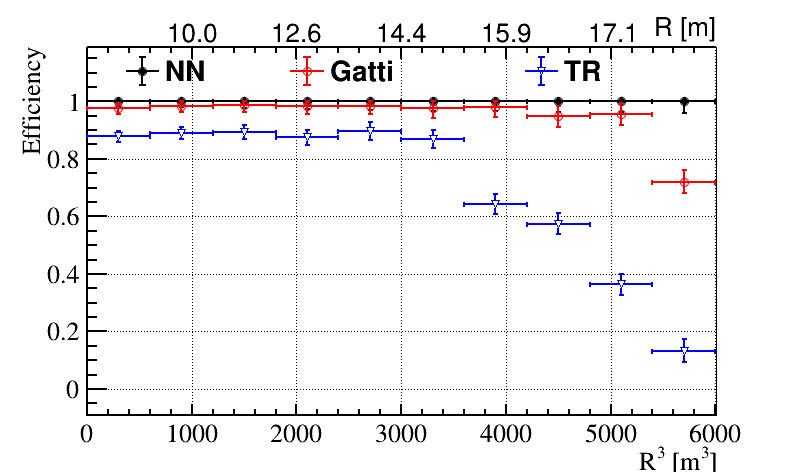}}\\
	\subfigure[Dataset 3] {\includegraphics[width=7cm]{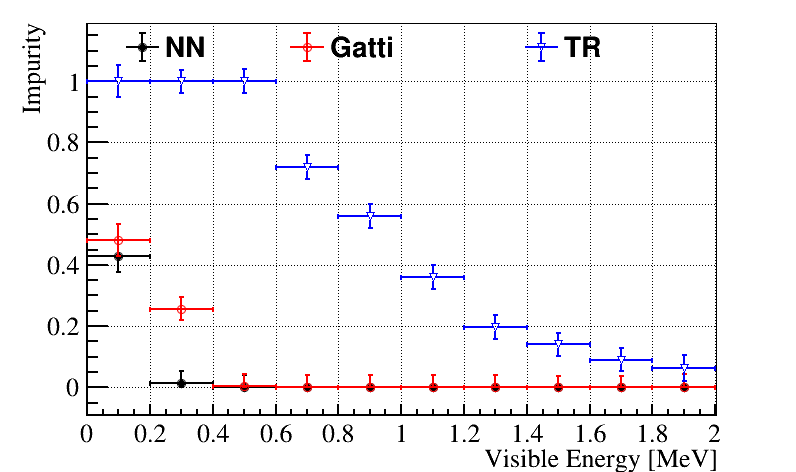}}\quad
	\subfigure[Dataset 3] {\includegraphics[width=7cm]{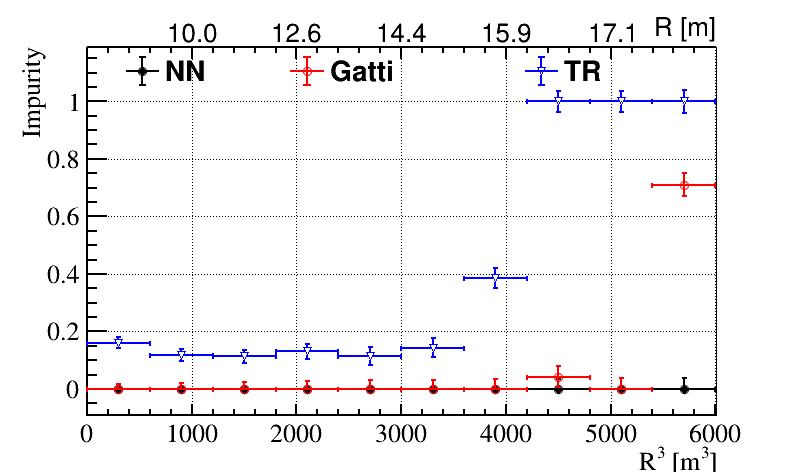}}
	\caption{Performance of the $\alpha/\beta$ and $p/\beta$ discrimination from all three methods. Impurity was obtained	at efficiency fixed to 90\% while efficiency was obtained at impurity fixed to 10\,\%. (a), (b), and (c) show results with Datasets 1, 2, and 3, respectively, as a function of visible energy. All three methods were used on events with detector radii between 9.5\,m and 10.5\,m. (d) shows the performance depending on the detector radius for events with visible energies between 1.5\,MeV and 2.0\,MeV.}
	\label{fig:ep_disc_energy}
\end{figure}

Alphas and protons, although showing individual quenching behaviour due to different charges, cause scintillation light to be emitted very similarly over time, which is reflected by almost identical time constants and weights in Table~\ref{tab:timeConstant}.
Accordingly, no differences are expected when comparing the results for discrimination against electrons on the basis of coinciding visible energy. 
Our results are indeed congruent within the tested energy ranges.
Here, we show the $p/\beta$ results which cover a wider energy range and point out that our conclusions equally apply to the respective plots for $\alpha/\beta$ appended to this paper (see Appendix \ref{app:alpha_beta}).

A discrimination between $\alpha$ and $p$ on the one hand and $\beta$ on the other can be considered a straightforward task due to the clear distinction features in the time profiles (Fig.~\ref{fig:timeProfiles} (a)) and also based on the experience in other experiments, as previously discussed. 
In our study, electron events were in either case treated as signal. 
The obtained level of background impurity was plotted as a function of signal efficiency in Fig.~\ref{fig:ep_efficiency} with the Gatti (a), NN (b), and TR (c) method. 
The events have visible energies between 1.5\,MeV and 2.0\,MeV. 
The detector region was limited to detector radii between 9.5\,m and 10.5\,m. 
The plots for the Gatti and NN method represent the whole central detector. 
Each plot contains three curves representing the different datasets. 
The Gatti and NN method have no apparent difficulty in event classification. 
In the NN case, only the data point at very high efficiency above 95\% registers a non-zero background contamination for Dataset 3.  
The TR parameter, which was designed and optimised for $e^+/e^-$ discrimination, is also sensible to $p/\beta$ discrimination, however, it performs weaker than the direct methods.
This is related to a known feature during the TR iteration process, which is the tendency of intense topology regions to attract the probability contributions that would technically correlate best with less pronounced regions.
Since the most striking differences in pulse shape appear at late times, where the pulse is low, the TR shows only weak sensitivity here.
Vanishing impurities below 50\% efficiency and a steep rise at high efficiencies show that the TR parameter is usable for picking pure signal samples but, other than NN and Gatti, inappropriate for highly efficient background cuts. 
The deterioration is strongly being amplified by including TTS and vertex (Dataset 2) and the addition of dark noise (Dataset3).

A direct comparison between all three methods is demonstrated in Fig.~\ref{fig:ep_disc_energy}. 
Panels (a), (b), and (c) show impurity and efficiency as a function of energy for Datasets 1, 2, and 3, respectively.
Again, the results refer to events with detector radii between 9.5\,m and 10.5\,m.
Panel (d) shows the radius dependence for Dataset 3, evaluated for events with visble energies between 1.5\,MeV and 2.0\,MeV.   
Efficiency was determined at a fixed impurity level of 10\%. 
Impurity was determined with the required efficiency set to 90\%.
It can be observed as a general trend that higher energies, which imply an increase in p.e.~statistics, favour the prediction power in PID. Even in the ideal Dataset 1 a clean data cut is achieved only above 1\,MeV. Impurities below this value rise fast towards lower energies, while a fixed level of impurity would go along with an according drop in efficiency. 
The transitions from Datasets 1 to 3 shift this edge to higher energies.
In direct comparison, the NN results mildly exceed those achieved with the Gatti analysis.
The TR method is suffering more than the direct methods from the lack of p.e.~statistics at low energies.
In Dataset 3, which represents the most realistic data, the TR method loses its prediction power below 0.6\,MeV.
Here, the contribution of $\sim$200 dark hits within the critical 400\,ns of pulse shape weighs heavy compared to the $\sim$1200\,p.e./MeV from the actual signal, more than half of which having lost their direct correlation with the event vertex due to attenuation effects like Rayleigh scattering.
We see that also the direct methods suffer from dark noise, but less severely.
At higher energies, the data points from TR approach their equivalents from the direct methods.  

The position dependence of the cut purity shows a stable behaviour throughout the CD for the direct methods NN and Gatti.
Both reach their best values between 1000\,m$^3$ and 3500\,m$^3$. 
A look at the p.e.~yield over detector radius displayed in Fig.~\ref{fig:nPEoverRadius} reveals that in fact least light is expected from the innermost and outermost detector regions, meaning less statistics for the analysis and having a similar effect as observed at lower energies.
In contrast to the direct methods, the TR approach loses all prediction power above $R^3\approx3600\,\textnormal m^3$, corresponding to $R\approx15.3\,$m.
Difficulties in the TR near the detector edge result in a high number of badly resolved and sometimes dislocated topologies which lack the characteristic distinction features. 
Our results for $p/\beta$ discrimination have important implications also for fast neutron background in DSNB.
Fast neutrons provoke proton signals as a result of recoil processes in the LS. 
Although the studied energies do not touch the range above 11\,MeV, which is relevant for DSNB, one can fairly assume a highly efficient suppression of fast neutron background from the trend in Fig.~\ref{fig:ep_disc_energy} (c).

\subsection{$e^+/e^-$ Discrimination}

\begin{figure}[hbt]
\centering
	\subfigure[Gatti] {\includegraphics[width=7cm]{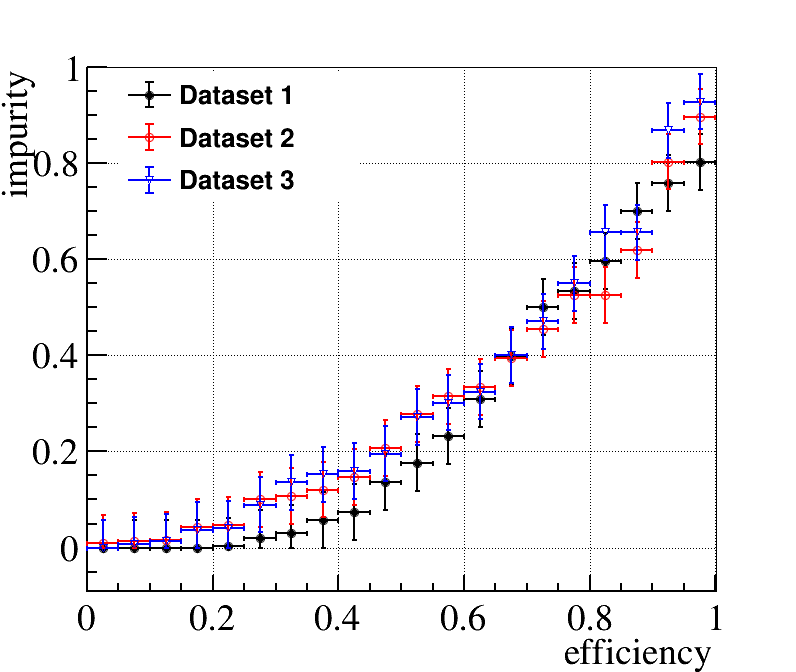}}\quad
	\subfigure[NN] {\includegraphics[width=7cm]{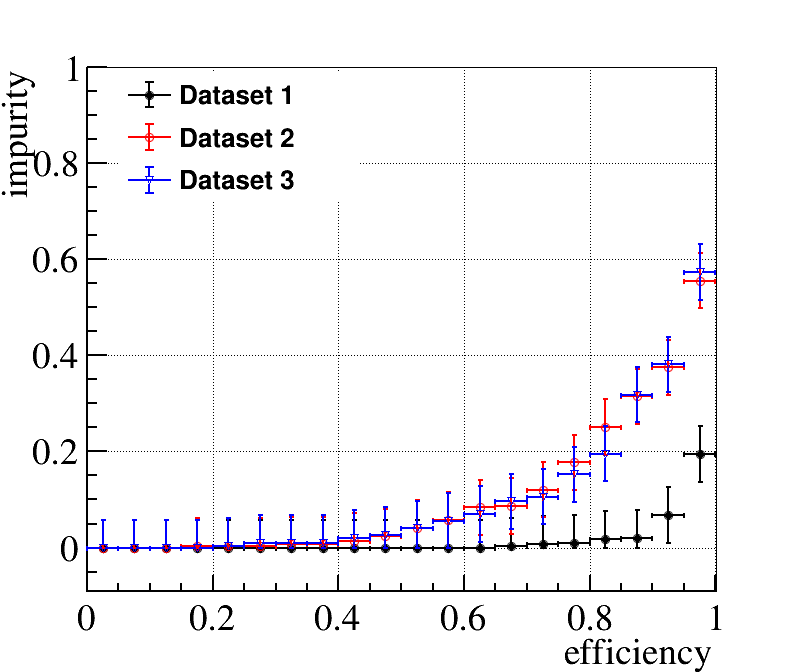}}\quad
	\subfigure[TR] {\includegraphics[width=7cm]{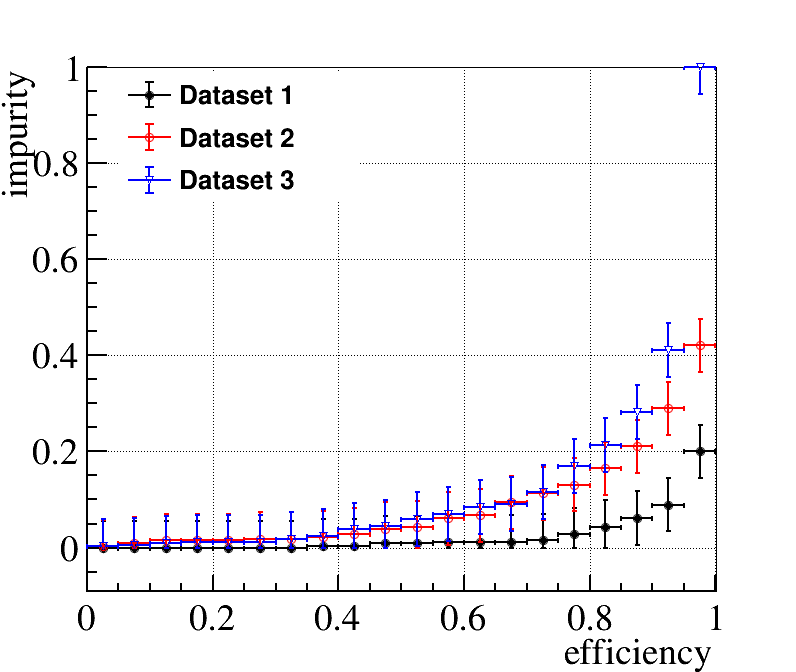}}
\caption{Impurity as a function of efficiency for $e^+/e^-$ discrimination. The results were obtained for visible energies between 2.75\,MeV and 3.25\,MeV. All three methods were used on events with detector radii between 9.5\,m and 10.5\,m.}
\label{fig:elecposi_efficiency}
\end{figure}

\begin{figure}[hbt]
	\centering
	\subfigure {\includegraphics[width=7cm]{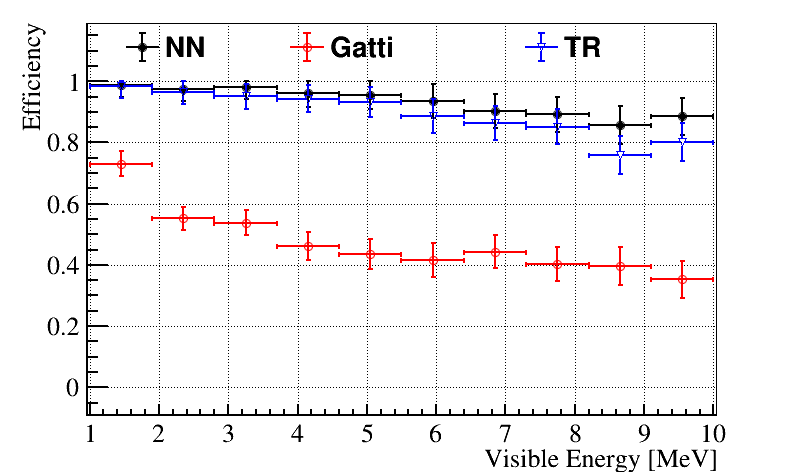}}\quad
	\subfigure {\includegraphics[width=7cm]{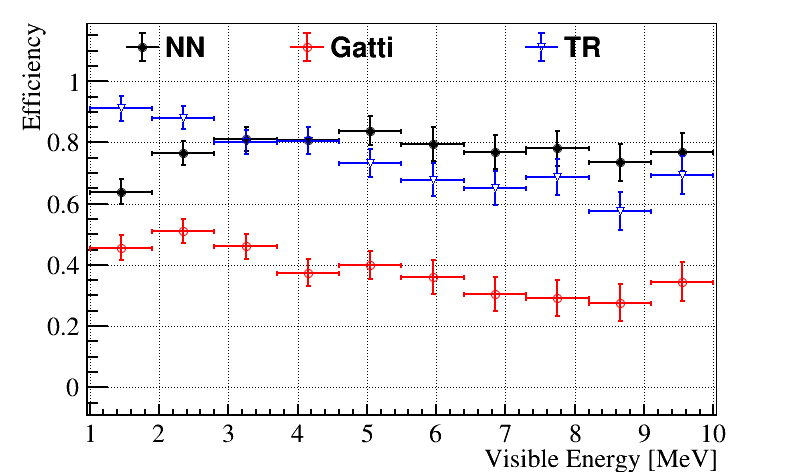}}\\
	\subfigure[Dataset 1]{\includegraphics[width=7cm]{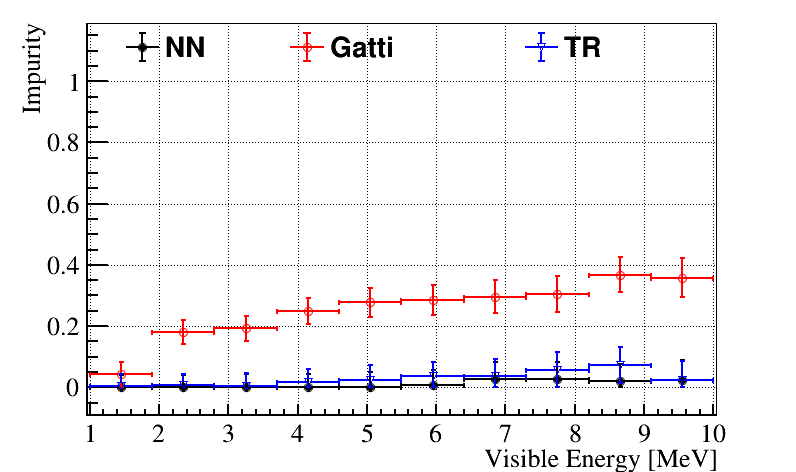}}\quad
	\subfigure[Dataset 2] {\includegraphics[width=7cm]{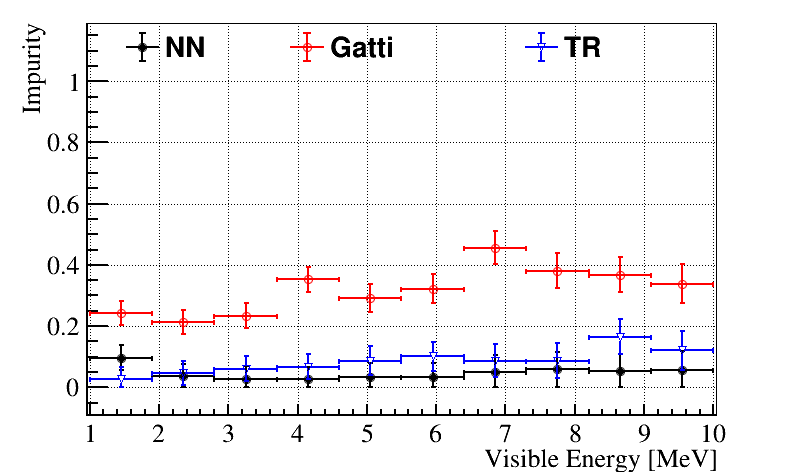}}\\
	\subfigure {\includegraphics[width=7cm]{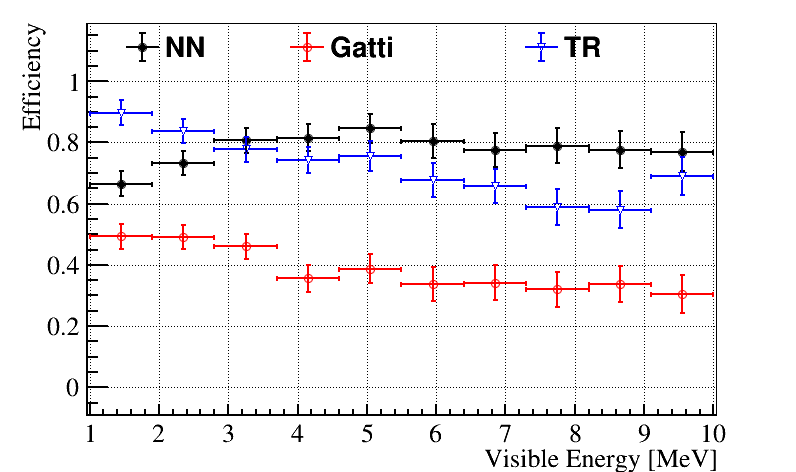}}\quad
	\subfigure {\includegraphics[width=7cm]{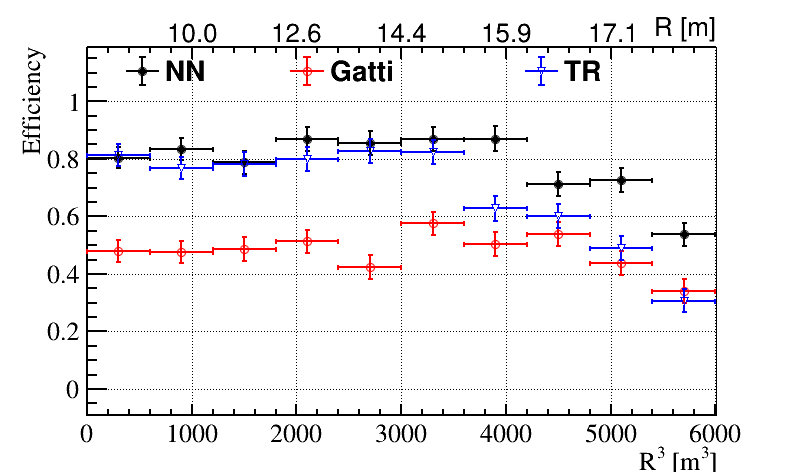}}\\
	\subfigure[Dataset 3] {\includegraphics[width=7cm]{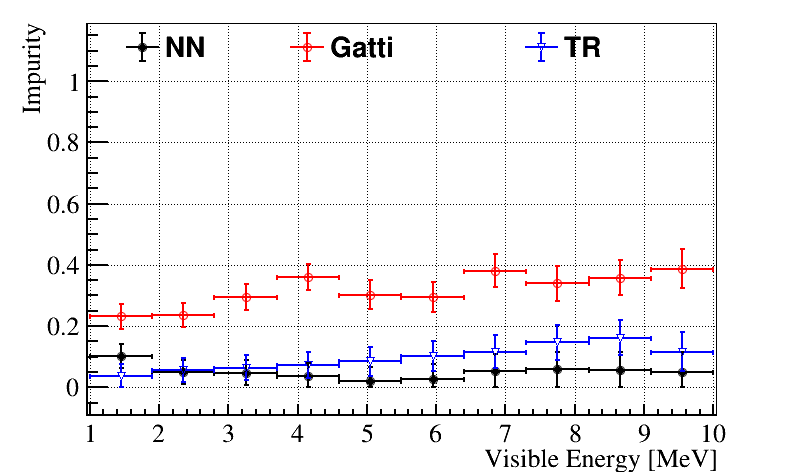}}\quad
	\subfigure[Dataset 3] {\includegraphics[width=7cm]{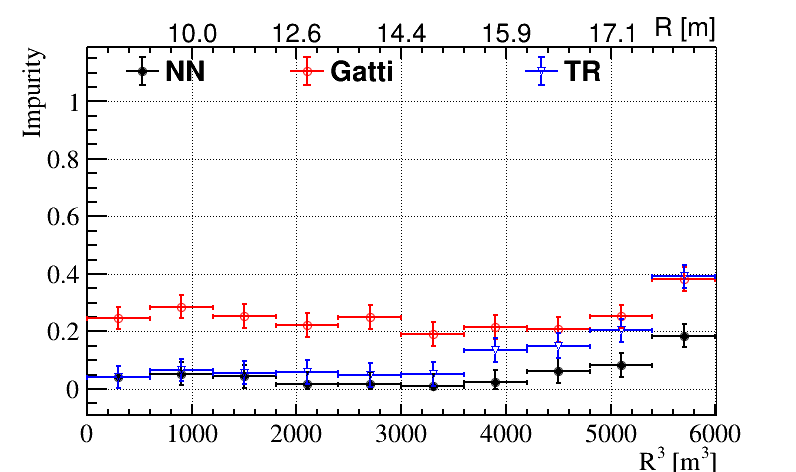}}
	\caption{Performance of the $e^+/e^-$ discrimination from all three methods. Impurity was obtained at efficiency fixed to 50\%, while efficiency was obtained at impurity fixed to 20\,\%. (a), (b), and (c) show results with Datasets 1, 2, and 3, respectively, as a function of visible energy. All three methods were used on events with detector radii between 9.5\,m and 10.5\,m. (d) shows the performance depending on the detector radius for events with visible energies between 2.75\,MeV and 3.25\,MeV.}
	\label{fig:elecposi_disc_energy}
\end{figure}

This analysis regards electron events as signal and positron events as background.
However, depending on the physics case the requirement can be vice versa.

The behaviour of the $e^+/e^-$ discrimination differs substantially from the previous results, which build on features in the light emission curves rather than on characteristic topologies.
The differences between $e^+$ and $e^-$ pulse shapes are much less pronounced as can be spotted in Fig.~\ref{fig:timeProfiles}. 
Accordingly, the depicted impurities in Fig.~\ref{fig:elecposi_efficiency} which were obtained for events with visible energies between 2.75\,MeV and 3.25\,MeV are on a higher level than for $\alpha/\beta$ and $p/\beta$. 
Especially the Gatti analysis turns out to be inappropriate for the task: the slight sensitivity observable in Dataset~1 is almost lost when going to realistic data, expressed by near-linear curves with a slope close to 1.
However, the similar results for the NN and TR method prove that a considerable amount of background can be removed by the cost of much less signal, e.g.~around 30\% impurity at 90\% efficiency in both methods for Dataset 3. 
Impurity rises faster towards high efficiencies.
Thus, for physics studies with high sample rates at hand, it could pay off to lower the efficiency requirements in favour of an increased signal-to-noise ratio. 
A large gap appears between the points for Datasets 1 and 2. 
This shows that a future experiment with reduced timing uncertainty has the potential to reduce the background by a whole order of magnitude while reaching very high efficiencies. 

Figure~\ref{fig:elecposi_disc_energy} (a), (b), and (c) show the energy of efficiency and impurity. 
The data points for efficiency were obtained at a fixed impurity of 20\,\%. The impurity was determined at 50\% efficiency. 
We observe a general impurity increase with energy, best to be seen in the upper row showing the ideal Dataset 1, which opposes to all other discussed event pairings. 
The reason for this is that with the increase of the $e^+$ kinetic energy causing the central ionisation, the relative weight of the off-center energy deposition of the two annihilation gammas of 1.022\,MeV total energy, decreases.
As soon as vertex and TTS smearing as well as dark noise enter the data (Datasets 2 and 3, respectively), the NN performance becomes worse also towards the low end of the energy spectrum. 
Apparently, the deterioration of data quality cannot fully be compensated for by statistics at these energies. 
As a result, the most sensitive region lies around 3\,MeV.
This actually meets the experimental focus for solar $^8$B neutrinos which lies between 2\,MeV and 5\,MeV. 
In absolute numbers, the impurities obtained for Dataset 3 do not fall below 5\% at 50\% efficiency, which rules out an event-by-event discrimination.
TR and NN produce very similar results between 2\,MeV and 3\,MeV.
With rising energy, the TR values depart further from the NN values.

Concerning the impact of dark noise, it can be concluded that the effect is less serious than found for the preceding event categories.
This can be traced back to fact that here the differences mainly are encrypted in the early part of the hit spectrum where the pulse peaks. 
The same reason explains why the TR method shows overall better results and is compatible with the NN.
The Gatti parameter on the other hand turns out to be unsuitable for $e^+/e^-$ discrimination. 
The method relies on the averaged time profiles presented in Fig.~\ref{fig:timeProfiles} (b).
The curves are very similar, in the peak region being shifted by only one or two 1-ns-bins.
Apparently, the mere comparison of the single event pulse shapes to the profiles is insufficient, whereas the more elaborated methods TR and NN prove to be more powerful.

The radius dependence shown in Fig.~\ref{fig:elecposi_disc_energy}(d) for visible energies between 2.75\,MeV and 3.25\,MeV is consistent with the previously discussed event categories.

\subsection{$e^-/ \gamma$ Discrimination}

\begin{figure}[hbt]
\centering
	\subfigure[Gatti] {\includegraphics[width=7cm]{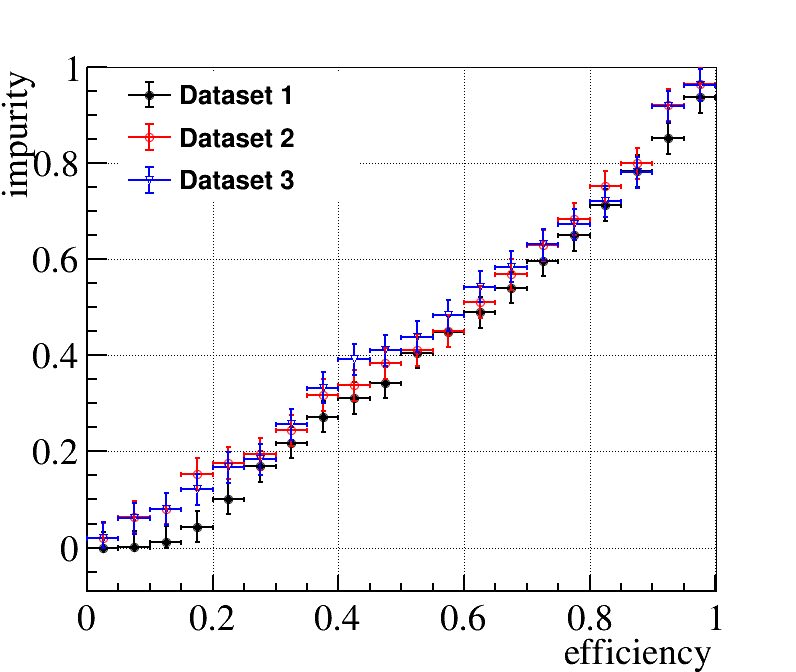}}\quad
	\subfigure[NN] {\includegraphics[width=7cm]{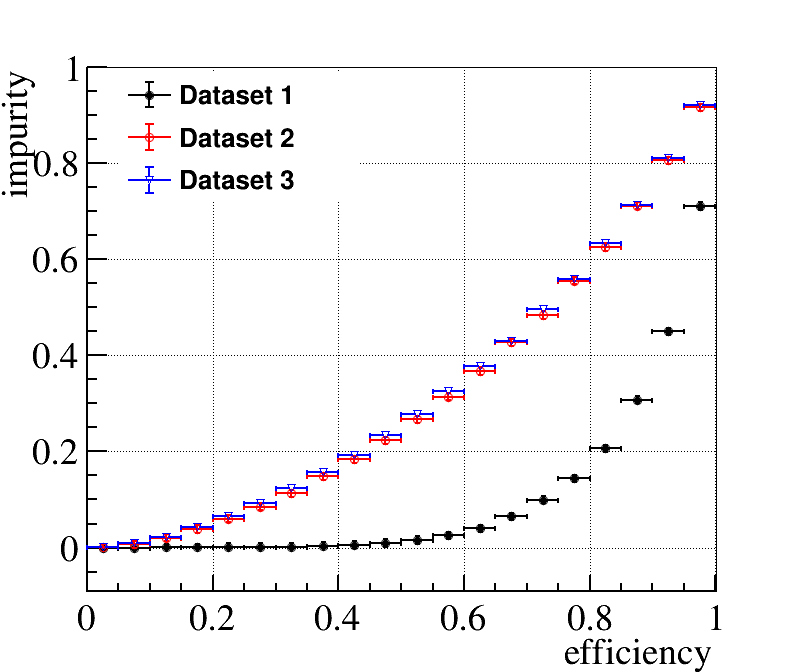}}\quad
	\subfigure[TR] {\includegraphics[width=7cm]{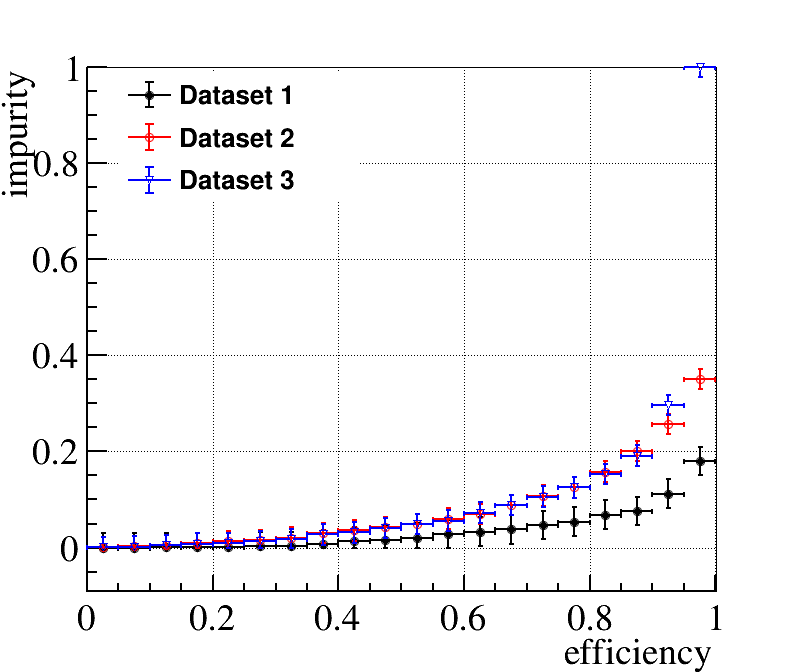}}
\caption{Impurity as a function of efficiency for $e^-/\gamma$ discrimination. The results were obtained for visible energies between 2.0\,MeV and 2.5\,MeV. All three methods were used on events with detector radii between 9.5\,m and 10.5\,m.}
\label{fig:egamma_efficiency}
\end{figure}

\begin{figure}[hbt]
	\centering
	\subfigure {\includegraphics[width=7cm]{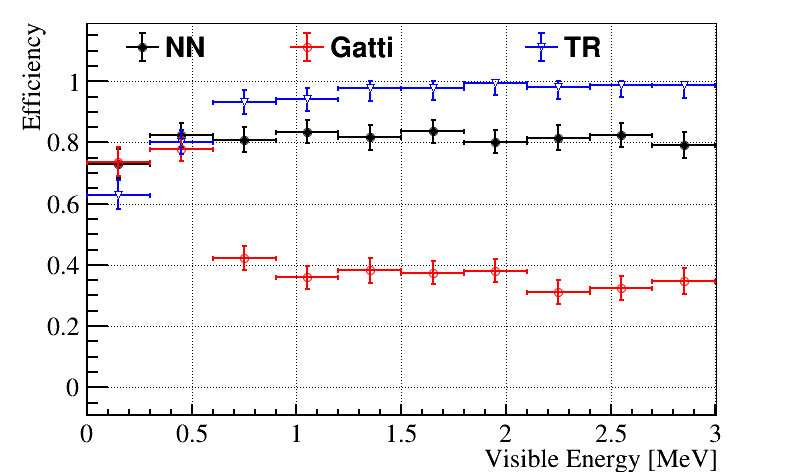}}\quad
	\subfigure {\includegraphics[width=7cm]{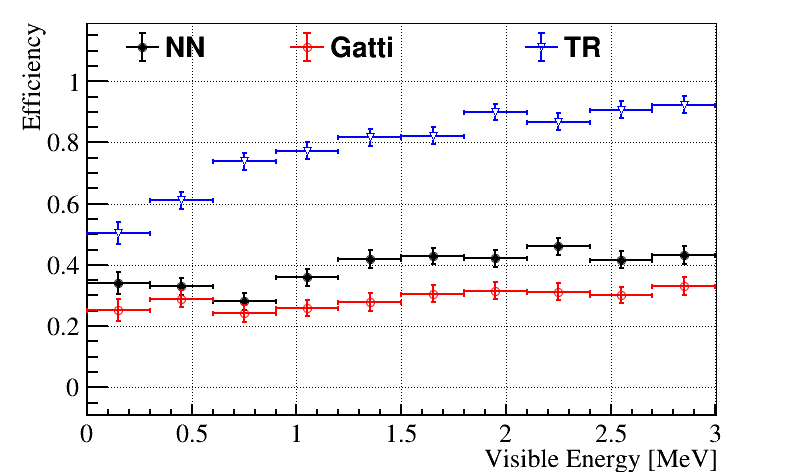}}\\
	\subfigure[Dataset 1]{\includegraphics[width=7cm]{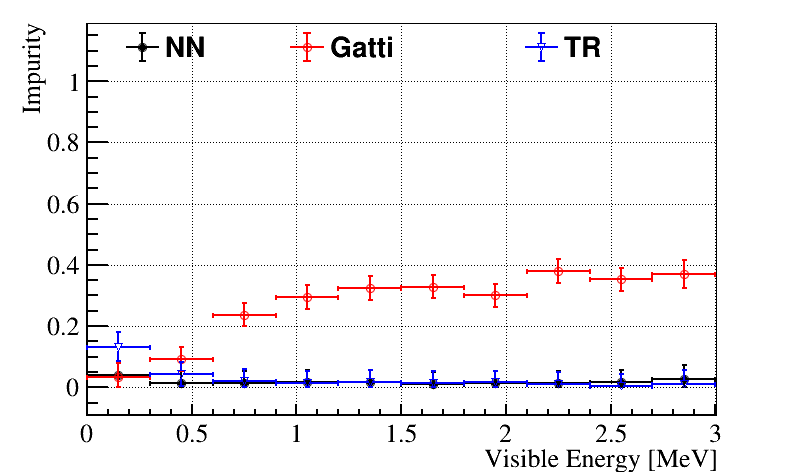}}\quad
	\subfigure[Dataset 2] {\includegraphics[width=7cm]{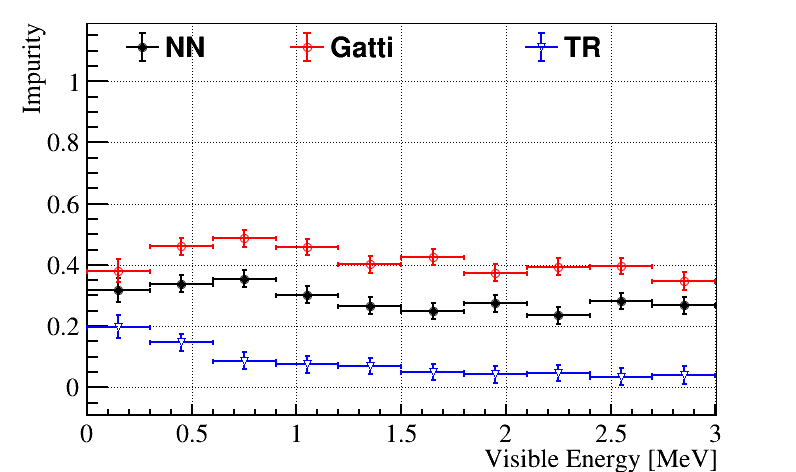}}\\
	\subfigure {\includegraphics[width=7cm]{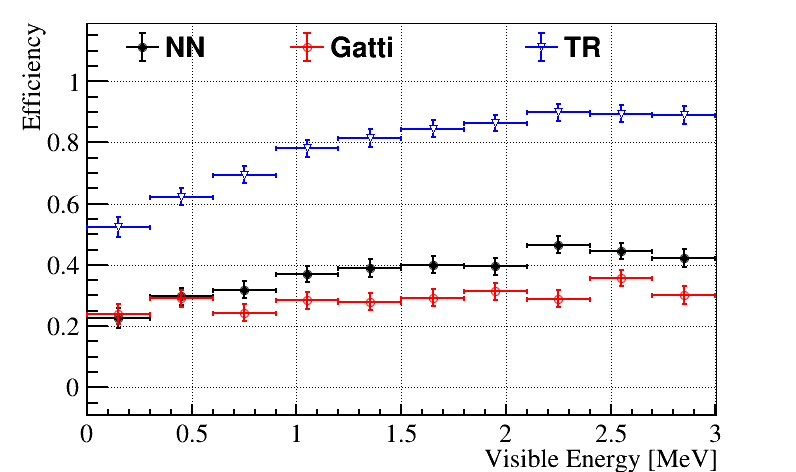}}\quad
	\subfigure {\includegraphics[width=7cm]{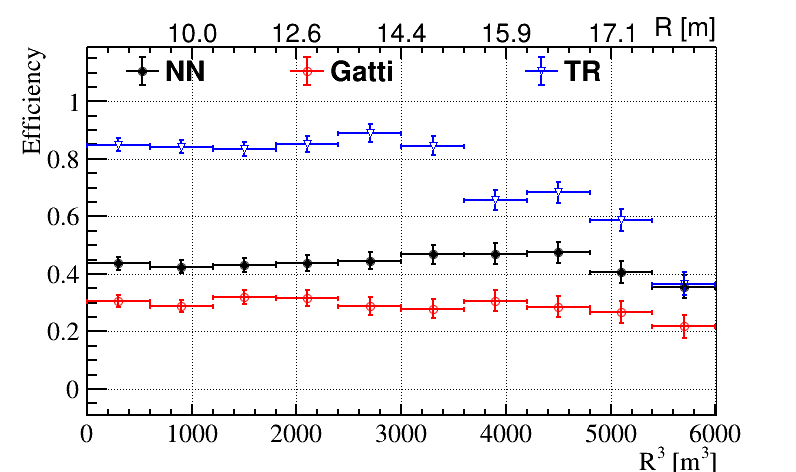}}\\
	\subfigure[Dataset 3] {\includegraphics[width=7cm]{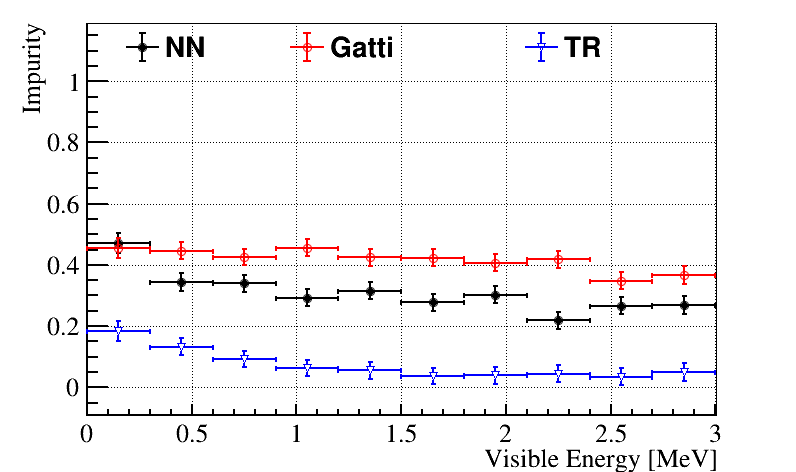}}\quad
	\subfigure[Dataset 3] {\includegraphics[width=7cm]{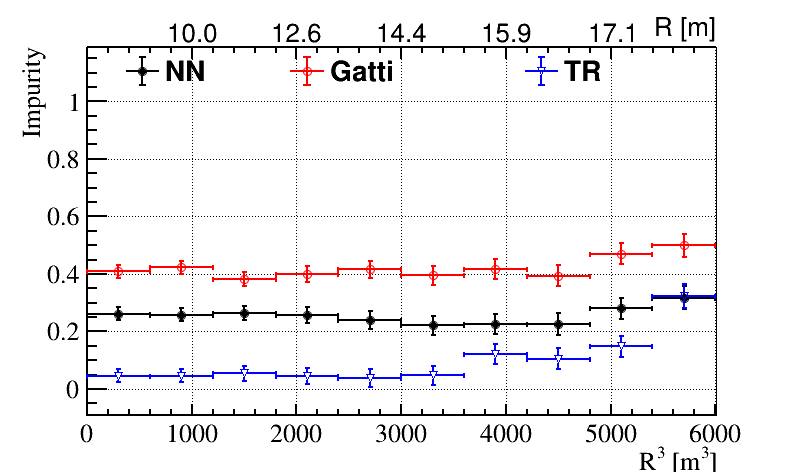}}
	\caption{Performance of the $e^-/\gamma$ discrimination from all three methods. Impurity was obtained at efficiency fixed to 50\%, while efficiency was obtained at impurity fixed to 20\,\%. (a), (b), and (c) show results with Datasets 1, 2, and 3, respectively, as a function of visible energy. All three methods were used on events with detector radii between 9.5\,m and 10.5\,m. (d) shows the performance depending on the detector radius for events with visible energies between 2.0\,MeV and 2.5\,MeV.}
	\label{fig:egamma_disc_energy}
\end{figure}

The discrimination of $e^-$ signal against $\gamma$ background was expected to be the most challenging of the investigated categories. 
The efficiency scan is shown in Fig.~\ref{fig:egamma_efficiency} for visible energies between 2.0\,MeV and 2.5\,MeV. 
Like for $e^+/e^-$, the NN and TR method prove to be sensitive to the task.
Again, a large gap between Datasets 1 and 2 indicates that in future detectors much potential can still be exploited by more accurate light sensors. 
The Gatti parameter, on the other hand, is hardly able to discriminate at all.

Energy and radius dependence were studied at 50\% efficiency and 10\% impurity. 
The results are shown in Fig.~\ref{fig:egamma_disc_energy} (a), (b), and (c) for energy dependence, and (d) for radius dependence at visible energies between 2.0\,MeV and 2.5\,MeV.
The data situation is similar to the $e^+/e^-$ case because the critical features in the time profiles arise at early times.
Within the energy range shared by the $e^+/e^-$ and $e^-/\gamma$ study, i.e.~between 1\,MeV and 3\,MeV, the TR parameter reaches comparable results. 
Below 1\,MeV, low statistics deteriorate the performance.
The NN is less powerful in $e^-/\gamma$ but approaches the TR results with rising energy. 
Since gammas spread their energy over a wider region with rising energy, a continuous decrease in impurity is expected even beyond the investigated energy region.
This actually does apply for certain gammas not descending from the natural decay chains, e.g.~6\,MeV and 8.5\,MeV gammas from neutron captures in the stainless steel surrounding JUNO's acrylic CD sphere.


\section{Conclusion}
\label{sec:conclusion}

The potential for event discrimination in JUNO was extensively studied on the basis of three distinct methods and different event pairings. 
Our studies concerning $\alpha/\beta$ and $p/\beta$ promise very clean and reliable cuts even at very low energies around 0.2\,MeV. 
A high p.e.~yield brings also $e^+/e^-$ discrimination into reach, although higher impurity rates only allow for statistical classification here. 
We point out that better results can be achieved for cosmogenic $^{10}$C background, whose $\beta^+$-decay is followed with 1\,ns delay by a 718\,keV $\gamma$ transition~\cite{HenningPhdthesis}. 
A separation between $e^-$ and $\gamma$ events was considered more challenging but was actually found to be feasible for energies above 1\,MeV. 
It needs to be investigated in a dedicated study how our discrimination would influence JUNO's sensitivities in the solar neutrino sector.
While the gamma contamination is expected to grow exponentially with detector radius, the usable volume is linked with its third power, meaning that already small expansions in fiducial radius would lead to a massive gain in the amount of data.
However, the accuracy is not high enough to expand JUNO's fiducial volume significantly in solar neutrino studies.

Except for $e^+/e^-$, all event pairings showed a continuous trend to gain in discrimination performance with visible energy. 
The former case however differentiates from the others since the decisive $\gamma$ component from $e^+$ annihilation is constant in terms of energy deposition and recedes behind the contribution from kinetic energy.
The examination of radius dependence revealed a correlation between cut performance and number of detected photons. 
This is in accordance with the observed energy dependence and causes the best results to show up at detector radii between 10\,m and 16\,m.

Apart from the fraction of direct light and the absolute p.e.~yield, also the technical equipment influences data quality. 
We found that timing uncertainties of PMTs, in turn being related to the vertex resolution, have a strong impact on the discrimination. 
Dark noise affects the results particularly at low energies, where they significantly reduce the relative fraction of direct photon signals.
The potential lying in an optimised detector can be learned from the big gap which still exists between our results with the ideal Dataset 1 and the more realistic Datsets 2 and 3.
This is valuable input for the design of future detectors like THEIA \cite{theia19}, where photo sensors with $\sim100\,$ps resolution represent a design option.

A direct comparison between the discrimination methods shows the power of the applied NN in spite of its simple architecture, as it proves to be sensitive to all studied cases.
Note that the 1D input to the network already means a reduction of the available event data. 
Therefore, it is not unexpected that TR can return better results at some points, e.g.~in $e^-/\gamma$ discrimination.
A NN considering hit information in three spatial dimensions plus hit time should in principle be able to outperform any classic approach.
However, this would also require a deeper and much more elaborated network architecture and a significantly more extensive training effort.
The Gatti and TR method played out their strengths in different disciplines. Gatti returns good results in $\alpha/\beta$ and $e^-/p$ discrimination, both relying on characteristics in the time spectrum of scintillation which show up especially in the tail region of the time profiles. The TR performs to its full potential in $e^+/e^-$ and $e^-/\gamma$ discrimination, where the distinction features manifest themselves around the profile peak.
Further efforts in the development of the TR need to focus on the performance towards the detector edge, where distortions momentarily impedes PID. 

\section*{Acknowledgements}

This work was supported by the Deutsche Forschungsgemeinschaft and the Helmholtz Association. Furthermore, we would like to thank the JUNO collaboration for providing the Monte Carlo software with which we carried out the simulations for this study.

\clearpage

\appendix
\section{Full collection of plots for $\alpha/\beta$ discrimination}
\label{app:alpha_beta} 

\begin{figure}[htb]
\centering
	\subfigure[Gatti]{\includegraphics[width=4.5cm]{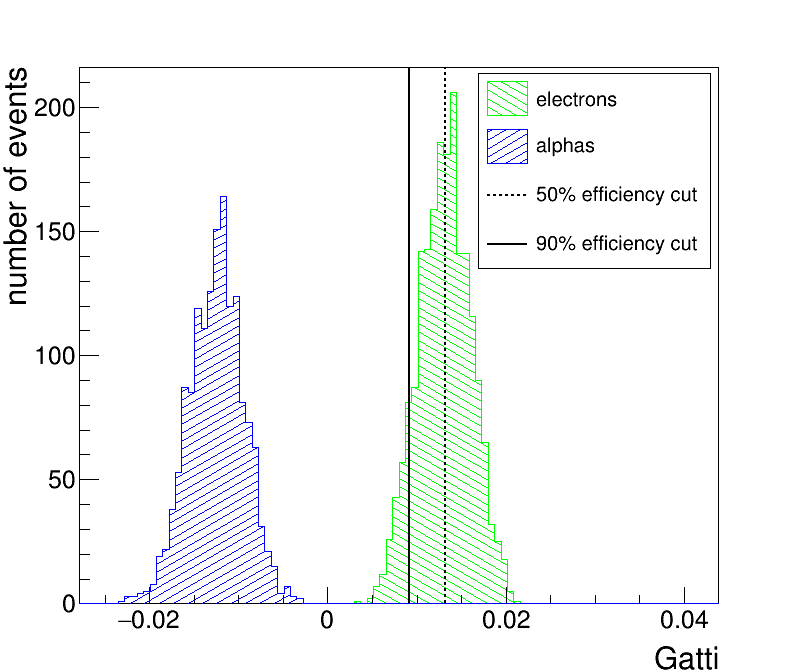}}\quad
	\subfigure[NN] {\includegraphics[width=4.5cm]{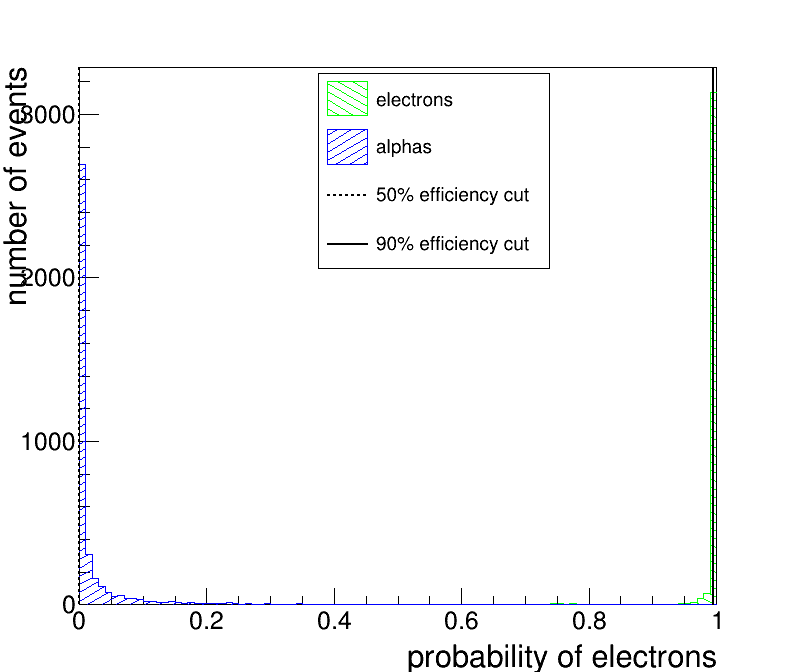}}\quad
	\subfigure[TR] {\includegraphics[width=4.5cm]{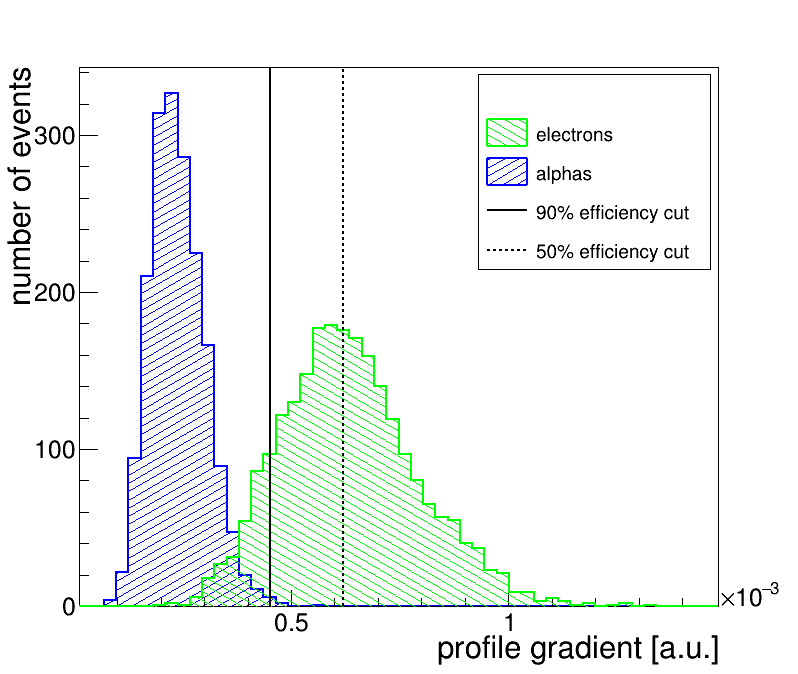}}
\caption{Distribution of the discrimination parameters in all three methods for $\alpha$ and $\beta$ events with visible energies between 1.0\,MeV and 1.5\,MeV. All three methods were used on events with detector radii between 9.5\,m and 10.5\,m.}
\label{fig:app_paramDistribution_alphabeta}
\end{figure}

\begin{figure}[hbt]
\centering
	\subfigure[Gatti] {\includegraphics[width=4.5cm]{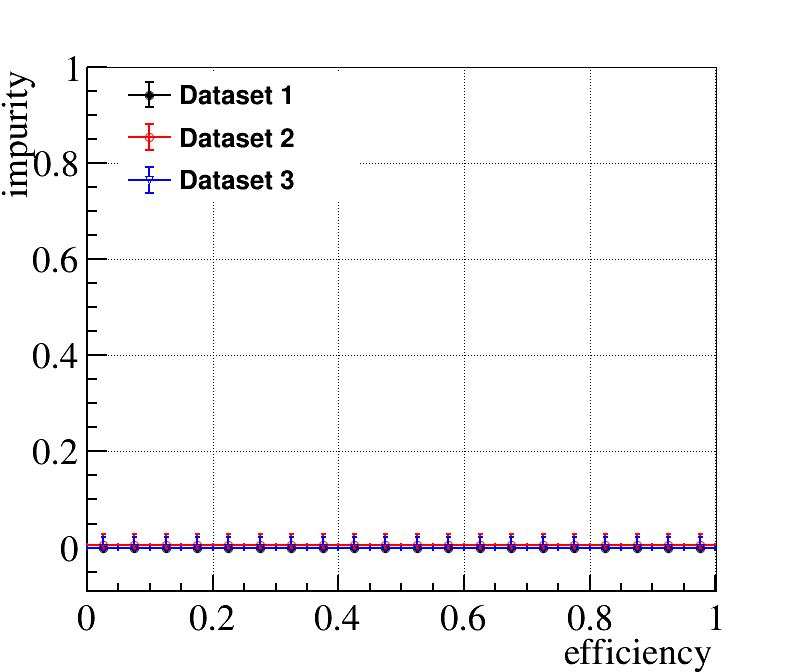}}\quad
	\subfigure[NN] {\includegraphics[width=4.5cm]{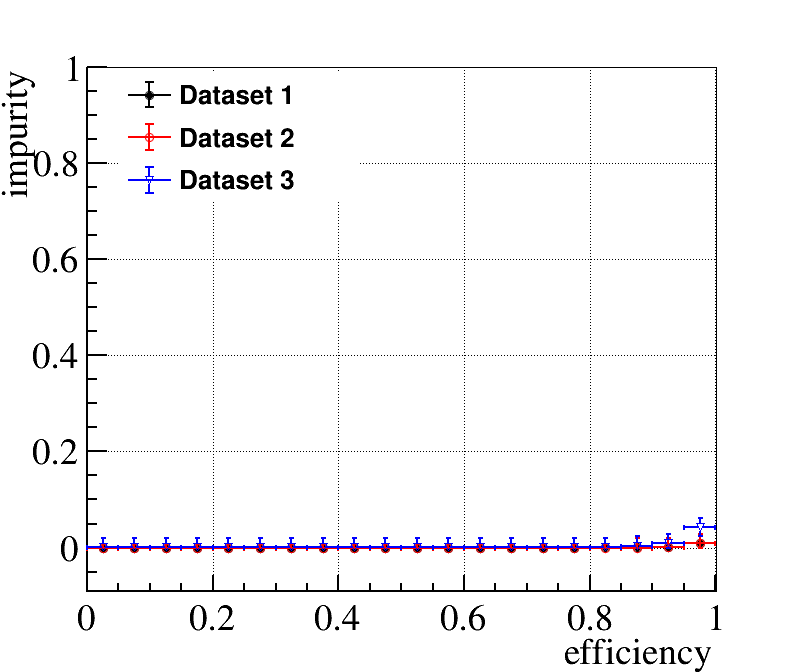}}\quad
	\subfigure[TR] {\includegraphics[width=4.5cm]{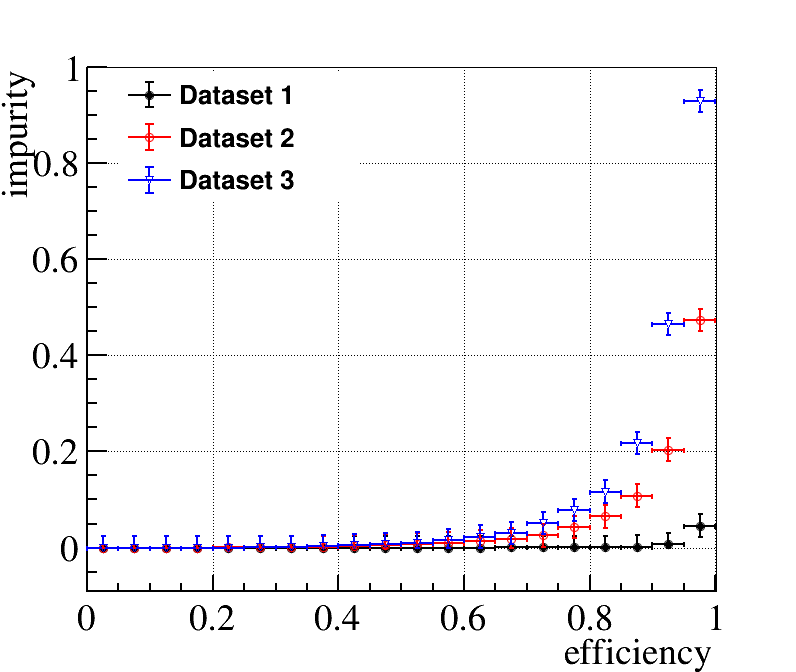}}
\caption{Impurity as a function of efficiency for $\alpha/\beta$ discrimination. The results were obtained for visible energies between 1.0\,MeV and 1.5\,MeV. All three methods were used on events with detector radii between 9.5\,m and 10.5\,m. Dataset 1 was used, i.e.~TTS, vertex smearing, and dark noise were not considered.}
\label{fig:app_egamma_efficiency}
\end{figure}

\clearpage

\begin{figure}[hbt]
	\centering
	\subfigure {\includegraphics[width=4.5cm]{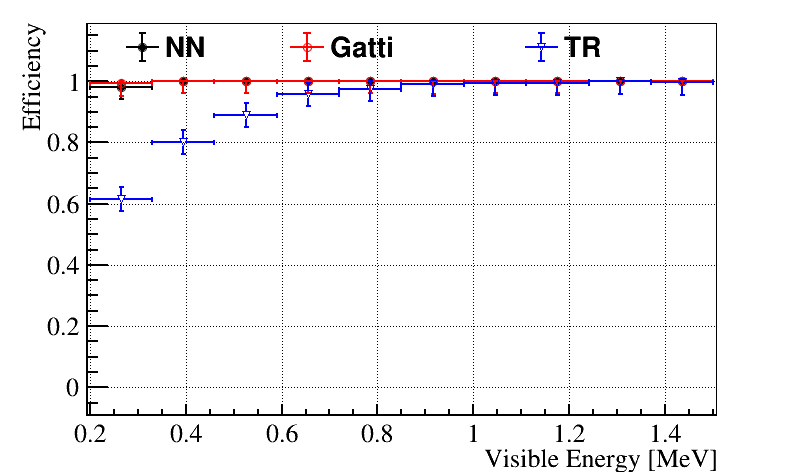}}\quad
	\subfigure {\includegraphics[width=4.5cm]{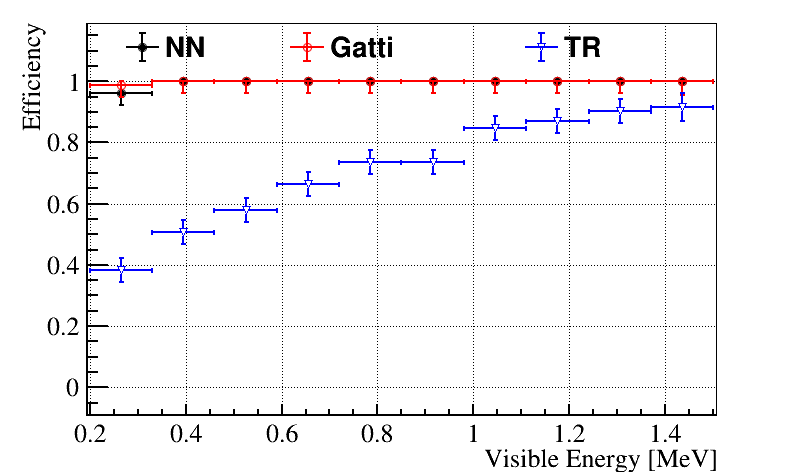}}\quad	
	\subfigure {\includegraphics[width=4.5cm]{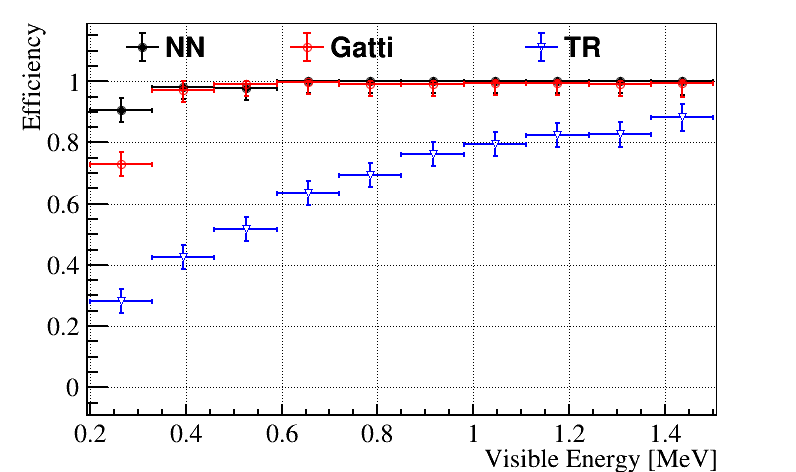}}\\

	\subfigure[Dataset 1]{\includegraphics[width=4.5cm]{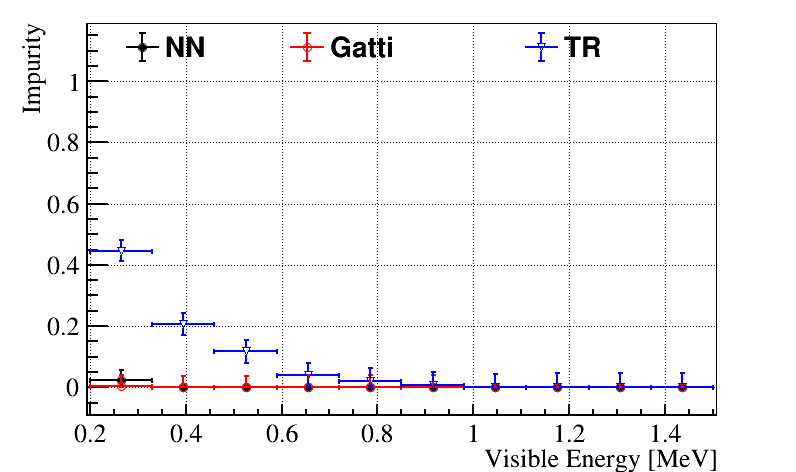}}\quad
	\subfigure[Dataset 2] {\includegraphics[width=4.5cm]{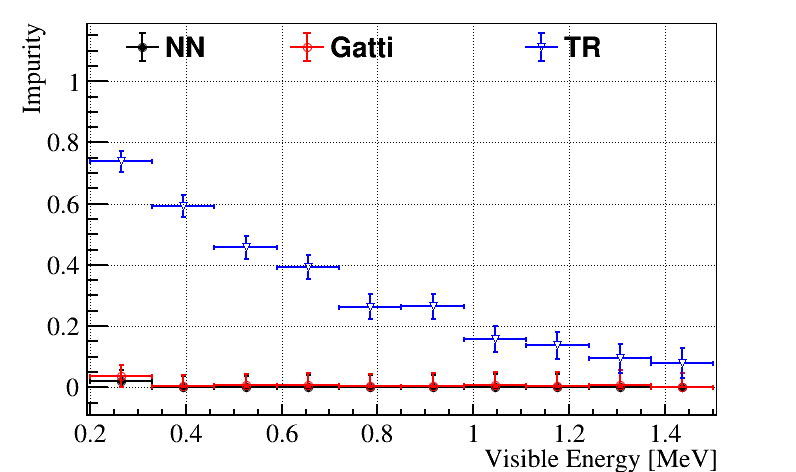}}\quad
	\subfigure[Dataset 3] {\includegraphics[width=4.5cm]{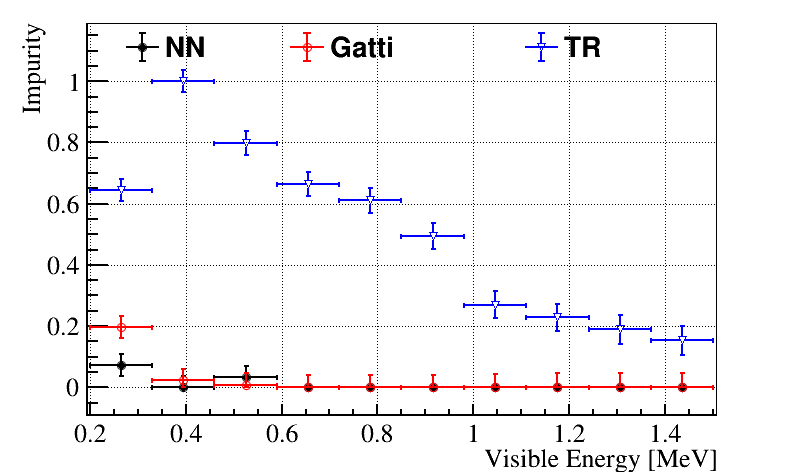}}


	\caption{Energy dependence of the $\alpha/\beta$ discrimination from all three methods. Impurity was obtained at efficiency fixed to 90\%, while efficiency was obtained at impurity fixed to 10\,\%. All three methods were used on events with detector radii between 9.5\,m and 10.5\,m.}
	\label{fig:app_egamma_disc_energy}
\end{figure}

\begin{figure}[hbt]
	\centering
	\subfigure {\includegraphics[width=4.5cm]{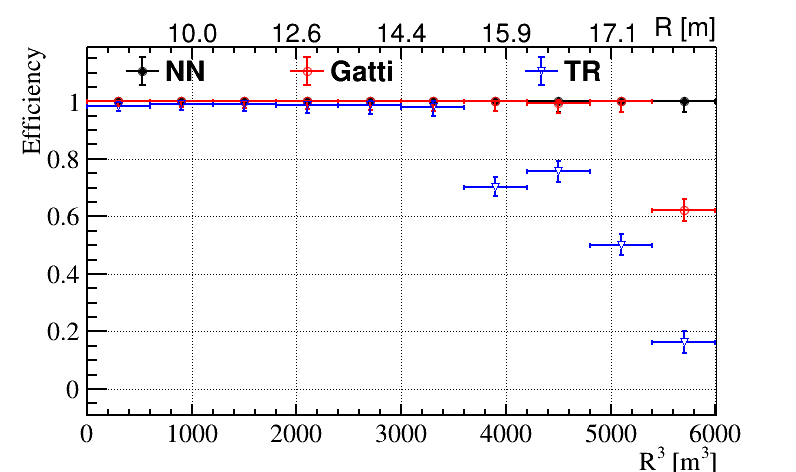}}\quad
	\subfigure {\includegraphics[width=4.5cm]{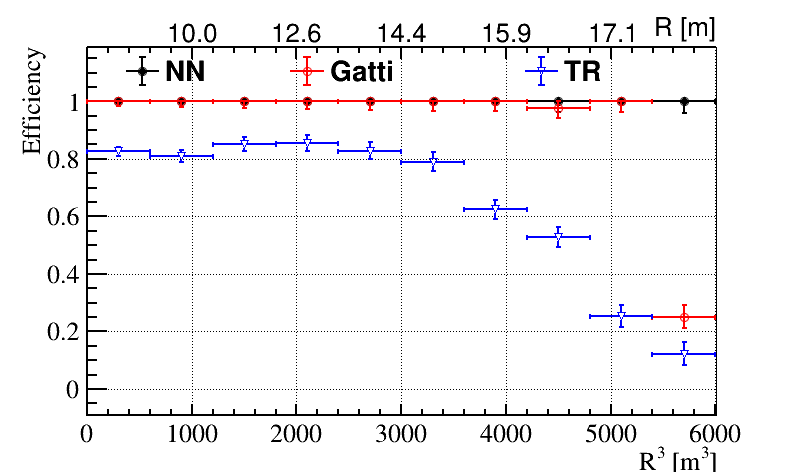}}\quad	
	\subfigure {\includegraphics[width=4.5cm]{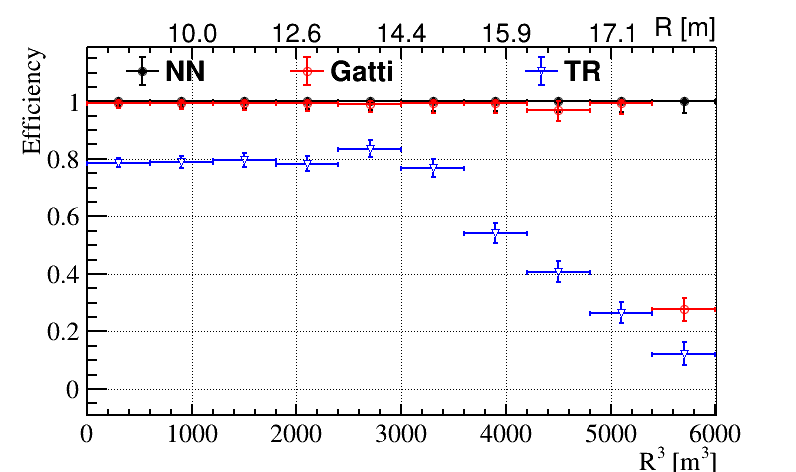}}\\

	\subfigure[Dataset 1]{\includegraphics[width=4.5cm]{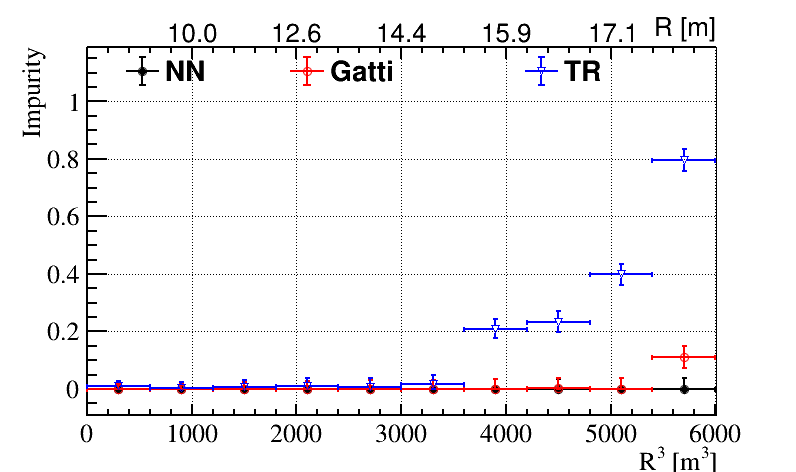}}\quad
	\subfigure[Dataset 2] {\includegraphics[width=4.5cm]{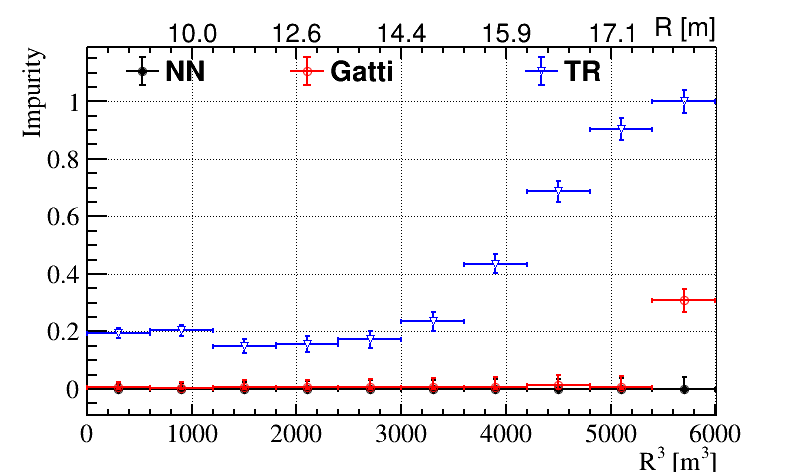}}\quad
	\subfigure[Dataset 3] {\includegraphics[width=4.5cm]{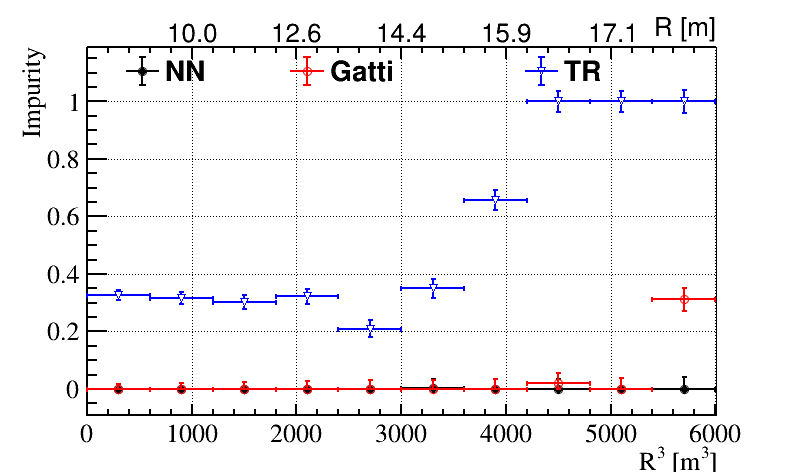}}


	\caption{Radius dependence of the $\alpha/\beta$ discrimination from all three methods for events with visible energies between 1.0\,MeV and 1.5\,MeV. Impurity was obtained at efficiency fixed to 90\%, while efficiency was obtained at impurity fixed to 10\,\%.}
	\label{fig:app_egamma_disc_radius}
\end{figure}

\clearpage

\section{Full collection of plots for $p/\beta$ discrimination}
\label{app:ep}

\begin{figure}[htb]
\centering
	\subfigure[Gatti]{\includegraphics[width=4.5cm]{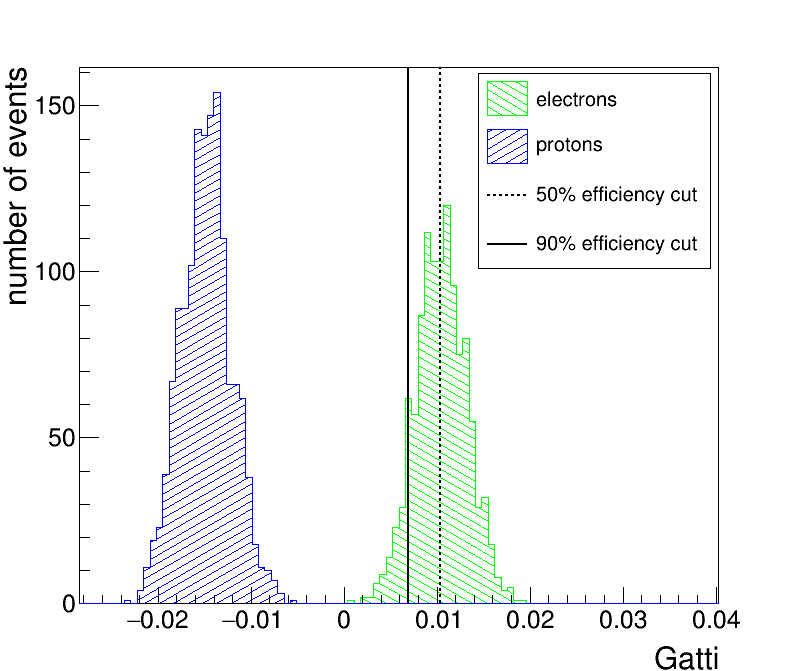}}\quad
	\subfigure[NN] {\includegraphics[width=4.5cm]{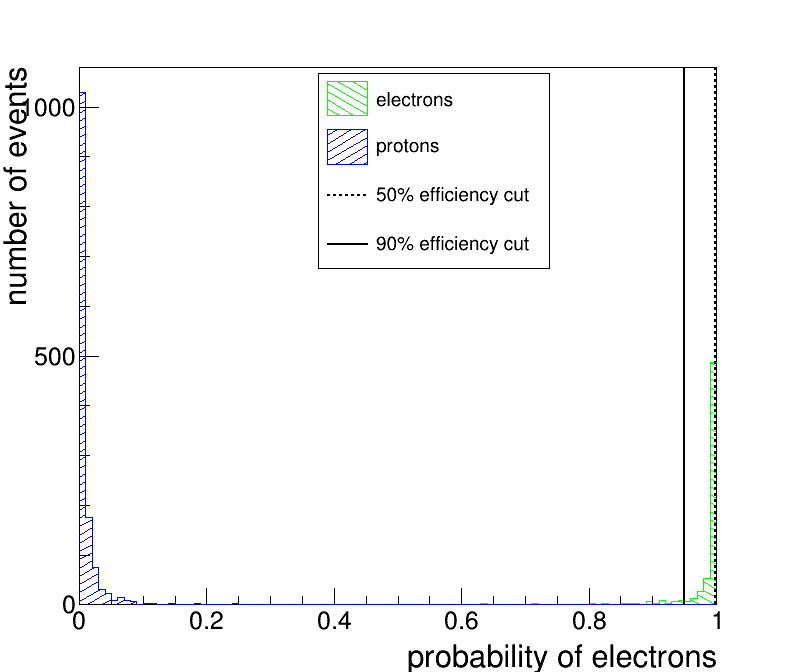}}\quad
	\subfigure[TR] {\includegraphics[width=4.5cm]{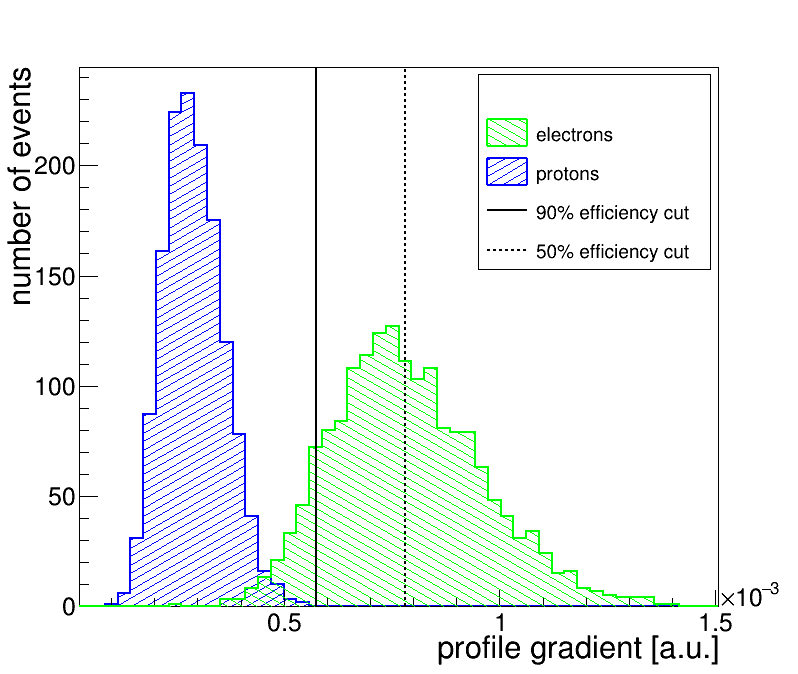}}
\caption{Distribution of the discrimination parameters in all three methods for $p$ and $e^-$ events with visible energies between 1.25\,MeV and 1.75\,MeV. All three methods were used on events with detector radii between 9.5\,m and 10.5\,m.}
\label{fig:app_paramDistribution_ep}
\end{figure}

\begin{figure}[hbt]
\centering
	\subfigure[Gatti] {\includegraphics[width=4.5cm]{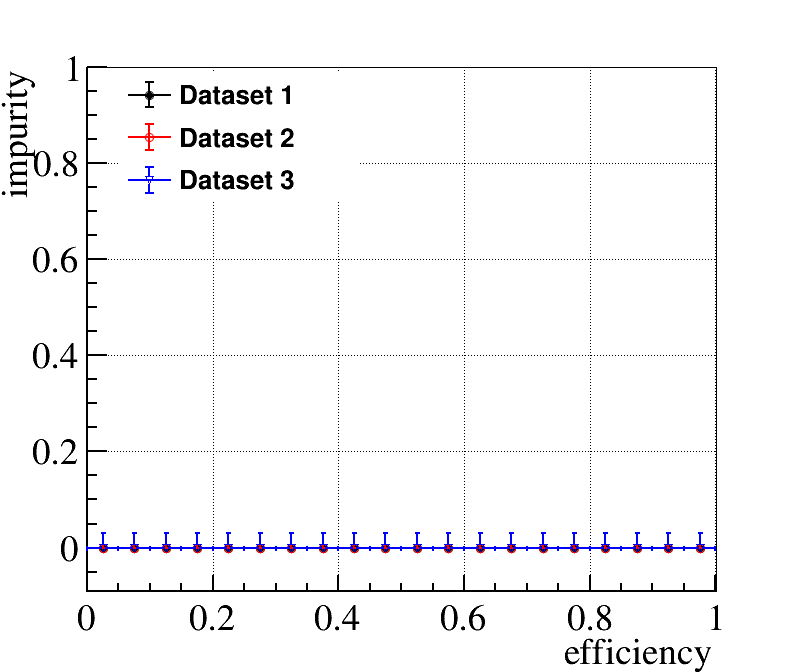}}\quad
	\subfigure[NN] {\includegraphics[width=4.5cm]{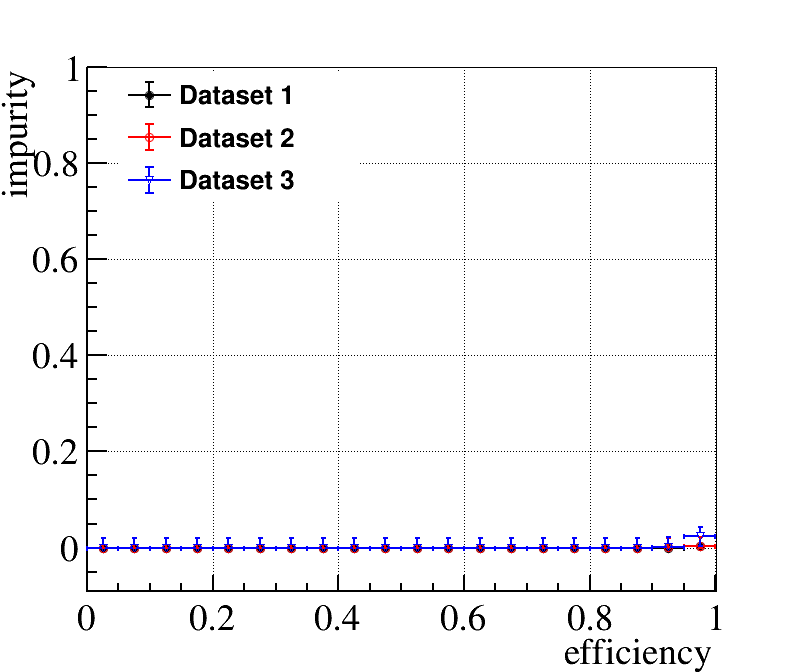}}\quad
	\subfigure[TR] {\includegraphics[width=4.5cm]{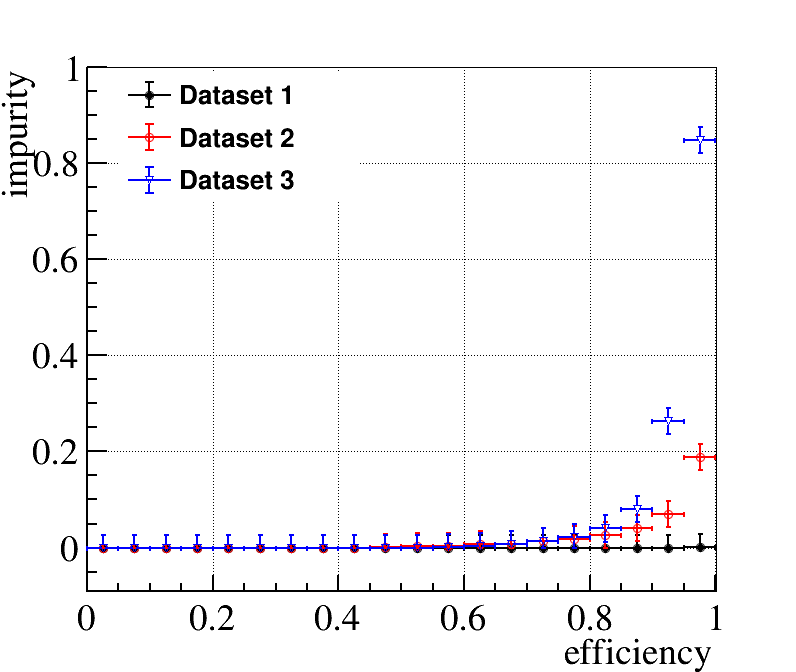}}
\caption{Impurity as a function of efficiency for $p/\beta$ discrimination. The results were obtained for visible energies between 1.5\,MeV and 2.0\,MeV. All three methods were used on events with detector radii between 9.5\,m and 10.5\,m. Dataset 1 was used, i.e.~TTS, vertex smearing, and dark noise were not considered.}
\label{fig:app_egamma_efficiency}
\end{figure}

\clearpage

\begin{figure}[hbt]
	\centering
	\subfigure {\includegraphics[width=4.5cm]{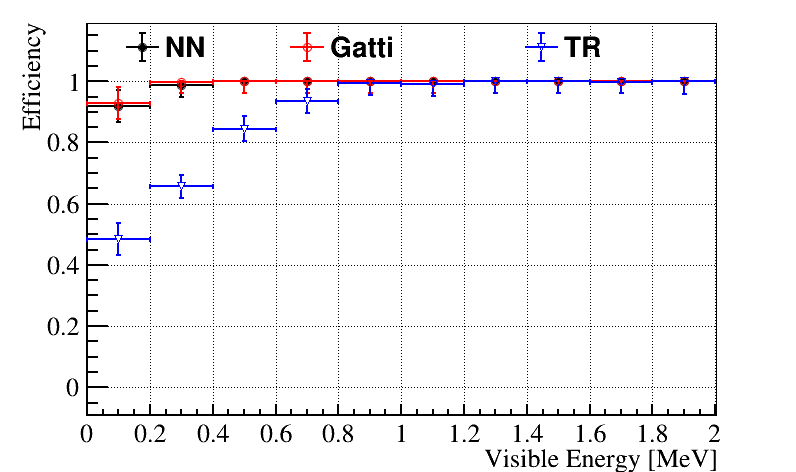}}\quad
	\subfigure {\includegraphics[width=4.5cm]{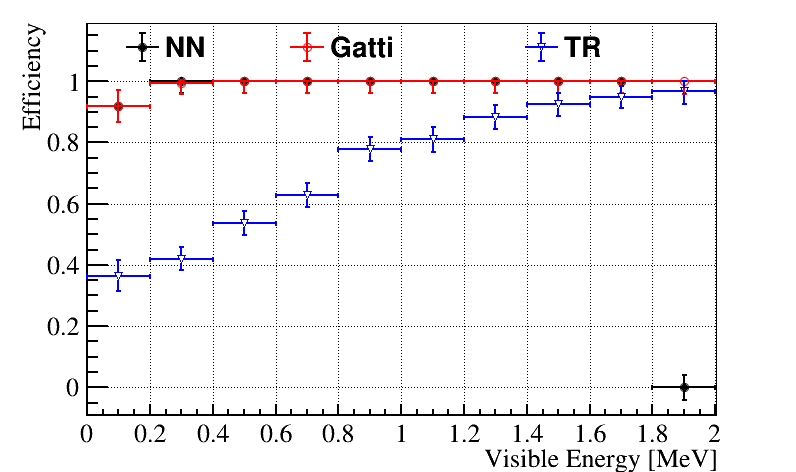}}\quad	
	\subfigure {\includegraphics[width=4.5cm]{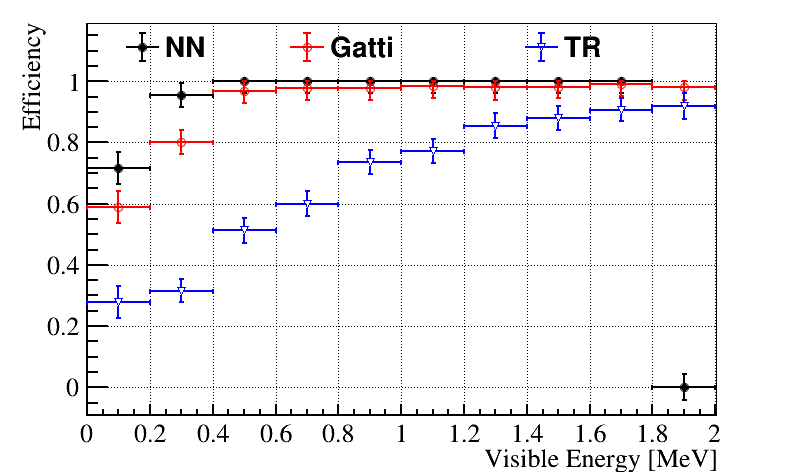}}\\

	\subfigure[Dataset 1]{\includegraphics[width=4.5cm]{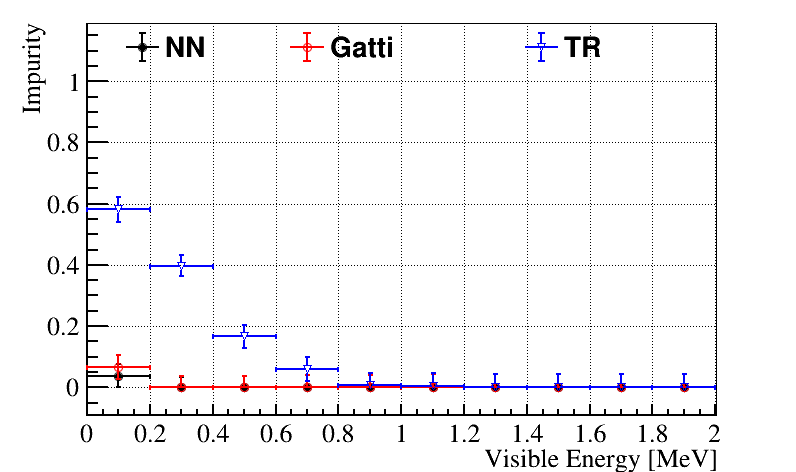}}\quad
	\subfigure[Dataset 2] {\includegraphics[width=4.5cm]{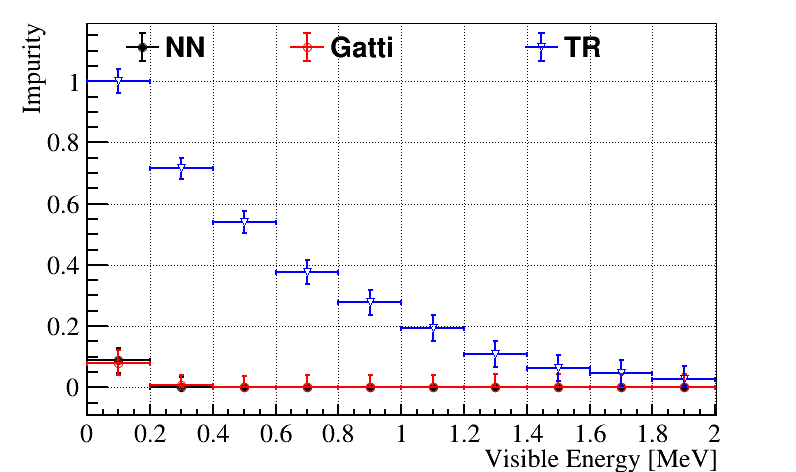}}\quad
	\subfigure[Dataset 3] {\includegraphics[width=4.5cm]{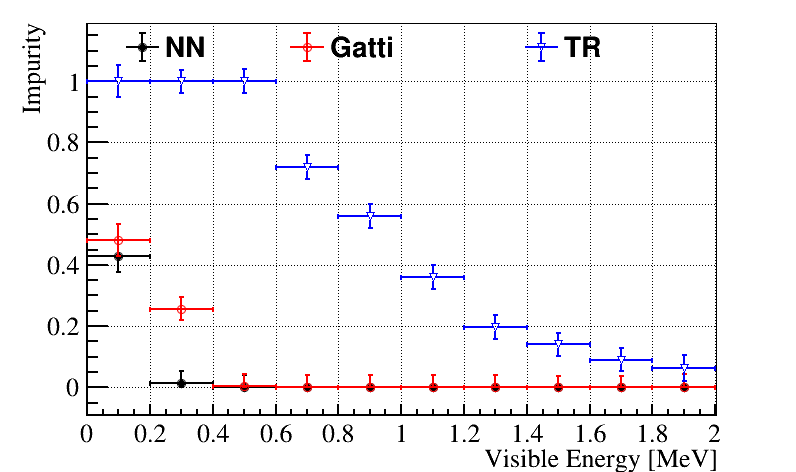}}


	\caption{Energy dependence of the $p/\beta$ discrimination from all three methods. Impurity was obtained at efficiency fixed to 90\%, while efficiency was obtained at impurity fixed to 10\,\%. All three methods were used on events with detector radii between 9.5\,m and 10.5\,m.}
	\label{fig:app_egamma_disc_energy}
\end{figure}

\begin{figure}[hbt]
	\centering
	\subfigure {\includegraphics[width=4.5cm]{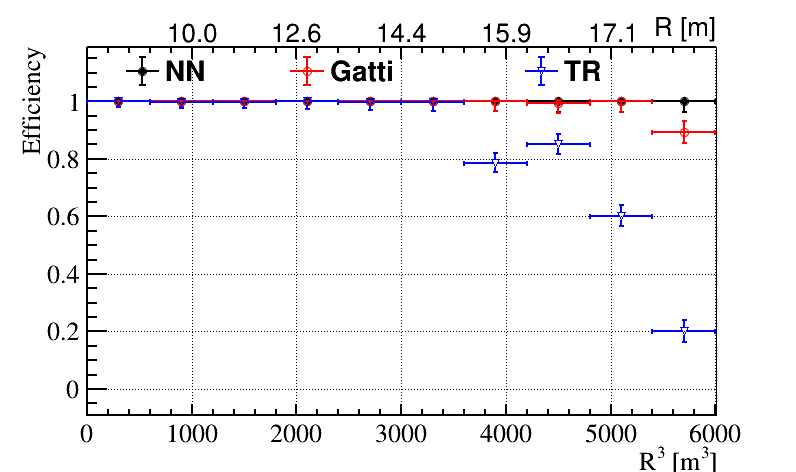}}\quad
	\subfigure {\includegraphics[width=4.5cm]{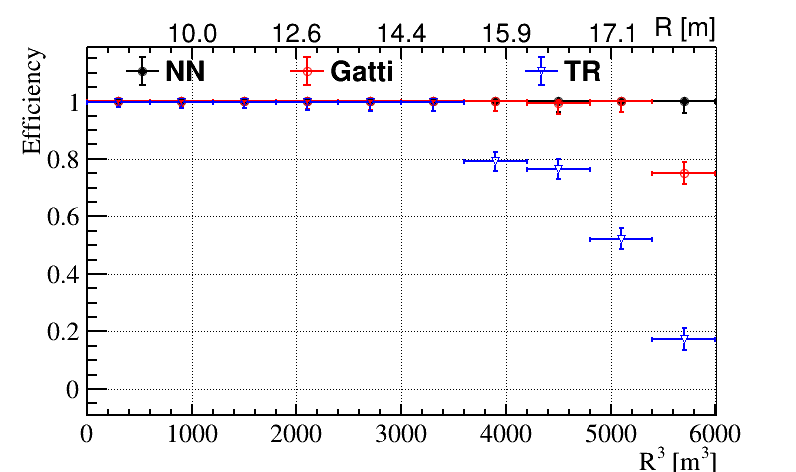}}\quad	
	\subfigure {\includegraphics[width=4.5cm]{resultsShared/ep_noise_radius_eff.png}}\\

	\subfigure[Dataset 1]{\includegraphics[width=4.5cm]{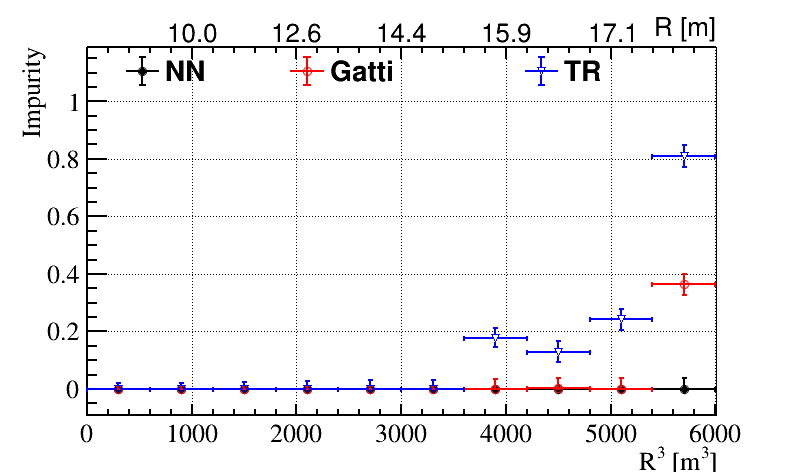}}\quad
	\subfigure[Dataset 2] {\includegraphics[width=4.5cm]{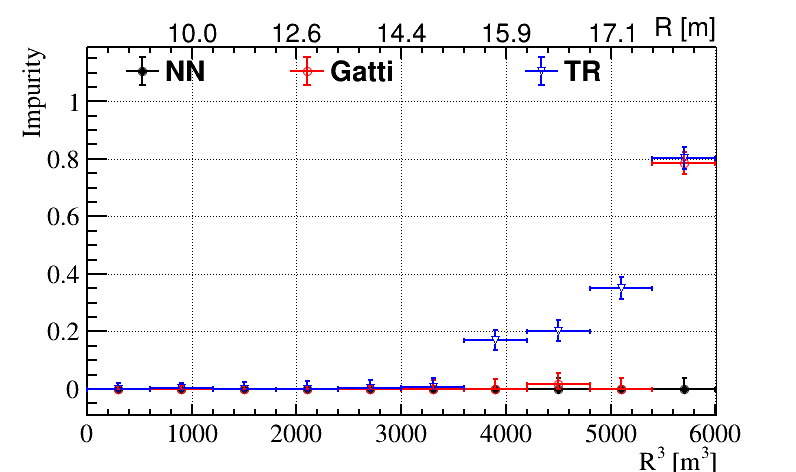}}\quad
	\subfigure[Dataset 3] {\includegraphics[width=4.5cm]{resultsShared/ep_noise_radius_imp.png}}


	\caption{Radius dependence of the $p/\beta$ discrimination from all three methods for events with visible energies between 1.5\,MeV and 2.0\,MeV. Impurity was obtained at efficiency fixed to 90\%, while efficiency was obtained at impurity fixed to 10\,\%.}
	\label{fig:app_egamma_disc_radius}
\end{figure}

\clearpage

\section{Full collection of plots for $e^+/e^-$ discrimination}
\label{app:elecposi}

\begin{figure}[htb]
\centering
	\subfigure[Gatti]{\includegraphics[width=4.5cm]{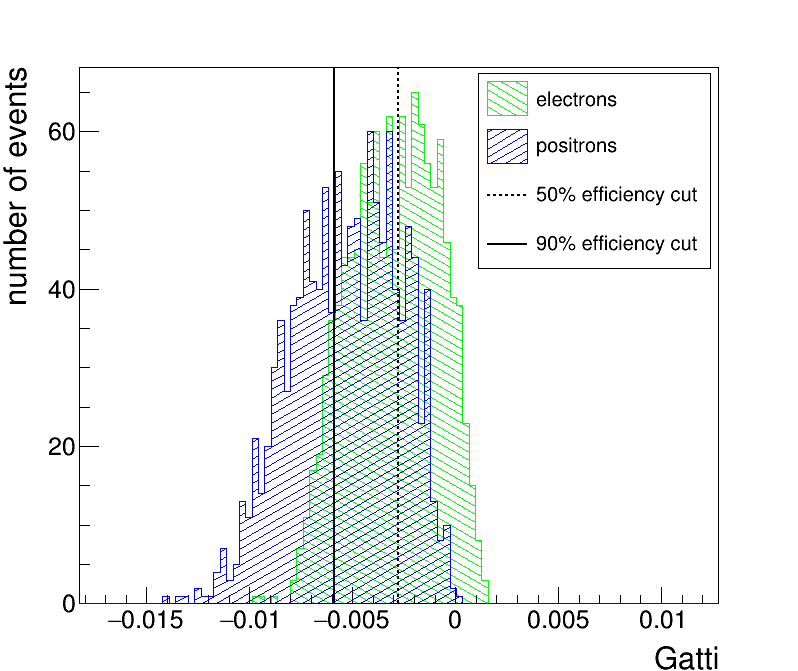}}\quad
	\subfigure[NN] {\includegraphics[width=4.5cm]{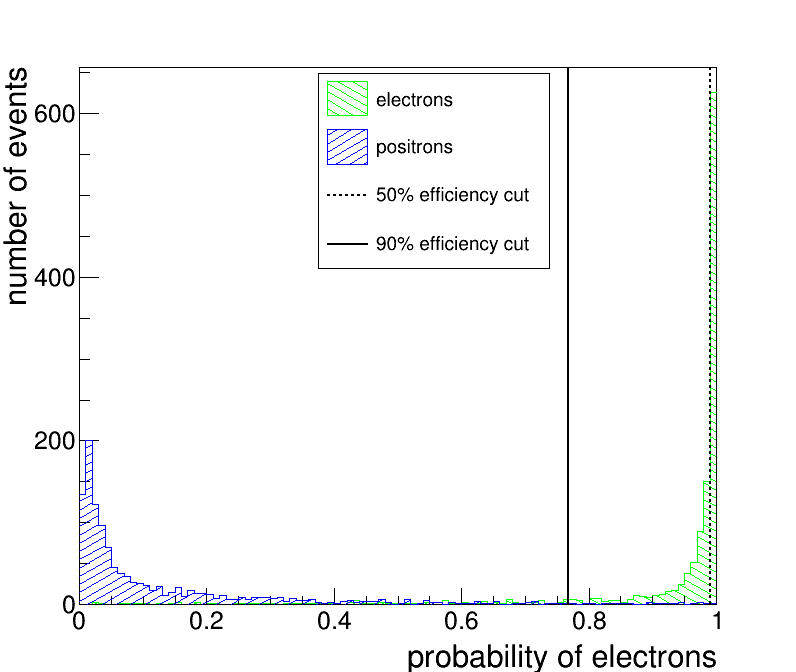}}\quad
	\subfigure[TR] {\includegraphics[width=4.5cm]{resultsTR/TR_dist_elecposi_3000keV.png}}
\caption{Distribution of the discrimination parameters in all three methods for $e^+$ and $e^-$ events with visible energies between 2.0\,MeV and 4.0\,MeV. All three methods were used on events with detector radii between 9.5\,m and 10.5\,m.}
\label{fig:app_paramDistribution_elecposi}
\end{figure}

\begin{figure}[hbt]
\centering
	\subfigure[Gatti] {\includegraphics[width=4.5cm]{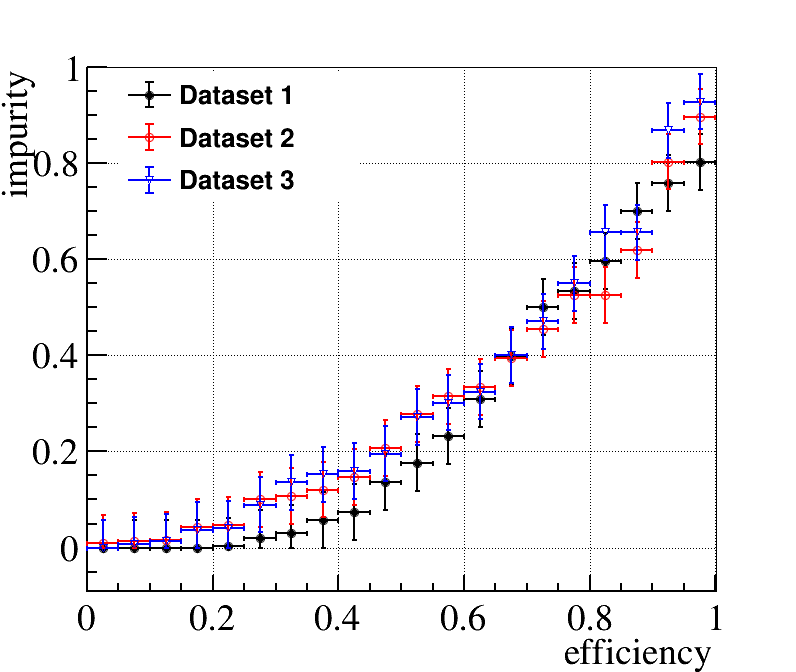}}\quad
	\subfigure[NN] {\includegraphics[width=4.5cm]{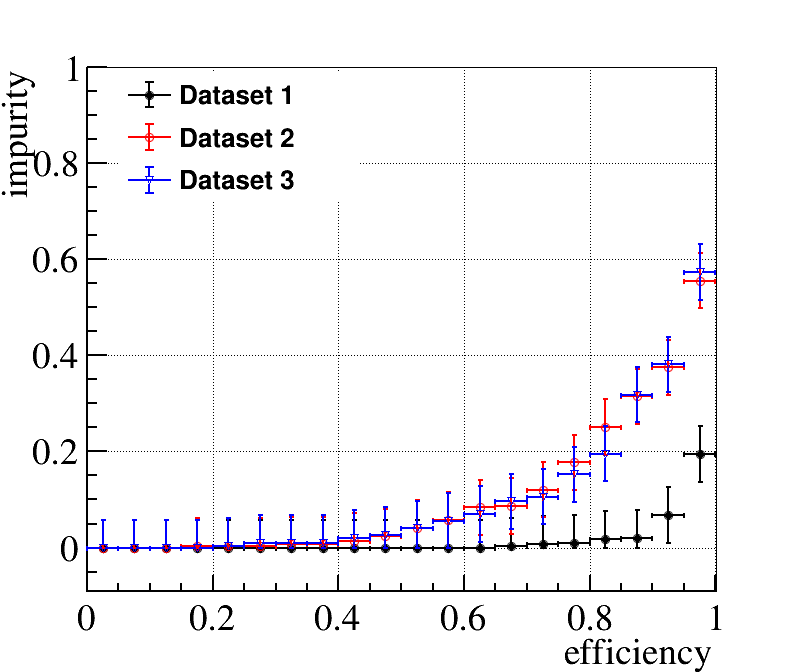}}\quad
	\subfigure[TR] {\includegraphics[width=4.5cm]{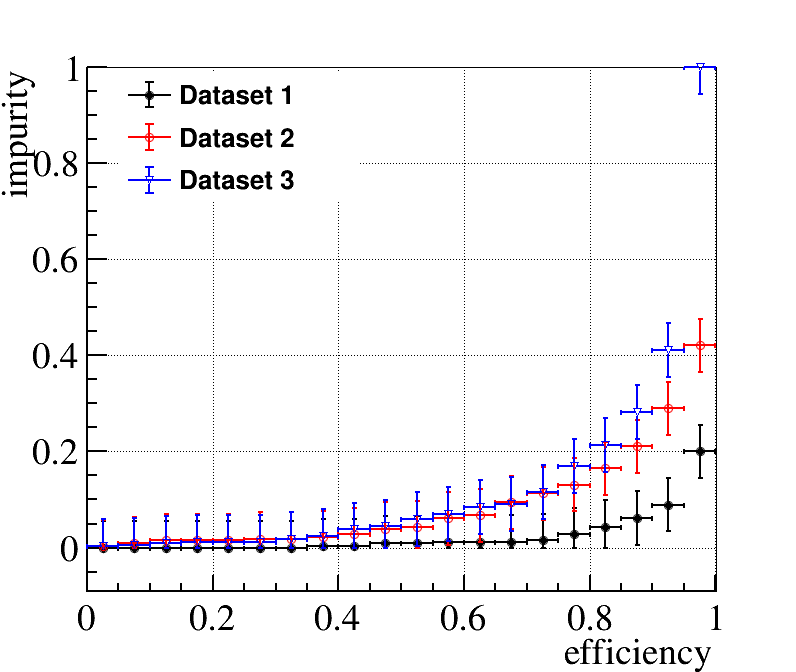}}
\caption{Impurity as a function of efficiency for $e^+/e^-$ discrimination. The results were obtained for visible energies between 2.75\,MeV and 3.25\,MeV. All three methods were used on events with detector radii between 9.5\,m and 10.5\,m. Dataset 1 was used, i.e.~TTS, vertex smearing, and dark noise were not considered.}
\label{fig:app_egamma_efficiency}
\end{figure}

\clearpage

\begin{figure}[hbt]
	\centering
	\subfigure {\includegraphics[width=4.5cm]{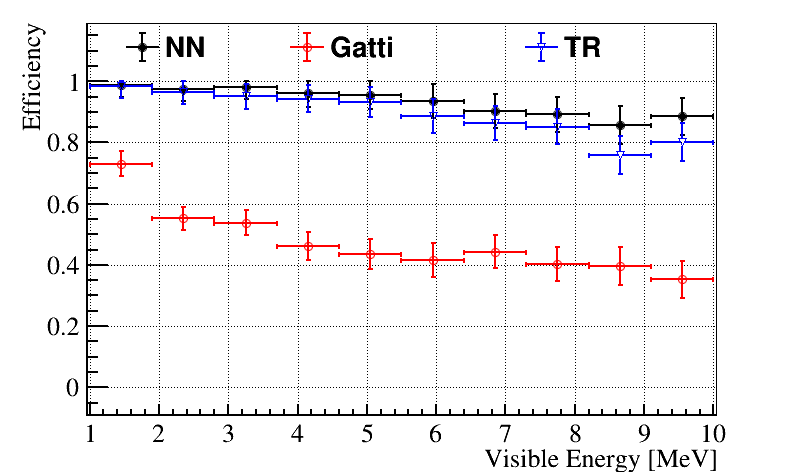}}\quad
	\subfigure {\includegraphics[width=4.5cm]{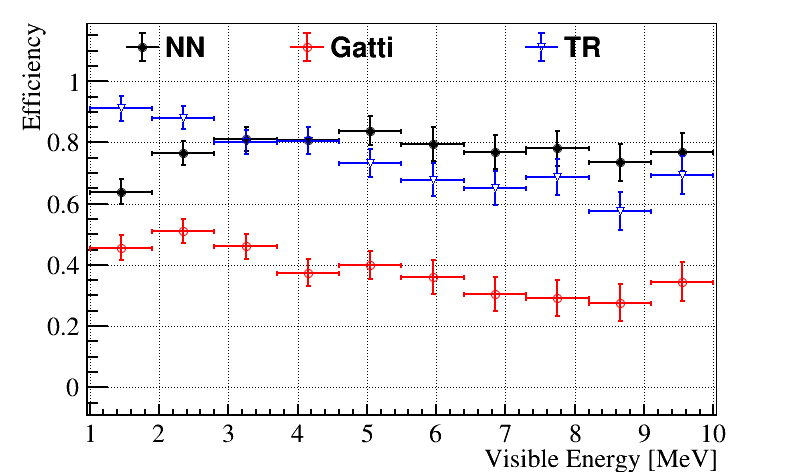}}\quad	
	\subfigure {\includegraphics[width=4.5cm]{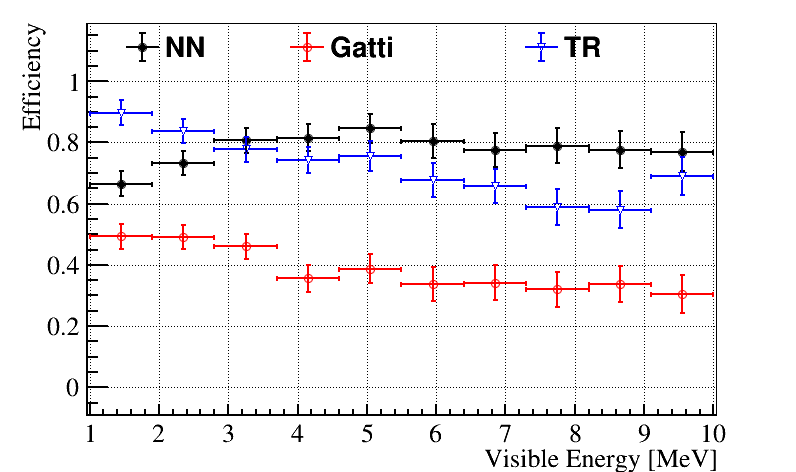}}\\

	\subfigure[Dataset 1]{\includegraphics[width=4.5cm]{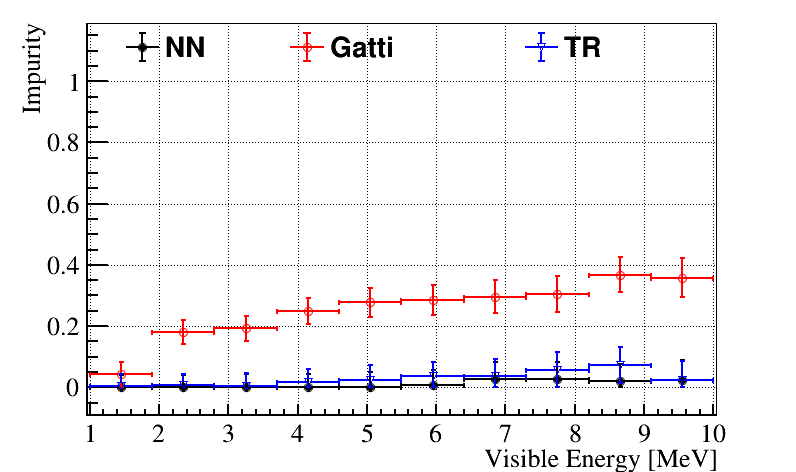}}\quad
	\subfigure[Dataset 2] {\includegraphics[width=4.5cm]{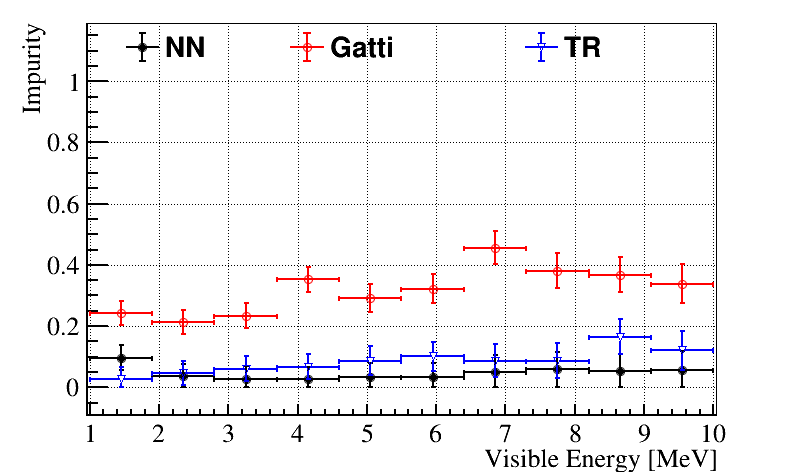}}\quad
	\subfigure[Dataset 3] {\includegraphics[width=4.5cm]{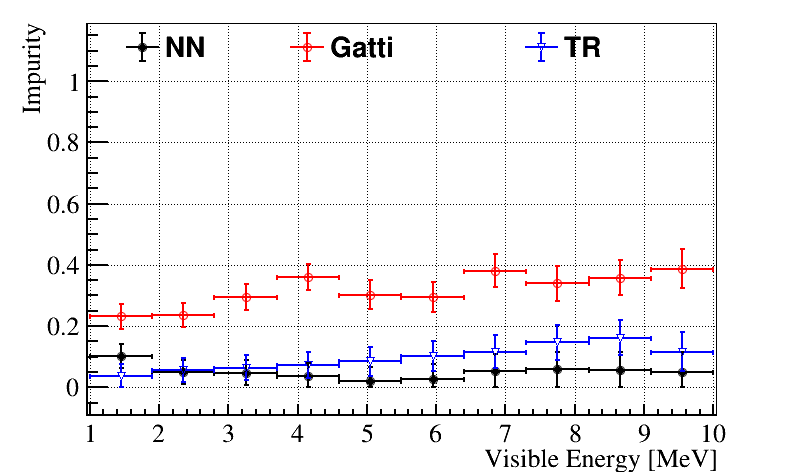}}


	\caption{Energy dependence of the $e^+/e^-$ discrimination from all three methods. Impurity was obtained at efficiency fixed to 50\%, while efficiency was obtained at impurity fixed to 20\,\%. All three methods were used on events with detector radii between 9.5\,m and 10.5\,m.}
	\label{fig:app_egamma_disc_energy}
\end{figure}

\begin{figure}[hbt]
	\centering
	\subfigure {\includegraphics[width=4.5cm]{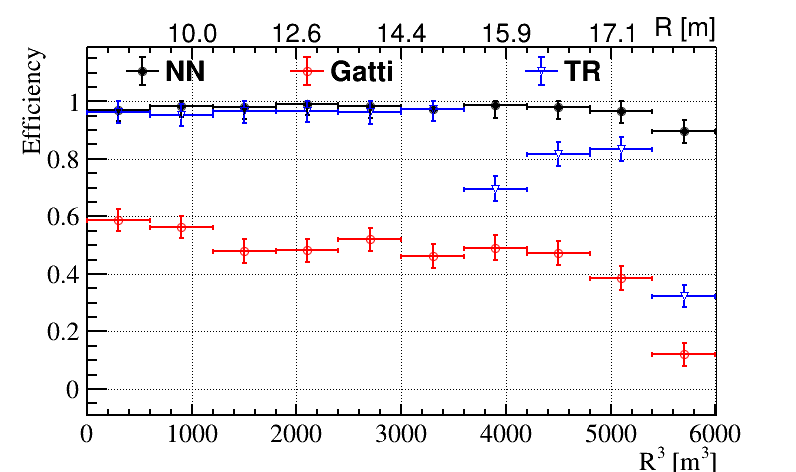}}\quad
	\subfigure {\includegraphics[width=4.5cm]{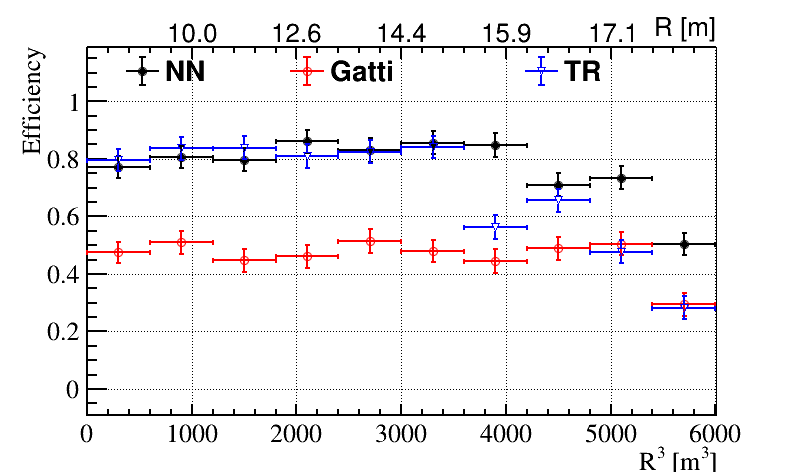}}\quad	
	\subfigure {\includegraphics[width=4.5cm]{resultsShared/elecposi_noise_radius_eff.png}}\\

	\subfigure[Dataset 1]{\includegraphics[width=4.5cm]{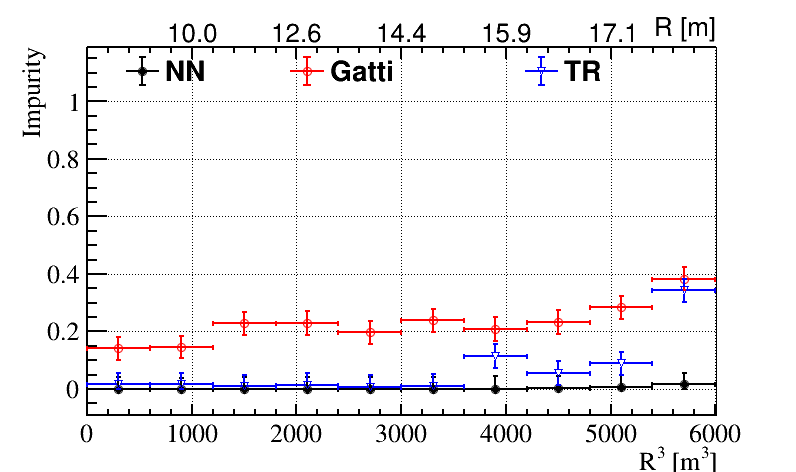}}\quad
	\subfigure[Dataset 2] {\includegraphics[width=4.5cm]{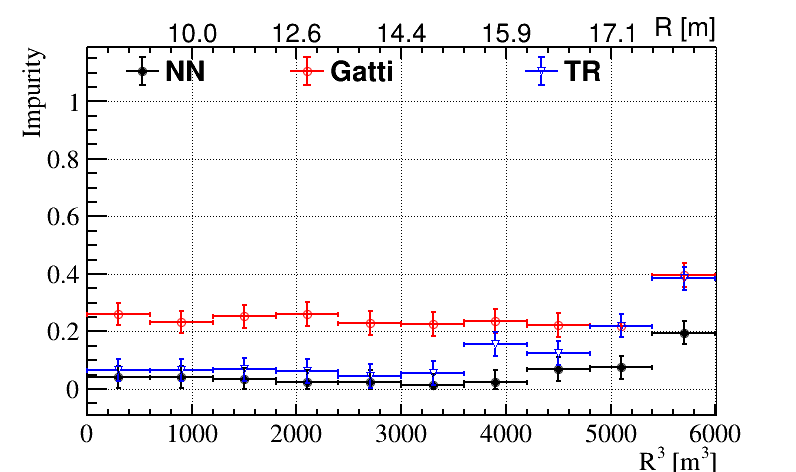}}\quad
	\subfigure[Dataset 3] {\includegraphics[width=4.5cm]{resultsShared/elecposi_noise_radius_imp.png}}


	\caption{Radius dependence of the $e^+/e^-$ discrimination from all three methods for events with visible energies between 2.75\,MeV and 3.25\,MeV. Impurity was obtained at efficiency fixed to 50\%, while efficiency was obtained at impurity fixed to 20\,\%.}
	\label{fig:app_egamma_disc_radius}
\end{figure} 

\clearpage

\section{Full collection of plots for $e^-/\gamma$ discrimination}
\label{app:egamma}

\begin{figure}[htb]
\centering
	\subfigure[Gatti]{\includegraphics[width=4.5cm]{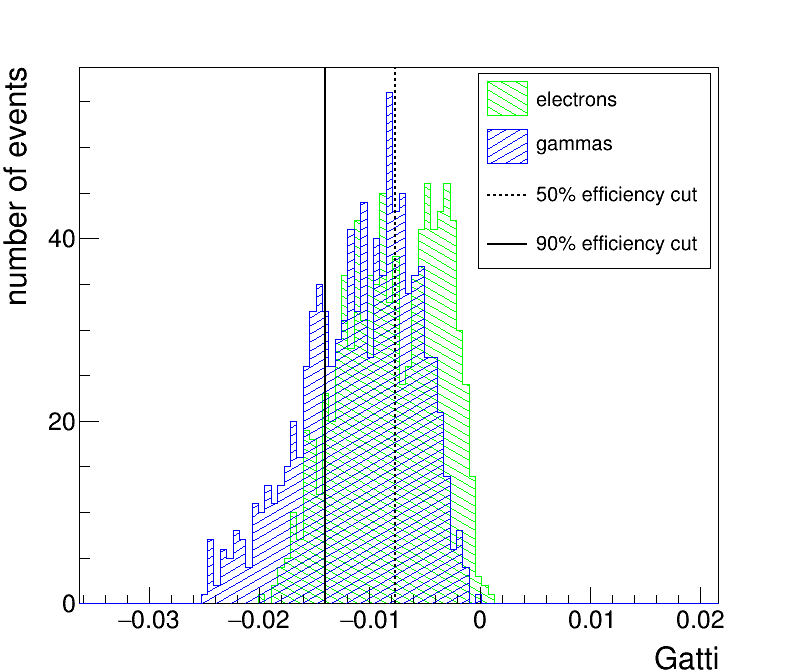}}\quad
	\subfigure[NN] {\includegraphics[width=4.5cm]{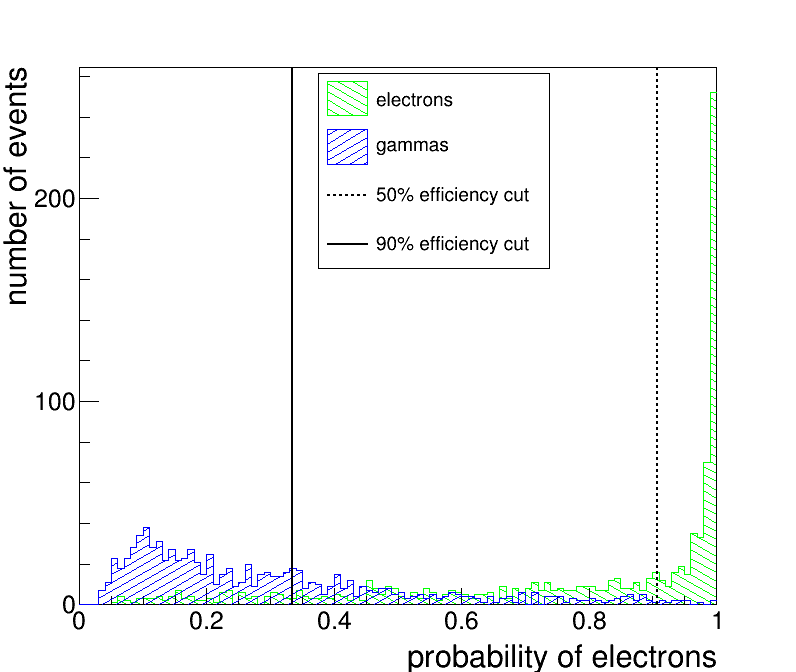}}\quad
	\subfigure[TR] {\includegraphics[width=4.5cm]{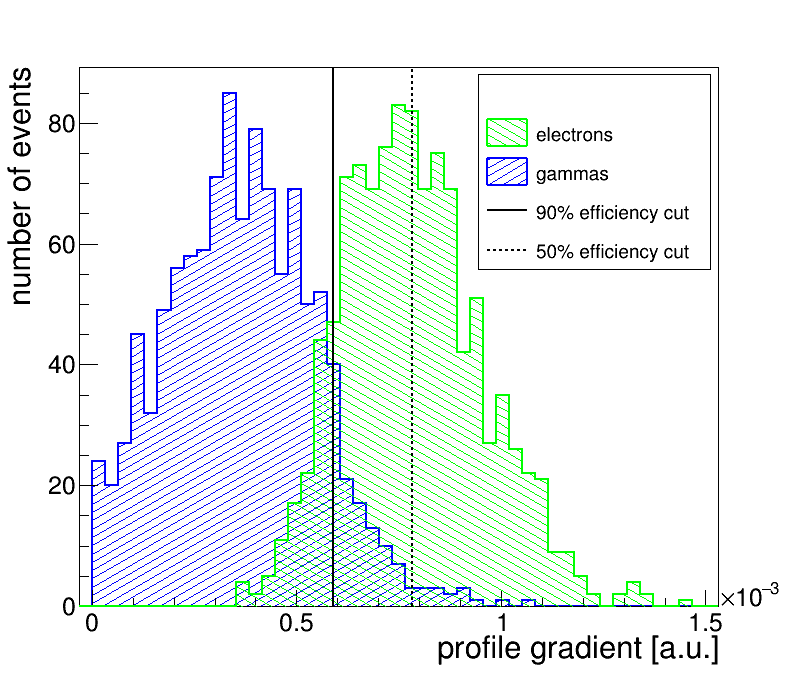}}
\caption{Distribution of the discrimination parameters in all three methods for $e^-$ and $\gamma$ events with visible energies between 1.25\,MeV and 1.75\,MeV. All three methods were used on events with detector radii between 9.5\,m and 10.5\,m.}
\label{fig:app_paramDistribution_egamma}
\end{figure}

\begin{figure}[hbt]
\centering
	\subfigure[Gatti] {\includegraphics[width=4.5cm]{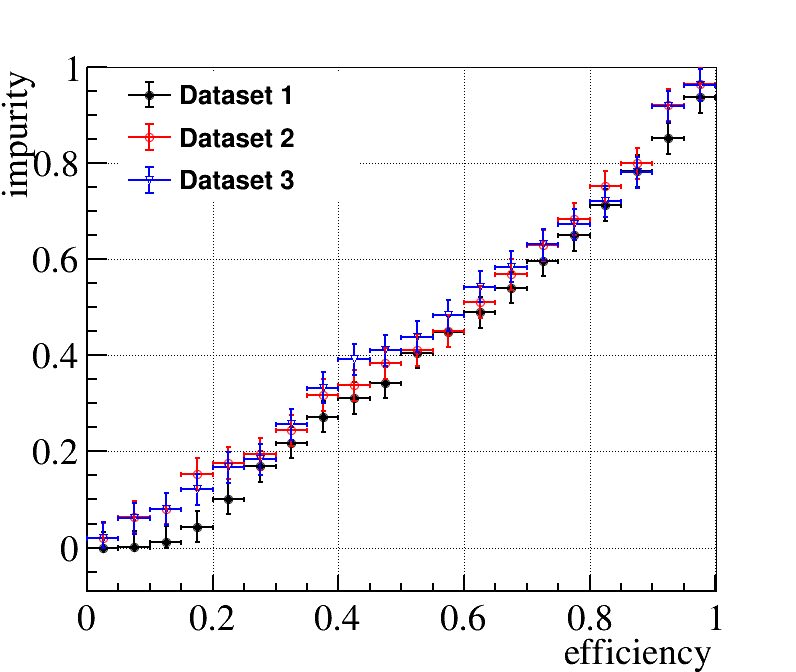}}\quad
	\subfigure[NN] {\includegraphics[width=4.5cm]{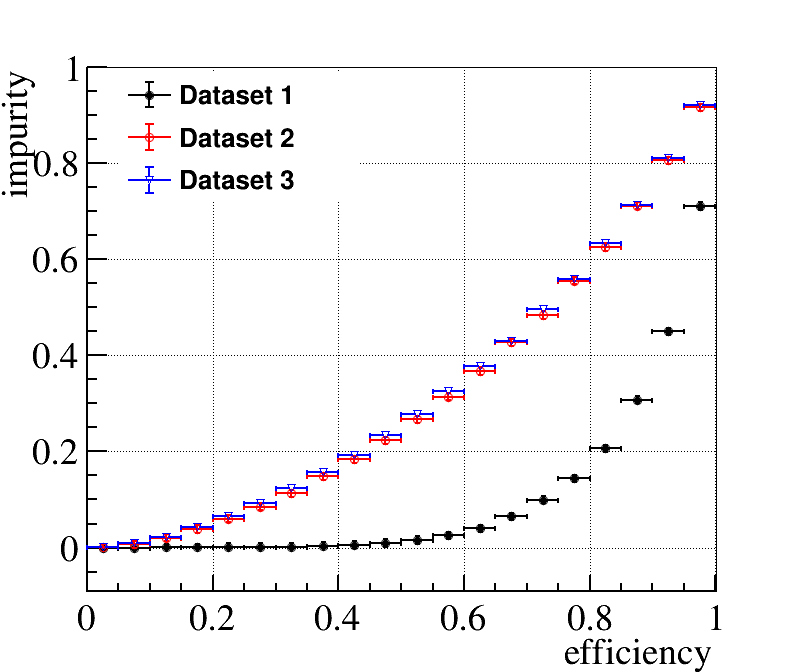}}\quad
	\subfigure[TR] {\includegraphics[width=4.5cm]{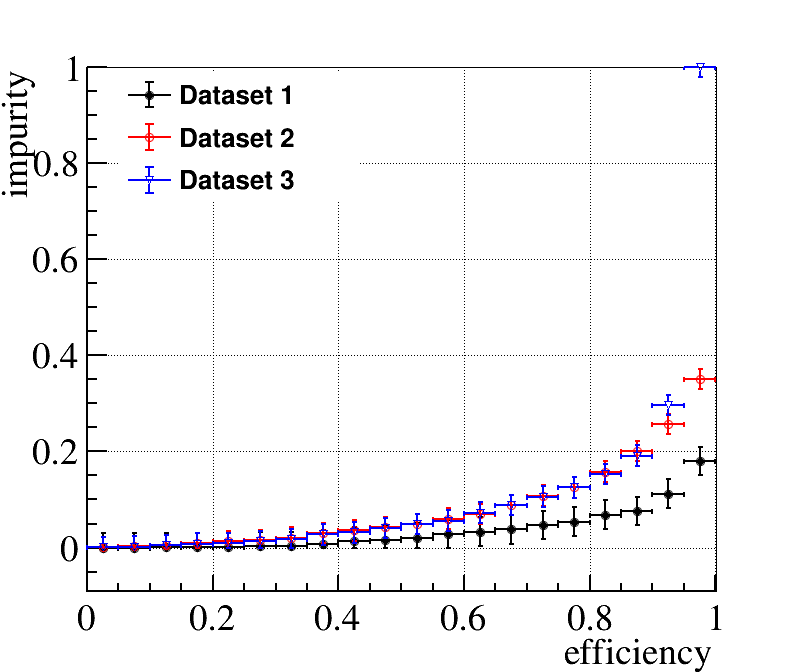}}
\caption{Impurity as a function of efficiency for $e^-/\gamma$ discrimination. The results were obtained for visible energies between 2.0\,MeV and 2.5\,MeV. All three methods were used on events with detector radii between 9.5\,m and 10.5\,m. Dataset 1 was used, i.e.~TTS, vertex smearing, and dark noise were not considered.}
\label{fig:app_egamma_efficiency}
\end{figure}

\clearpage

\begin{figure}[hbt]
	\centering
	\subfigure {\includegraphics[width=4.5cm]{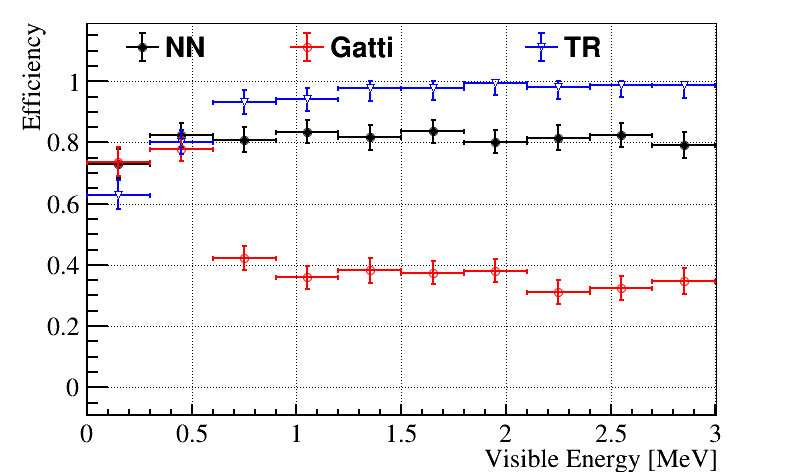}}\quad
	\subfigure {\includegraphics[width=4.5cm]{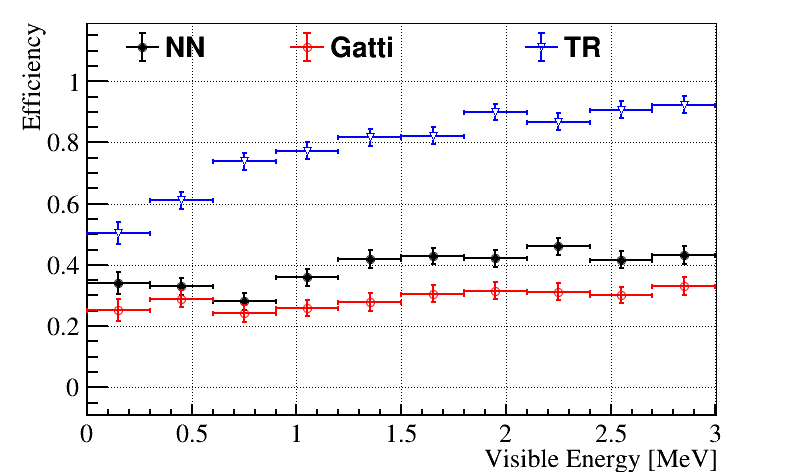}}\quad	
	\subfigure {\includegraphics[width=4.5cm]{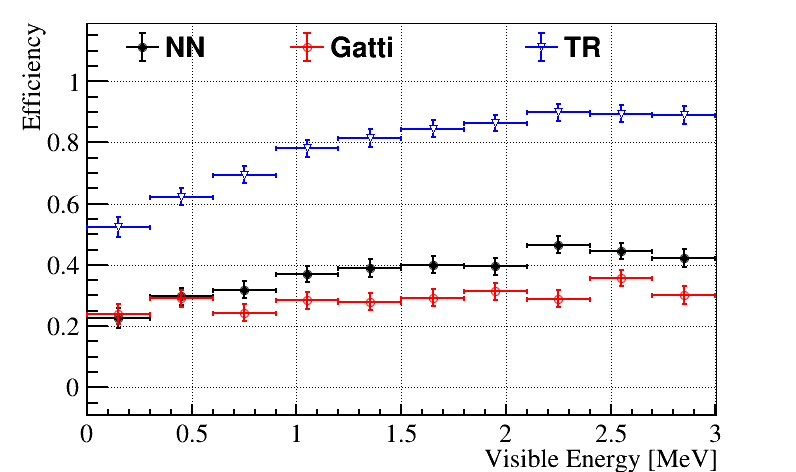}}\\

	\subfigure[Dataset 1]{\includegraphics[width=4.5cm]{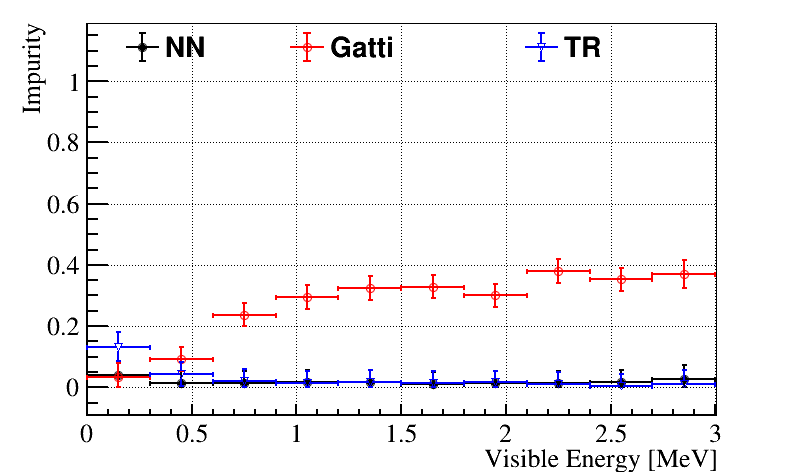}}\quad
	\subfigure[Dataset 2] {\includegraphics[width=4.5cm]{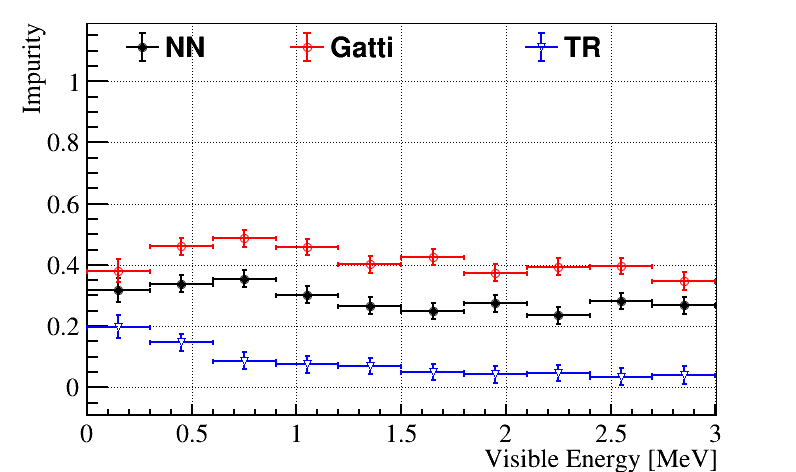}}\quad
	\subfigure[Dataset 3] {\includegraphics[width=4.5cm]{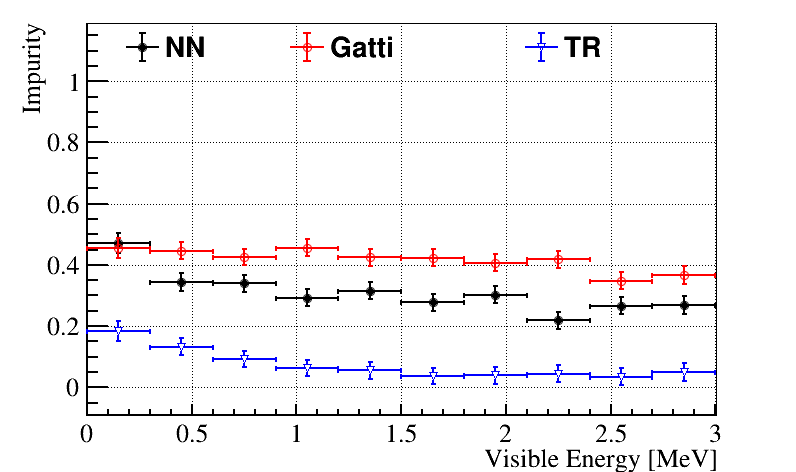}}


	\caption{Energy dependence of the $e^-/\gamma$ discrimination from all three methods. Impurity was obtained at efficiency fixed to 50\%, while efficiency was obtained at impurity fixed to 20\,\%. All three methods were used on events with detector radii between 9.5\,m and 10.5\,m.}
	\label{fig:app_egamma_disc_energy}
\end{figure} 

\begin{figure}[hbt]
	\centering
	\subfigure {\includegraphics[width=4.5cm]{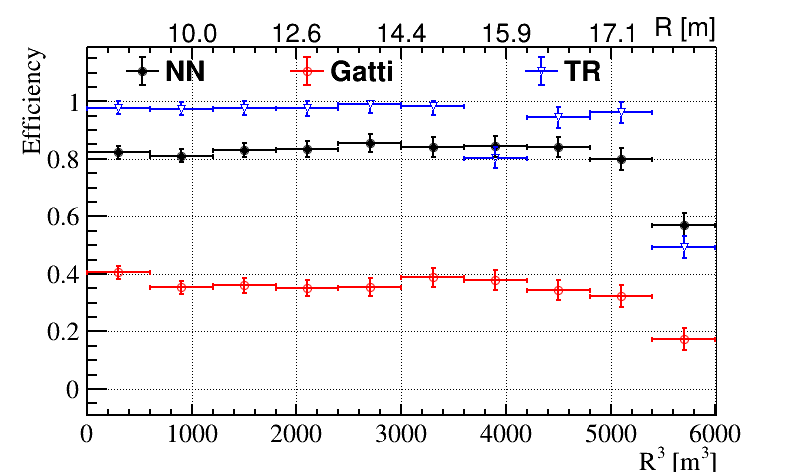}}\quad
	\subfigure {\includegraphics[width=4.5cm]{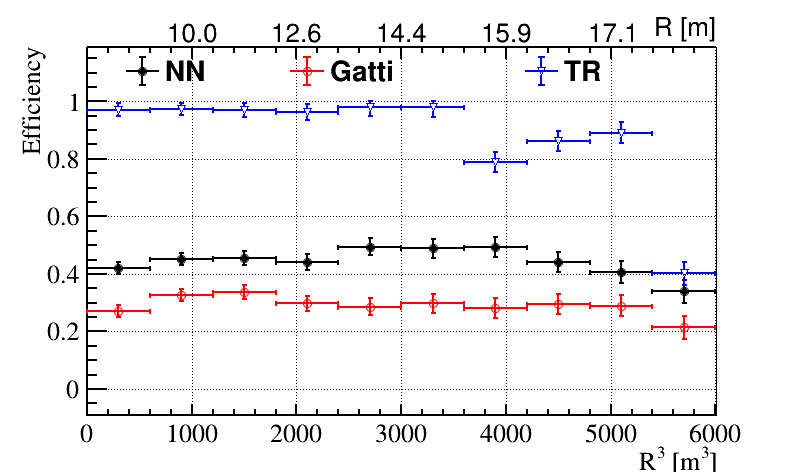}}\quad	
	\subfigure {\includegraphics[width=4.5cm]{resultsShared/egamma_noise_radius_eff.png}}\\

	\subfigure[Dataset 1]{\includegraphics[width=4.5cm]{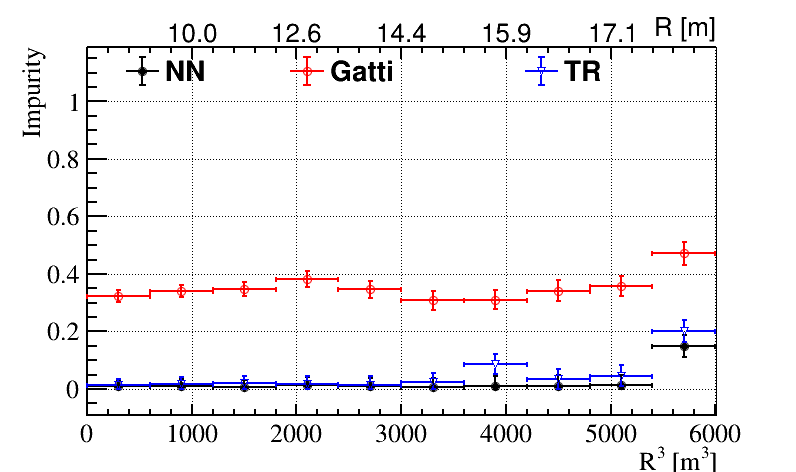}}\quad
	\subfigure[Dataset 2] {\includegraphics[width=4.5cm]{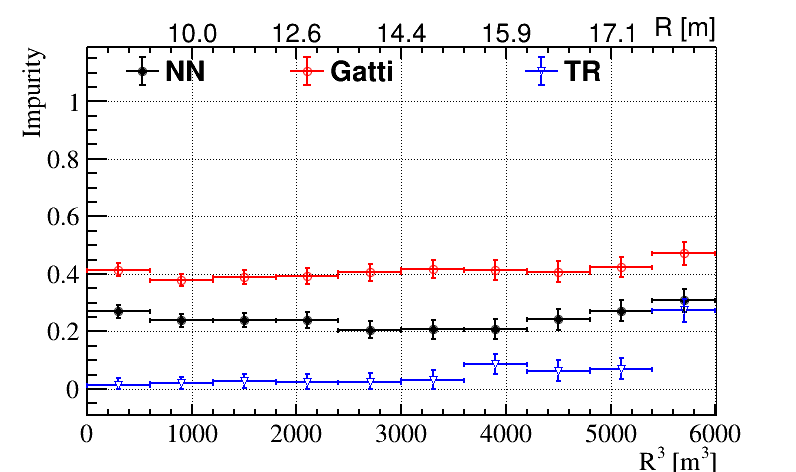}}\quad
	\subfigure[Dataset 3] {\includegraphics[width=4.5cm]{resultsShared/egamma_noise_radius_imp.png}}


	\caption{Radius dependence of the $e^-/\gamma$ discrimination from all three methods for events with visible energies between 2.0\,MeV and 2.5\,MeV. Impurity was obtained at efficiency fixed to 50\%, while efficiency was obtained at impurity fixed to 20\,\%.}
	\label{fig:app_egamma_disc_radius}
\end{figure}

\bibliographystyle{unsrt}
\bibliography{referencesNew}
\end{document}